\documentclass[11pt]{article}

\usepackage{verbatim}
\usepackage{amsmath}
\usepackage{titling}
\usepackage{cancel}
\usepackage{enumitem} % Provides the ability to easily modify the amount of white space in a list
\setlist{noitemsep,topsep=0pt,parsep=0pt} % Modify the amount of white space for all lists

\usepackage[margin=1cm]{caption} % Adds an addition margin on either side of figure captions. Must come before "\usepackage{subfig}" (otherwise there are errors).
\usepackage{subfig}
\usepackage{mathtools} % Adds cases* and dcases envirnoments
\usepackage{graphicx}

\usepackage{tikz}
\usetikzlibrary{arrows,backgrounds,calc,fit,decorations.pathreplacing,decorations.markings,shapes.geometric}

\tikzset{every fit/.append style=text badly centered}

\usepackage{tyson}

\usepackage[bookmarks=true,hypertexnames=false]{hyperref}
\hypersetup{colorlinks=true, citecolor=blue, linkcolor=red, urlcolor=blue}

\newcommand{\Holant}{\operatorname{Holant}}

\newcommand{\holant}[2]{\ensuremath{\Holant\left(#1\mid #2\right)}}

\newcommand{\CSP}{\operatorname{\#CSP}}

{\par\medskip}

\def\borderColor{blue!60}
\def\scale{0.6}
\def\nodeDist{1.4cm}
\tikzstyle{internal} = [draw, fill, shape=circle]
\tikzstyle{external} = [shape=circle]
\tikzstyle{square}   = [draw, fill, rectangle]
\tikzstyle{triangle} = [draw, fill, regular polygon, regular polygon sides=3, inner sep=3pt]
\tikzstyle{pentagon} = [draw, fill, regular polygon, regular polygon sides=5, inner sep=2pt, minimum size=14pt]

\begin{document}
\title{{\bf Complexity Classification of the Eight-Vertex Model}}

\vspace{0.3in}

\author{Jin-Yi Cai\thanks{Computer Sciences Department, University of Wisconsin, Madison, USA.
 {\tt jyc@cs.wisc.edu}}
\and Zhiguo Fu\thanks{School of Mathematics, Jilin University, Changchun, China. {\tt
fuzg@jlu.edu.cn}}}

\date{}
\maketitle

\bibliographystyle{plain}

\begin{abstract}
We prove a complexity dichotomy theorem for the eight-vertex model. For every setting of the parameters of the model, we prove that computing the partition function is either solvable in polynomial time or \#P-hard. The dichotomy criterion is explicit. For tractability, we find some new classes
of problems computable in polynomial time.
For \#P-hardness, we employ M\"{o}bius transformations to prove the
success of interpolations.
 \end{abstract}
\newpage
\setcounter{page}{1}

\section{Introduction}\label{sec:intro}

There are two complementary motivations for this work, one
from physics, and one from the classification program for counting problems
in complexity theory.
From physics, there is a long history in the study of
various elegant models which define partition functions
that capture physical properties.
The eight-vertex model is one such model,
and it generalizes the six-vertex model. From
complexity theory,
we have made substantial progress in classifying counting
problems expressed as  \emph{sum-of-product}
computations in  all three frameworks: graph homomorphisms (GH),
counting constraint satisfaction problems (\#CSP),
and Holant problems. However,
the advances for GH and \#CSP have been far more conclusive
than for Holant problems: On the Boolean domain (where variables take
0-1 values), the  known complexity dichotomy for \#CSP
applies to all complex-valued
 constraint functions which need not be symmetric~\cite{cailuxia-2014},
but currently  the strongest Holant dichotomy without auxiliary
functions can only handle symmetric constraints~\cite{caiguowilliams13}.
%{cai-fu-guo-williams} for planar.
To classify Holant problems without the symmetry assumption,
currently we have to assume the presence of auxiliary functions.
 E.g., assuming all unary functions are present,
called Holant$^*$ problems, we have a dichotomy
that applies to symmetric as well as asymmetric
constraint functions~\cite{cailuxiastar}.
Beckens~\cite{Beckens-arXiv} recently proved an extension to a dichotomy
for Holant$^+$ problems, which assume the presence of
four unary functions including the  pinning functions
{\sc Is-Zero} and {\sc Is-One}
(which set a variable to 0 or 1).
If one only assumes the presence of
the two pinning functions, this is called  the Holant$^c$
problems.  The strongest known  Holant$^c$ dichotomies
are for symmetric complex-valued constraints~\cite{cai-huang-lu}, or
for  real-valued
constraints without symmetry assumption~\cite{cailuxia-2017}.
If one considers what tractable problems emerge
on planar graphs, again we have a full dichotomy
for Pl-\#CSP~\cite{cai:fu:plcsp}, but only for symmetric
constraints concerning Pl-Holant problems~\cite{cai-fu-guo-williams}.
%%
%% also GH Cai-Chen-Lu, general #CSP Cai-Chen, planar: cai-lu-xia, Guo-williams
% Cai-Fu... anything else?  spin systems?  Dyer-Greenhill, Bulatov-Grohe,
%Bulatov, Dyer-Richerby, Cai-Chen-Lu for nonnegative general domain #CSP
% check the Cai-Chen paper for others. Jerrum, Goldberg etc.
%for \#CSP cover for symmetric as well as asymmetric
%constraint functions, while the dichotomy for Holant problems
%on the Boolean domain only applies to symmetric constraints.
There are also several known dichotomies for GH and \#CSP
on domain size greater than $2$~\cite{Dyer-Greenhill,Bulatov-Grohe,Goldberg-Jerrum-Grohe-Thurley,Cai-Chen-Lu-forGH,
1-below-Bulatov-Dalmau,Bulatov,Dyer-Richerby,2-below,Cai-Chen-Lu-for-nonnegative-general-domain-CSP,Cai-Chen-generalCSP},
but very little is known for Holant problems.

% "1-below"
%A. A. Bulatov and V. Dalmau, Towards a dichotomy theorem for the counting constraint satisfaction problem, Inform. and Comput., 205 (2007), pp. 651--678
% "2-below"
%A. Bulatov, M. Dyer, L. A. Goldberg, M. Jalsenius, M. Jerrum, and D. Richerby, The complexity of weighted and unweighted #CSP, J. Comput. System Sci., 78 (2012), pp. 681--688,

Generally speaking, to handle constraint functions that are
not necessarily symmetric seems to be very challenging
for Holant problems.  The
%six-vertex model  and the
eight-vertex model can be
viewed as fundamental building blocks toward a full Holant dichotomy
on the Boolean domain without the symmetry restrictions.
Not only they are small arity cases
in such a theorem, they also present a pathway to overcome
some technical obstacles.
%such as the parity condition.

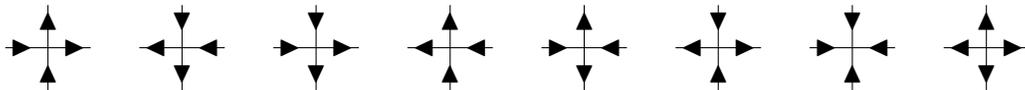
\begin{figure}[h]
\centering
\subfloat{%%%%%%%%1
\begin{tikzpicture}[scale=0.35]
\node [external, scale=0.7] (1) at (0, 2) {};
\node [external, scale=0.7] (2) at (2, 0) {};
\node [external, scale=0.7] (3) at (4, 2) {};
\node [external, scale=0.7] (4) at (2, 4) {};
%\node at (2, 1.2) {$u'$};
\node at (2, 1){$\blacktriangle$};
\node at (2, 3){$\blacktriangle$};
\node at (1, 2){$\blacktriangleright$};
\node at (3, 2){$\blacktriangleright$};
\draw (1)--(3);
\draw (2)--(4);
\end{tikzpicture}
}
\subfloat{%%%%%%%%%%%%%%%%2
\begin{tikzpicture}[scale=0.35]
\node [external, scale=0.7] (1) at (0, 2) {};
\node [external, scale=0.7] (2) at (2, 0) {};
\node [external, scale=0.7] (3) at (4, 2) {};
\node [external, scale=0.7] (4) at (2, 4) {};
%\node at (2, 1.2) {$u'$};
\node at (2, 1){$\blacktriangledown$};
\node at (2, 3){$\blacktriangledown$};
\node at (1, 2){$\blacktriangleleft$};
\node at (3, 2){$\blacktriangleleft$};
\draw (1)--(3);
\draw (2)--(4);
\end{tikzpicture}
}
\subfloat{%%%%%%%%%3
\begin{tikzpicture}[scale=0.35]
\node [external, scale=0.7] (1) at (0, 2) {};
\node [external, scale=0.7] (2) at (2, 0) {};
\node [external, scale=0.7] (3) at (4, 2) {};
\node [external, scale=0.7] (4) at (2, 4) {};
%\node at (2, 1.2) {$u'$};
\node at (2, 1){$\blacktriangledown$};
\node at (2, 3){$\blacktriangledown$};
\node at (1, 2){$\blacktriangleright$};
\node at (3, 2){$\blacktriangleright$};
\draw (1)--(3);
\draw (2)--(4);
\end{tikzpicture}
}
\subfloat{%%%%%%4
\begin{tikzpicture}[scale=0.35]
\node [external, scale=0.7] (1) at (0, 2) {};
\node [external, scale=0.7] (2) at (2, 0) {};
\node [external, scale=0.7] (3) at (4, 2) {};
\node [external, scale=0.7] (4) at (2, 4) {};
%\node at (2, 1.2) {$u'$};
\node at (2, 1){$\blacktriangle$};
\node at (2, 3){$\blacktriangle$};
\node at (1, 2){$\blacktriangleleft$};
\node at (3, 2){$\blacktriangleleft$};
\draw (1)--(3);
\draw (2)--(4);
\end{tikzpicture}
}
\subfloat{%%%%%%%%%%5
\begin{tikzpicture}[scale=0.35]
\node [external, scale=0.7] (1) at (0, 2) {};
\node [external, scale=0.7] (2) at (2, 0) {};
\node [external, scale=0.7] (3) at (4, 2) {};
\node [external, scale=0.7] (4) at (2, 4) {};
%\node at (2, 1.2) {$u'$};
\node at (2, 1){$\blacktriangledown$};
\node at (2, 3){$\blacktriangle$};
\node at (1, 2){$\blacktriangleright$};
\node at (3, 2){$\blacktriangleleft$};
\draw (1)--(3);
\draw (2)--(4);
\end{tikzpicture}
}
\subfloat{%%%%%%%%%6
\begin{tikzpicture}[scale=0.35]
\node [external, scale=0.7] (1) at (0, 2) {};
\node [external, scale=0.7] (2) at (2, 0) {};
\node [external, scale=0.7] (3) at (4, 2) {};
\node [external, scale=0.7] (4) at (2, 4) {};
%\node at (2, 1.2) {$u'$};
\node at (2, 1){$\blacktriangle$};
\node at (2, 3){$\blacktriangledown$};
\node at (1, 2){$\blacktriangleleft$};
\node at (3, 2){$\blacktriangleright$};
\draw (1)--(3);
\draw (2)--(4);
\end{tikzpicture}
}
\subfloat{%%%%%%%%%%%7
\begin{tikzpicture}[scale=0.35]
\node [external, scale=0.7] (1) at (0, 2) {};
\node [external, scale=0.7] (2) at (2, 0) {};
\node [external, scale=0.7] (3) at (4, 2) {};
\node [external, scale=0.7] (4) at (2, 4) {};
%\node at (2, 1.2) {$u'$};
\node at (2, 1){$\blacktriangle$};
\node at (2, 3){$\blacktriangledown$};
\node at (1, 2){$\blacktriangleright$};
\node at (3, 2){$\blacktriangleleft$};
\draw (1)--(3);
\draw (2)--(4);
\end{tikzpicture}
}
\subfloat{%%%%%%%%%%8
\begin{tikzpicture}[scale=0.35]
\node [external, scale=0.7] (1) at (0, 2) {};
\node [external, scale=0.7] (2) at (2, 0) {};
\node [external, scale=0.7] (3) at (4, 2) {};
\node [external, scale=0.7] (4) at (2, 4) {};
%\node at (2, 1.2) {$u'$};
\node at (2, 1){$\blacktriangledown$};
\node at (2, 3){$\blacktriangle$};
\node at (1, 2){$\blacktriangleleft$};
\node at (3, 2){$\blacktriangleright$};
\draw (1)--(3);
\draw (2)--(4);
\end{tikzpicture}
}
\caption{\scriptsize{Valid configurations of the eight-vertex model.}}
\label{figure:8-vertexxx}
\end{figure}

In physics, the eight-vertex model  is
%considers a large number of oxygen and hydrogen atoms, and is
a generalization of
the six-vertex model, including models for water ice,
potassium dihydrogen phosphate KH$_2$PO$_4$ (KDP model of a ferroelectric)
and the Rys $F$ model of an antiferroelectric.
One can mathematically describe the  eight-vertex model as an orientation
problem on 4-regular graphs: Given a 4-regular graph $G = (V, E)$,
an orientation is an assignment of a direction to every $e \in E$.
An orientation is valid for the eight-vertex model
iff at every vertex the in-degree (and out-degree) is even.
This generalizes the six-vertex model where the in-degree (and out-degree)
is two and thus the orientation
is \emph{Eulerian}.  One can think of the valid configurations
in a eight-vertex model
as Eulerian orientations with possible sources and sinks.
The valid local configurations are illustrated in Figure~\ref{figure:8-vertexxx}.
The  energy $E$  of the system is determined by
eight parameters $\epsilon_1, \epsilon_2, \ldots, \epsilon_8$
associated with each type of the local configuration,
and $w_j = \exp\left(-\frac{\epsilon_j}{k_B T}\right)$
is called the  Boltzmann weight ($k_B$
 is Boltzmann's constant, and $T$ is the system's temperature).
If there are  $n_i$ sites in  local configuration type $i$,
 then
$E = n_1 \epsilon_1 + \ldots +  n_8  \epsilon_8$
is the total energy,
and the  partition function is $Z_{\rm Eight} = \sum e^{-E/k_BT}
= \sum \prod w_j^{n_j}$,
where the sum is  over all valid configurations.
This is a sum-of-product computation.
In our more general
definition (see Subsection~\ref{trac}) the 8 possible weights $w_j$
can be zero, and thus the six-vertex model is the special case
with $w_7=w_8=0$, disallowing the configurations 7 and 8.

Compared to the six-vertex model,  there are more
non-trivial tractable problems. Partly this is
 because the support of a
constraint function in the eight-vertex model can be an affine subspace
of dimension 3 (over ${\mathbb Z}_2$).
Some tractable problems are only revealed to be
so after surprising holographic transformations
$\left[\begin{smallmatrix}
1 & \frak{i} \\
\frak{i} & 1
\end{smallmatrix}\right]$
or
$\left[\begin{smallmatrix}
1 & 0 \\
0 & \sqrt[4]{\frak{i}}
\end{smallmatrix}\right]$. No previously known tractable classes
required such transformations.
More  tractable problems usually mean that
it is more challenging to prove a dichotomy.
Such a theorem says that there are no other tractable problems
beyond the ones already discovered (if \#P does not
collapse to P.)

We discover a connection for a class of
8-vertex models with \#CSP$^2$ problems,
which are a variant of  \#CSP
where every variable appears an even number of times.
%This is forced by the support structure of the
%constraint function in the 8-vertex model,
%and being an  \#CSP$^2$ problem there are more
Compared to \#CSP, there are more tractable problems
for \#CSP$^2$.  A crucial ingredient in our proof
is a recent \#CSP$^2$ dichotomy~\cite{cailuxia-2017} that is valid for
asymmetric signatures.
Our new tractable families for the 8-vertex model
also give new tractable families for the
so-called 2,4-spin Ising model on the lattice graph,
where the $(+/-)$ spins are on square faces, and
local interactions are among horizontal, vertical, two diagonals, and
all 4 neighbors.

A new contribution of this work is to use
M\"{o}bius transformations
 ${\mathfrak z} \mapsto
\frac{a {\mathfrak z} + b}{c {\mathfrak z} + d}$
to prove \#P-hardness.
Typically to prove some problem \#P-hard by
interpolation, we want to prove that certain
quantities (such as eigenvalues)
are not roots of unity, lest the iteration repeat after
a bounded number of steps.
We usually establish this property by showing that
we can produce these quantities of norm $\ne 1$.
However in this paper, there are settings where this is
impossible. In this case we prove that
the constraint functions define certain M\"{o}bius transformations
that map the unit circle to
 unit circle on $\mathbb{C}$. By exploiting the mapping
properties we can
obtain a suitable M\"{o}bius transformation which generates a group of
infinite order. Hence even though they only produce
quantities of complex norm 1, they nevertheless can be guaranteed
not to repeat.
This allows us to show that
our interpolation proof succeeds.

\section{Preliminaries}\label{trac}
In the present paper, $\frak{i}$ denotes a square root of $-1$, i.e., $\frak{i}^2=-1$.
$\alpha$ denotes a square root of $\frak{i}$, i.e., $\alpha^2=\frak{i}$.
Let $\mathbb{N} = \{0,1,2,\ldots\}$,
$H=\frac{1}{\sqrt{2}}
\left[\begin{smallmatrix}
1 & 1 \\
 1 & -1
\end{smallmatrix}\right]
$, $Z=\frac{1}{\sqrt{2}} \left[\begin{smallmatrix}
1 & 1 \\
\mathfrak{i} & -\mathfrak{i}
\end{smallmatrix}\right]$.

\subsection{Background}
A constraint function $f$ of arity $k$
is a map $\{0,1\}^k  \rightarrow \mathbb{C}$.
Fix a set of constraint functions $\mathcal{F}$. A signature grid
$\Omega=(G, \pi)$
 is a tuple, where $G = (V,E)$
is a graph, $\pi$ labels each $v\in V$ with a function
$f_v\in\mathcal{F}$ of arity ${\operatorname{deg}(v)}$,
and the incident edges
$E(v)$ at $v$ with input variables of $f_v$.
% Each
%$f_v$ maps
%$\{0, 1\}^{\operatorname{deg}(v)}$ to $\mathbb{C}$.
 We consider all 0-1 edge assignments $\sigma$,
each gives an evaluation
$\prod_{v\in V}f_v(\sigma|_{E(v)})$, where $\sigma|_{E(v)}$
denotes the restriction of $\sigma$ to $E(v)$. The counting problem on the instance $\Omega$ is to compute
\[\operatorname{Holant}_{\Omega}(\mathcal{F})=\sum_{\sigma:E\rightarrow\{0, 1\}}\prod_{v\in V}f_v(\sigma|_{E(v)}).\]
The Holant problem parameterized by the set $\mathcal{F}$ is denoted by Holant$(\mathcal{F})$.
Replacing $f$ by $c\cdot f$ for any $c \not =0$
only changes the value $\operatorname{Holant}_{\Omega}(\mathcal{F})$
by $c^n$ where $n$ is the number of times $f$ appears in $\Omega$.
Thus it does not change its complexity, therefore
we can ignore such constant factors.
We also write  Holant$(\mathcal{F}, f)$
for Holant$(\mathcal{F} \cup \{f\})$.
We use Holant$(\mathcal{F}|\mathcal{G})$ to denote the Holant problem over signature grids with a bipartite graph $H = (U,V,E)$,
where each vertex in $U$ or $V$ is assigned a signature in $\mathcal{F}$ or $\mathcal{G}$
respectively.

Given an instance $\Omega=(G, \pi)$ of Holant$(\mathcal{F})$, we add a middle point on each edge as a new vertex to $G$, then each edge becomes a path of length two through the new vertex. Extend $\pi$ to label a
binary
function $g$ to each new vertex. This gives a bipartite Holant problem Holant$(g\mid \mathcal{F})$. It is obvious  that Holant$(=_2\mid \mathcal{F})$ is
the same problem as Holant$(\mathcal{F})$.

A constraint function is also called a signature.
A function $f$ of arity $k$ can be represented by listing its values in lexicographical order as in a truth table, which
is a vector in
$\mathbb{C}^{2^{k}}$, or as a tensor in $(\mathbb{C}^2)^{\otimes k}$, or as a matrix in $\mathbb{C}^{2^{k_1}}\times \mathbb{C}^{2^{k_2}}$
if we partition the $k$ variables to two parts,  where $k_1+k_2=k$.
A function is symmetric if its value depends only
on the Hamming weight of its input.
A symmetric function $f$ on $k$ Boolean variables can
be expressed as
$[f_0, f_1, \ldots, f_k]$,
where $f_w$ is the value of $f$ on inputs of Hamming weight $w$.
For example, $(=_k)$ is the {\sc Equality} signature $[1, 0, \ldots, 0, 1]$
(with $k-1$ 0's) of arity $k$.
We use $\neq_2$ to denote binary {\sc Disequality} function $[0,1,0]$. Note that
$N=
\left[\begin{smallmatrix}
0 & 1 \\
 1 & 0
\end{smallmatrix}\right]
\otimes
\left[\begin{smallmatrix}
0 & 1 \\
 1 & 0
\end{smallmatrix}\right]
=
\left[\begin{smallmatrix}
 0 & 0 & 0 & 1 \\
 0 & 0 & 1 & 0 \\
 0 & 1 & 0 & 0 \\
 1 & 0 & 0 & 0 \\
\end{smallmatrix}\right]$, i.e.,
$N$ is the double {\sc Disequality} in parallel,
which is the function of connecting two pairs of edges by $(\not =_2)$.
The support of a function $f$ is the set of inputs on which $f$ is nonzero.
%We can treat the support of $f$ as a relation or a $0,1$-valued function.
%%% JYC not used. i think

Counting constraint satisfaction problems (\#CSP)
can be defined as a special case of Holant problems.
An instance of $\CSP(\mathcal{F})$ is presented
as a bipartite graph.
There is one node for each variable and for each occurrence
of constraint functions respectively.
Connect a constraint node to  a variable node if the
variable appears in that occurrence
of constraint, with a labeling on the edges
for the order of these variables.
This bipartite graph is also known as the \emph{constraint graph}.
If we attach each variable node with an \textsc{Equality} function,
and consider every edge as a variable, then
the Holant problem on this bipartite graph is just this \#CSP.
Thus
$\CSP(\mathcal{F}) \equiv^p_T \holant{\mathcal{EQ}}{\mathcal{F}}$,
where $\mathcal{EQ} = \{{=}_1, {=}_2, {=}_3, \dotsc\}$ is the set of \textsc{Equality} signatures of all arities.

For any positive integer $d$,
the problem $\CSP^d(\mathcal{F})$ is the same as $\CSP(\mathcal{F})$ except that every variable appears a multiple of $d$ times.
Thus,
$\#\operatorname{CSP}^d(\mathcal{F}) \equiv^p_T \operatorname{Holant}(\mathcal{EQ}_d|{\mathcal{F}})$,
where $\mathcal{EQ}_d = \{{=}_d, {=}_{2 d}, {=}_{3 d}, \dotsc\}$ is the set of \textsc{Equality} signatures of arities that are multiples of $d$.
For $d=1$, we have just $\CSP$ problems. For $d=2$,
these are $\CSP$ problems where every variable appears an even number of times.

A spin system on $G = (V, E)$ has a variable for every $v \in V$
and a binary function $g$ for every edge $e \in E$.
The partition function is $\sum_{\sigma: V \rightarrow\{0, 1\}}
\prod_{(u, v) \in E} g(\sigma(u), \sigma(v))$.
Spin systems are special cases of \#CSP$(\mathcal{F})$
where $\mathcal{F}$ consists of a single binary function.

%In this paper we deal with orientation problems, which
%can be expressed as bipartite Holant problems Holant$(\ne_2 \mid \mathcal{F})$.

For $T \in {\rm GL}_2({\mathbb{C}})$
 and a signature $f$ of arity $n$, written as
a column vector  $f \in \mathbb{C}^{2^n}$, we denote by
$T^{-1}f = (T^{-1})^{\otimes n} f$ the transformed signature.
  For a signature set $\mathcal{F}$,
define $T^{-1} \mathcal{F} = \{T^{-1}f \mid  f \in \mathcal{F}\}$.
For signatures written as
 row vectors we define $\mathcal{F} T$ similarly.
%Whenever we write $T^{-1} f$ or $T^{-1} \mathcal{F}$,
%we view the signatures as column vectors;
%similarly for $f T$ or $\mathcal{F} T$ as row vectors.
%
%
%Let $T \in {\rm GL}_2({\mathbb{C}})$.
The holographic transformation defined by $T$ is the following operation:
given a signature grid $\Omega = (H, \pi)$ of $\holant{\mathcal{F}}{\mathcal{G}}$,
for the same bipartite graph $H$,
we get a new signature grid $\Omega' = (H, \pi')$ of $\holant{\mathcal{F} T}{T^{-1} \mathcal{G}}$ by replacing each signature in
$\mathcal{F}$ or $\mathcal{G}$ with the corresponding signature in $\mathcal{F} T$ or $T^{-1} \mathcal{G}$.

A signature $f$ of arity 4 has the signature matrix
\[M = M_{x_1x_2, x_3x_4}(f)=\left[\begin{smallmatrix}
f_{0000} & f_{0001} & f_{0010} & f_{0011}\\
f_{0100} & f_{0101} & f_{0110} & f_{0111}\\
f_{1000} & f_{1001} & f_{1010} & f_{1011}\\
f_{1100} & f_{1101} & f_{1110} & f_{1111}
\end{smallmatrix}\right]
.\]
We will use $M(f)$ to denote $M_{x_1x_2, x_3x_4}(f)$.
If $\{i,j,k,\ell\}$ is a permutation of $\{1,2,3,4\}$,
then the $4 \times 4$ matrix $M_{x_ix_j, x_kx_{\ell}}(f)$ lists the 16 values
 with row  index $x_ix_j \in\{0, 1\}^2$
and column index
$x_kx_{\ell} \in\{0, 1\}^2$ in lexicographic order.
A binary signature $g$ has the signature matrix
$M(g)=\left[\begin{smallmatrix}
 g_{00} & g_{01}\\
g_{10} & g_{11}
\end{smallmatrix}\right]
.$

The eight-vertex model is the Holant problem Holant$(\neq_2 \mid f)$ where
$f$ is a 4-ary signature with the signature matrix
$M(f)=\left[\begin{smallmatrix}
a & 0 & 0 & b\\
0 & c & d & 0\\
0 & w & z & 0\\
y & 0 & 0 & x\\
\end{smallmatrix}\right]$. This
%problem set
 corresponds to a set of (non-bipartite)
 Holant problems by a holographic reduction \cite{Val08}.  The matrix form of $(\neq_2)$ is $\left[\begin{smallmatrix}
0 & 1 \\
 1 & 0
\end{smallmatrix}\right]=Z^{\tt T}Z$.
Under a holographic transformation by $Z^{-1}$, Holant$(\neq_2 \mid f)$ becomes Holant$(=_2 \mid Z^{\otimes 4}f)$, where $Z^{\otimes 4}f$ is a column vector $f$ multiplied by the matrix tensor power $Z^{\otimes 4}$.
%Function $Z^{\otimes 4} f$ is symmetric iff $f$ is symmetric.
%Although we focus on
The bipartite Holant problems of the form Holant$(\neq_2 \mid f)$
%, they
naturally correspond to the non-bipartite Holant problems Holant$(Z^{\otimes 4}f)$.
In general $f$ and $Z^{\otimes 4}f$ are non-symmetric functions.
%\textcolor{red}{In the following of this paper, we only use diagonal holographic transformation.}
We often use the fact that for any diagonal $D =
\left[\begin{smallmatrix}
u & 0 \\
0 & v
\end{smallmatrix}\right]  \in {\rm GL}_2({\mathbb{C}})$,
the transformed signature $(\neq_2)D^{\otimes 2}$ is
just $(\neq_2)$ up to a nonzero factor $uv$.

%We also use a $4 \times 4$ matrix to denote a signature of arity 4,

%In the present paper,

%If $f$ and $g$ have signature matrices
% $ M_{x_ix_j, x_k x_{\ell}}(f)$ and $M_{x_{s}x_{t}, x_{u}x_{v}}(g)$,
%by connecting $x_k$ to $x_{s}$, $x_{\ell}$ to $x_t$,
%both with  {\sc Disequality}  $(\not =_2)$, we get a signature
%of arity 4 with the signature matrix
%%$M_{x_ix_j, x_kx_{\ell}}(f)N M_{x_{s}x_{t}, x_{u}x_{v}}(g)$,
%$M_{x_ix_j, x_k x_{\ell}}(f) N M_{x_{s}x_{t}, x_{u}x_{v}}(g)$
%by matrix product
%with row index $x_ix_j$ and column index $x_{u}x_{v}$.

\paragraph{Symmetry among the three inner pairs}
We call $(a, x)$ the outer pair of $f$, $ (b, y), (c, z), (d, w)$ the inner pairs of $f$,
$\left[\begin{smallmatrix}
a & b\\
y & x\\
\end{smallmatrix}\right]$ the outer matrix of $f$ and
$\left[\begin{smallmatrix}
 c & d\\
 w & z
\end{smallmatrix}\right]$ the inner matrix of $f$.
Denote the 3 inner pairs of
 ordered complementary strings
%%%%%%%%%%JYC, since in lm 4.1 we use alpha ...
%for actual entry values. i change them. al--> la, be --> mu , gamma --> nu
%%%%%%%%%
% by $\alpha = 0011, \overline{\alpha} = 1100$,
%$\beta = 0110, \overline{\beta} = 1001$, and $\gamma = 0101, \overline{\gamma}
%= 1010$.
%The support of $f$ is the union of the pairs $(\alpha, \overline{\alpha})$,
%$(\beta, \overline{\beta})$ and $(\gamma, \overline{\gamma})$,
%on which $f$ has values
%$(a,x), (b,y)$ and $(c,z)$.
%If $f$ has the same value in  a pair,
% say $a=x$ on $\alpha$ and $\overline{\alpha}$,
% we say it is a twin.
%%%%%%%%%%%%%%%%%%%%%%
 by $\lambda = 0011, \overline{\lambda} = 1100$,
$\mu = 0110, \overline{\mu} = 1001$, and $\nu = 0101, \overline{\nu}
= 1010$.
The permutation group $S_4$ on $\{x_1, x_2, x_3, x_4\}$
induces a group action on $\{s \in \{0, 1\}^4
\mid {\rm wt}(s) =2\}$ of size 6.
This is a faithful representation of $S_4$ in $S_6$.
% Every nontrivial $\sigma \in S_4$
%induces a nontrivial. For, if j=sigma(i) \not =i, then there is
% a string s: s_i =0, s_j = 1, rest have equal wt, so total wt 2.
%%and sigma(s) \not = s.
Since the action of $S_4$ preserves complementary pairs,
%strings,
this group action has nontrivial blocks of imprimitivity,
%namely $\{\{0011,1100\}, \{0110,1001\}, \{0101,1010\}\}$.
namely $\{A,B,C\} =
\{\{\lambda, \overline{\lambda}\},
\{\mu, \overline{\mu}\}, \{\nu, \overline{\nu}\}\}$.
The action on the blocks is a homomorphism of $S_4$ onto
$S_3$, i.e., we can
permute the blocks arbitrarily by permuting the variables $\{x_1, x_2, x_3, x_4\},$
 with kernel $K = \{1, (12)(34), (13)(24), (14)(23)\}$.
In particular one can calculate that the subgroup
$S_{\{2,3,4\}} = \{1, (23), (34), (24), (243), (234)\}$
maps onto $\{1, (AC), (BC), (AB), (ABC), (ACB)\}$.
%%% actually even more explicit:
% 1, (ac)(a'c'), (bc)(b'c'), (ab)(a'b'), (abc)(a'b'c'), (acb)(a'c'b')
%To summerize,
By a permutation from $S_4$,
we may permute the matrix $M(f)$
by any permutation on the values
$\{b, c, d\}$ with the corresponding permutation on $\{y, z, w\}$,
and moreover we can further flip
%within
an even number
of pairs $(b, y)$, $(c, z)$
and $(d, w)$.
In particular, we can
arbitrarily reorder the  three rows in
$\left[\begin{smallmatrix}
b & y \\
c & z \\
d & w
\end{smallmatrix}\right]$, and
we can also reverse the order of
 arbitrary two rows together.
 In the following, when we say by the symmetry of $\{(b, y), (c, z), (d, w)\}$,
 it means this group action.

Let $f$ be a 4-ary signature with the signature matrix
$M(f)=
\left[\begin{smallmatrix}
f_{0000} & 0 & 0 & f_{0011}\\
0 & f_{0101} & f_{0110} & 0\\
0 & f_{1001} & f_{1010} & 0\\
f_{1100} & 0 & 0 & f_{1111}
\end{smallmatrix}\right]$.
When we permute the variables of $f$,
 the entries of the signature matrix will
move correspondingly. This is a group action of $S_4$ on the set
of $4 \times 4$ signature matrices.
For any permutation $\sigma \in S_4$,
we denote by $M^{{\sf R}_{\sigma}}(f)$
the permuted matrix obtained from $M(f) = M_{x_1x_2,x_3x_4}(f)$
by applying the  permutation $\sigma$ on the variables
$x_1,x_2,x_3,x_4$.
For example, $M^{{\sf R}_{(12)}}(f)$ is  obtained from $M(f)$
by exchanging $x_1$ and $x_2$,
and corresponds to switching the middle two rows of $M(f)$.
This results in  the matrix
$M_{x_2x_1, x_3x_4}(f)=
\left[\begin{smallmatrix}
f_{0000} & 0 & 0 & f_{0011}\\
0 & f_{1001} & f_{1010} & 0\\
0 & f_{0101} & f_{0110} & 0\\
f_{1100} & 0 & 0 & f_{1111}
\end{smallmatrix}\right]$.
Some frequently used simple operations ${\sf R}_{\sigma}$ are
illustrated in Figure~\ref{fig:permutation}.
These include:
\begin{itemize}
\item
${\sf R}_{(12)}$ which switches the middle two rows;
\item
${\sf R}_{(34)}$ which switches the middle two columns;
\item
${\sf R}_{(23)}$ which effects the movement of entries in
Figure~\ref{fig:permutation}(c);
\item
${\sf R}_{(24)}$ which effects the movement of entries in
Figure~\ref{fig:permutation}(d); and
\item
${\sf R}_{\sf T} = {\sf R}_{(13)(24)}$ which is taking the transpose.
\end{itemize}
Note that as a group action,
these operations can be applied in sequence, corresponding to group
multiplication (from left to right) in $S_4$.
For example, corresponding to the permutation \[(12)(34)(13)(24)=(14)(23)
=
\left(\begin{matrix}
1 & 2 & 3 & 4\\
4 & 3 & 2 & 1
\end{matrix}\right),\] we have
$M^{\sf R_{(12)}\sf R_{(34)}\sf R_{T}}(f)
 = M_{x_4x_3, x_2x_1}(f).$

%The movements of the entries corresponding to these operations are illustrated
%in Figure~\ref{fig:permutation}.
%For a particular operation $\sf R$, we denote by $M^{\sf R}(f)$ the respecting
%signature matrix.
%Note that these operations can be applied in sequence, corresponding to group
%multiplication (from left to right) in $S_4$.
%For example, corresponding to the permutation $(12)(34)(13)(14)=(14)(23)$,
%$M^{\sf R_{12}\sf R_{34}\sf R_{T}}(f)$ results in 
%the matrix $M_{x_4x_3, x_2x_1}(f)$.

 \input{permutation}

 \begin{definition}\label{redu-defi}
 A 4-ary signature $f$ is redundant iff in its  4 by 4 signature matrix
  the middle two rows are identical and the
 middle two columns are identical.
 We call $\left[\begin{smallmatrix}
f_{0000} & f_{0001} & f_{0011}\\
f_{0100} & f_{0101} & f_{0111}\\
f_{1100} & f_{1101} & f_{1111}
\end{smallmatrix}\right]$ the compressed signature matrix  of $f$.
 \end{definition}
\begin{theorem}\cite{caiguowilliams13}\label{redundant}
If $f$ is a redundant signature and its compressed signature matrix has full rank,
 then $\operatorname{Holant}(\neq_2|f)$ is \#$\operatorname{P}$-hard.
\end{theorem}

%%%%%%%%%%%%%%%%%%%%%%%%%%%%%%%%%%%%%%%%%%%%%%%%%%%%%%%%%%%%%%%%%%%%%%%%%%%%%%%%%%%%%%%%%%%%%%%%%%%%%%%%%%%%%%%%%%%%%%%%%%%%%%%%%%%%%%%%%%%%%%%%%%%%%%%%%%%%%%%
\subsection{Gadget Construction}
One basic notion used throughout the paper is realization.
We say a signature $f$ is \emph{realizable} or \emph{constructible} from a signature set $\mathcal{F}$
if there is a gadget with some dangling edges such that each vertex is assigned a signature from $\mathcal{F}$,
and the resulting graph,
when viewed as a black-box signature with inputs on the dangling edges,
is exactly $f$.
If $f$ is realizable from a set $\mathcal{F}$,
then we can freely add $f$ into $\mathcal{F}$ while preserving the complexity.
%add the figure to f-gate
\begin{figure}[t]
 \centering
 \begin{tikzpicture}[scale=\scale,transform shape,node distance=\nodeDist,semithick]
  \node[external]  (0)                     {};
  \node[internal]  (1) [below right of=0]  {};
  \node[external]  (2) [below left  of=1]  {};
  \node[internal]  (3) [above       of=1]  {};
  \node[internal]  (4) [right       of=3]  {};
  \node[internal]  (5) [below       of=4]  {};
  \node[internal]  (6) [below right of=5]  {};
  \node[internal]  (7) [right       of=6]  {};
  \node[internal]  (8) [below       of=6]  {};
  \node[internal]  (9) [below       of=8]  {};
  \node[internal] (10) [right       of=9]  {};
  \node[internal] (11) [above right of=6]  {};
  \node[internal] (12) [below left  of=8]  {};
  \node[internal] (13) [left        of=8]  {};
  \node[internal] (14) [below left  of=13] {};
  \node[external] (15) [left        of=14] {};
  \node[internal] (16) [below left  of=5]  {};
  \path let
         \p1 = (15),
         \p2 = (0)
        in
         node[external] (17) at (\x1, \y2) {};
  \path let
         \p1 = (15),
         \p2 = (2)
        in
         node[external] (18) at (\x1, \y2) {};
  \node[external] (19) [right of=7]  {};
  \node[external] (20) [right of=10] {};
  \path (1) edge                             (5)
            edge[bend left]                 (11)
            edge[bend right]                (13)
            edge node[near start] (e1) {}   (17)
            edge node[near start] (e2) {}   (18)
        (3) edge                             (4)
        (4) edge[out=-45,in=45]              (8)
        (5) edge[bend right, looseness=0.5] (13)
            edge[bend right, looseness=0.5]  (6)
        (6) edge[bend left]                  (8)
            edge[bend left]                  (7)
            edge[bend left]                 (14)
        (7) edge node[near start] (e3) {}   (19)
       (10) edge[bend right, looseness=0.5] (12)
            edge[bend left,  looseness=0.5] (11)
            edge node[near start] (e4) {}   (20)
       (12) edge[bend left]                 (16)
       (14) edge node[near start] (e5) {}   (15)
            edge[bend right]                (12)
       (16) edge[bend left,  looseness=0.5]  (9)
            edge[bend right, looseness=0.5]  (3);
  \begin{pgfonlayer}{background}
%    \node[draw=\borderColor,thick,rounded corners,fit = (3) (4) (9) (e1) (e2) (e3) (e4) (e5)] {};
   \node[draw=\borderColor,thick,rounded corners,fit = (3) (4) (9) (e1) (e2) (e3) (e4) (e5),inner sep=0pt,transform shape=false] {};
  \end{pgfonlayer}
 \end{tikzpicture}
 \caption{An $\mathcal{F}$-gate with 5 dangling edges.}
 \label{fig:Fgate}
\end{figure}
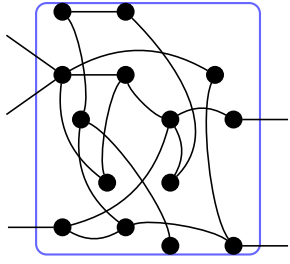

Formally,
this notion is defined by an $\mathcal{F}$-gate.
An $\mathcal{F}$-gate  $(G, \pi)$  is similar to a signature grid for $\Holant(\mathcal{F})$ except that $G = (V,E,D)$ is a graph with some dangling edges $D$.
The dangling edges define external variables for the $\mathcal{F}$-gate.
(See Figure~\ref{fig:Fgate} for an example.)
We name the regular edges in $E$ by $1, 2, \dotsc, m$ and the dangling edges in $D$ by $m+1, \dotsc, m+n$.
Then we can define a function $f$ for this $\mathcal{F}$-gate as
\[
f(y_1, \dotsc, y_n) = \sum_{x_1, \dots, x_m \in \{0, 1\}} H(x_1, \dotsc, x_m, y_1, \dotsc, y_n),
\]
where $(y_1, \dotsc, y_n) \in \{0, 1\}^n$ is an assignment on the dangling edges
and $H(x_1, \dotsc, x_m, y_1, \dotsc, y_n)$ is the value of the signature grid on an assignment of all edges in $G$,
which is the product of evaluations at all vertices in $V$.
We also call this function $f$ the signature of the $\mathcal{F}$-gate.

\paragraph{Binary modification}
A binary modification to the variable $x_i$ of $f$ using the binary signature
$g(x_1, x_2)=(g_{00}, g_{01}, g_{10}, g_{11})=(0, 1, t, 0)^T$
means connecting the variable $x_i$ of $f$ to the variable $x_2$ of $g$ by $\neq_2$.
Note that $g(x_1, x_2)= (0, 1, t, 0)^T$ is a weighted binary
{\sc Disequality}; a modification to $x_1$ of $f$ using
$g$ amounts to  multiplying $t$ to every entry of $f$ where
the index $x_i=1$.
For example, for a 4-ary signature with the signature matrix
$M(f)=\left[\begin{smallmatrix}
f_{0000} & 0 & 0 & f_{0011}\\
0 & f_{0101} & f_{0110} & 0\\
0 & f_{1001} & f_{1010} & 0\\
f_{1100} & 0 & 0 & f_{1111}\\
\end{smallmatrix}\right],$
 by a binary modification to the variable $x_1$ of $f$ using the binary signature
$(0, 1, t, 0)^T$ we get a signature $f'$ with the signature matrix
\[M(f')=\left[\begin{smallmatrix}
f_{0000} &  &  & f_{0011}\\
 & f_{0101} & f_{0110} & \\
 & tf_{1001} & tf_{1010} & \\
tf_{1100} &  &  & tf_{1111}\\
\end{smallmatrix}\right].\]
Similarly, modifications to $x_2$, $x_3$ and $x_4$ of $f$ using
$g$ give the following signatures respectively,
\[\left[\begin{smallmatrix}
f_{0000} &  &  & f_{0011}\\
 & tf_{0101} & tf_{0110} & \\
 & f_{1001} & f_{1010} & \\
tf_{1100} &  &  & tf_{1111}\\
\end{smallmatrix}\right];~~~
\left[\begin{smallmatrix}
f_{0000} &  &  & tf_{0011}\\
 & f_{0101} & tf_{0110} & \\
 & f_{1001} & tf_{1010} & \\
f_{1100} &  &  & tf_{1111}\\
\end{smallmatrix}\right];~~~
\left[\begin{smallmatrix}
f_{0000} &  &  & tf_{0011}\\
 & tf_{0101} & f_{0110} & \\
 & tf_{1001} & f_{1010} & \\
f_{1100} &  &  & tf_{1111}\\
\end{smallmatrix}\right].\]

\begin{figure}[h]
\centering
\begin{tikzpicture}[scale=0.4]
\node [internal, scale=0.5] (0) at (0, 2) {};
\node at (-0.5, 3) {\tiny{$x_{i}$}};
\node at (-0.5, 1) {\tiny{$x_{j}$}};
\node at (0.5, 3) {\tiny{$x_{k}$}};
\node at (0.5, 1) {\tiny{$x_{\ell}$}};
\node [external] (1) at (3, 2) {};
\node [external] (0a) at (-3, 3.2) {};
\node [external] (0b) at (-3, 0.8) {};
\node [triangle, scale=0.5] (2) at (6, 2) {};
\node at (5.5, 3) {\tiny{$x_{1}$}};
\node at (5.5, 1) {\tiny{$x_{2}$}};
\draw (0)  to [bend right=45] (2)[postaction={decorate, decoration={
                                        markings,
                                        mark=at position 0.5 with {\arrow[>=square,white, scale=0.7] {>}; },
                                        mark=at position 0.5 with {\arrow[>=open square, scale=0.7]  {>}; } } }];
\draw (0)  to [bend left=45] (2)[postaction={decorate, decoration={
                                        markings,
                                        mark=at position 0.5 with {\arrow[>=square,white, scale=0.7] {>}; },
                                        mark=at position 0.5 with {\arrow[>=open square, scale=0.7]  {>}; } } }];
\draw (0) to [bend right=15] (0a);
\draw (0) to [bend left=15] (0b);
\end{tikzpicture}
\caption{{
The circle is assigned $f$, the squares are assigned $\neq_2$ and the triangle is assigned $g$.}}
\label{loop}
\end{figure}
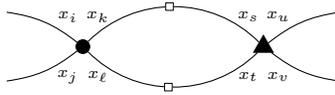

In the following, we often use the gadgets in Figure~\ref{loop-binary} and Figure~\ref{loop}.
We will not draw these gadgets every time.
When we say ``by connecting two 4-ary signatures $f$ and $g$, we get a
 signature $h$ with the signature matrix $M_{x_{i}x_{j},x_{u}x_{v}}(h)=M_{x_{i}x_{j}, x_{k}x_{\ell}}(f)NM_{x_{s}x_{t}, x_{u}x_{v}}(g)$",
this means that we connect the variables $x_{k}, x_{\ell}$ of $f$ to
 the variables $x_{s}, x_{t}$ of  $g$
respectively by $(\neq_2)$, as shown in Figure~\ref{loop-binary}.

\begin{figure}[h]
\centering
\begin{tikzpicture}[scale=0.4]
\node [internal, scale=0.5] (0) at (0, 2) {};
\node at (-0.5, 3) {\tiny{$x_{i}$}};
\node at (-0.5, 1) {\tiny{$x_{j}$}};
\node at (0.5, 3) {\tiny{$x_{k}$}};
\node at (0.5, 1) {\tiny{$x_{\ell}$}};
\node [external] (1) at (3, 2) {};
\node [external] (0a) at (-3, 3.2) {};
\node [external] (0b) at (-3, 0.8) {};
\node [triangle, scale=0.5] (2) at (6, 2) {};
\node at (5.5, 3) {\tiny{$x_{s}$}};
\node at (5.5, 1) {\tiny{$x_{t}$}};
\node at (6.5, 3) {\tiny{$x_{u}$}};
\node at (6.5, 1) {\tiny{$x_{v}$}};
\node [external] (2a) at (9, 3.2) {};
\node [external] (2b) at (9, 0.8) {};
\draw (0)  to [bend right=45] (2)[postaction={decorate, decoration={
                                        markings,
                                        mark=at position 0.5 with {\arrow[>=square,white, scale=0.7] {>}; },
                                        mark=at position 0.5 with {\arrow[>=open square, scale=0.7]  {>}; } } }];
\draw (0)  to [bend left=45] (2)[postaction={decorate, decoration={
                                        markings,
                                        mark=at position 0.5 with {\arrow[>=square,white, scale=0.7] {>}; },
                                        mark=at position 0.5 with {\arrow[>=open square, scale=0.7]  {>}; } } }];
\draw (0) to [bend right=15] (0a);
\draw (0) to [bend left=15] (0b);
\draw (2) to [bend right=15] (2b);
\draw (2) to [bend left=15] (2a);
\end{tikzpicture}
\caption{{
The circle is assigned $f$, the squares are assigned $\neq_2$ and the triangle is assigned $g$.}}
\label{loop-binary}
\end{figure}

When we say ``doing a loop to $f$ using the binary signature $g=(0, 1, t, 0)^T$, we get a signature
$h=M_{x_{i}x_{j}, x_{k}x_{\ell}}(f)Ng$",
this means that
we connect the variables $x_{k}, x_{\ell}$ of $f$ with the
variables $x_1, x_2$ of $g$ using $(\neq_2)$,
 as shown in Figure~\ref{loop}.

Now we give some lemmas.
% that we will use in the following.

\begin{lemma}\cite{cai:fu:plcsp}\label{equality-4-csp2}
For any set of signatures $\mathcal{F}$,
\[
\#\operatorname{CSP}^2(\mathcal{F})\leq^p_T\operatorname{Holant}(=_2|=_4, \mathcal{F}).
\]
\end{lemma}

\begin{lemma}\label{inverse}
Let $g=(0, 1, t, 0)^T$ be a binary signature, where $t\neq 0$.
Then for any signature set $\mathcal{F}$, we have
\[
\operatorname{Holant}(\neq_2|\mathcal{F}, g, (0, 1, t^{-1}, 0)^T)\leq_{T}\operatorname{Holant}(\neq_2|\mathcal{F}, g).
\]
\end{lemma}
\begin{proof}
If $g(x_1, x_2)=(0, 1, t, 0)^T$, then $h(x_1, x_2)=g(x_2, x_1)=(0, t, 1, 0)^T=t(0, 1, t^{-1}, 0)^T$.
After the nonzero scalar $t$, we have $(0, 1, t^{-1}, 0)^T$.
\end{proof}

\begin{lemma}\label{full:out:deg:inner}
Let $f$ be a 4-ary signature with the signature matrix $M(f)=\left[\begin{smallmatrix}
a & 0 & 0 & b\\
0 & c & d & 0\\
0 & w & z & 0\\
y & 0 & 0 & a\\
\end{smallmatrix}\right]$, where $czdw\neq 0$.
If
$f$ satisfies the following conditions,
\begin{itemize}
\item $\left[\begin{smallmatrix}
a & b\\
y & a\\
\end{smallmatrix}\right]$ has full rank;
\item $\left[\begin{smallmatrix}
c & d\\
 w & z
\end{smallmatrix}\right]$ is degenerate;
\item $c+d\neq 0$ or $c+w\neq 0$;
\end{itemize}
then $\operatorname{Holant}(\neq_2|f)$ is \#P-hard.
\end{lemma}
\begin{proof}
We prove the lemma for $c+d\neq 0$.
The case $c+w\neq 0$ can be obtained from
 $M_{x_3x_4,x_1x_2}(f)= M^{\sf R_T}(f)
=\left[\begin{smallmatrix}
a & 0 & 0 & y\\
0 & c & w & 0\\
0 & d & z & 0\\
b & 0 & 0 & a\\
\end{smallmatrix}\right]$.
% The proof  for the case that $c+w\neq0$ is same and we omit it here .

Since $\left[\begin{smallmatrix}
c & d\\
 w & z
\end{smallmatrix}\right]$ is degenerate and $czdw\neq 0$, there exists $k\neq 0$ such that
$\left[\begin{smallmatrix}
c & d\\
 w & z
\end{smallmatrix}\right]=\left[\begin{smallmatrix}
c & d\\
 ck & dk
\end{smallmatrix}\right]$.

By connecting the variables  $x_3$ and $x_4$ of $f$ using $\neq_2$, we
get a binary signature
$(c+d)(0, 1, k, 0)^T$.
By Lemma~\ref{inverse}, we have $(0, 1, k^{-1}, 0)^T$.
By doing a binary modification to the variable $x_1$ of $f$ using $(0, 1, k^{-1}, 0)^T$, we get a signature $f_1$ with the signature matrix
$M(f_1)=\left[\begin{smallmatrix}
a & 0 & 0 & b\\
0 & c & d & 0\\
0 & c & d & 0\\
\frac{y}{k} & 0 & 0 & \frac{a}{k}\\
\end{smallmatrix}\right]$.
Then by connecting the variables  $x_1$ and $x_2$ of $f_1$ using $\neq_2$, we get the binary signature
$2c(0, 1, \frac{d}{c}, 0)^T$.
By Lemma~\ref{inverse}, we have $(0, 1, \frac{c}{d}, 0)^T$.
By doing a binary modification to the variable $x_3$ of $f_1$ using $(0, 1, \frac{c}{d}, 0)^T$, we get a signature $f_2$ with the signature matrix
$M(f_2)=\left[\begin{smallmatrix}
a & 0 & 0 & \frac{bc}{d}\\
0 & c & c & 0\\
0 & c & c & 0\\
\frac{y}{k} & 0 & 0 & \frac{ac}{dk}\\
\end{smallmatrix}\right]$.
This signature is redundant.
Since
$\det \left[\begin{smallmatrix}
a & b\\
y & a\\
\end{smallmatrix}\right] \not = 0$,
%$a^2-by\neq 0$,
the  compressed signature matrix
of $f_2$ has full rank.
Thus $\operatorname{Holant}(\neq_2|f_2)$ is \#$\operatorname{P}$-hard by Theorem~\ref{redundant}.
So $\operatorname{Holant}(\neq_2|f)$ is \#$\operatorname{P}$-hard.
\end{proof}

\subsection{Tractable Signature Sets}
We define some sets of signatures that are known to be tractable.
%These form three families:
%product-type signatures,
%affine signatures,  and local affine signatures.

\paragraph{Affine Signatures $\mathscr{A}$}

%\begin{definition}
%For a signature $f$ of arity $n$,
%the support of $f$ is
%\[\operatorname{supp}(f)=\{(x_1, x_2, \ldots,  x_n)
%\in \mathbb{Z}_2^n \mid f(x_1, x_2, \ldots, x_n)\neq 0\}.\]
%\end{definition}

\begin{definition}
Let $f$ be a signature of arity $n$.
We say $f$ has affine support of dimension $k$ if
the support of $f$ is an affine subspace of
dimension $k$ over $\mathbb{Z}_2$, i.e.,
there is a matrix $A$  over $\mathbb{Z}_2$
such that
 $f(x_1, x_2, \ldots, x_n)\neq 0$ iff $AX=0$, where $X = (x_1, x_2, \ldots,
 x_n, 1)^T$ and
the affine space $\{(x_1, x_2, \ldots, x_n)
\in \mathbb{Z}_2^n \mid AX=0\}$ has dimension $k$.
\end{definition}

\begin{definition}\label{definition-affine}
 A signature $f(x_1, \ldots, x_n)$ of arity $n$
is \emph{affine} if it has the form
 \[
  \lambda \cdot \chi_{A X = 0} \cdot {\frak i} ^{Q(X)},
 \]
 where $\lambda \in \mathbb{C}$,
 $X = (x_1, x_2, \dotsc, x_n, 1)$,
 $A$ is a matrix over $\mathbb{Z}_2$,
 %$\mathbf{v}_j$ is a vector over $\mathbb{Z}_2$,
 $Q(x_1, x_2, \ldots, x_n)\in \mathbb{Z}_4[x_1, x_2, \ldots, x_n]$
is a quadratic (total degree at most 2) multilinear polynomial
 with the additional requirement that the coefficients of all
 cross terms are even, i.e., $Q$ has the form
 \[Q(x_1, x_2, \ldots, x_n)=a_0+\displaystyle\sum_{k=1}^na_kx_k+\displaystyle\sum_{1\leq i<j\leq n}2b_{ij}x_ix_j,\]
 and $\chi$ is a 0-1 indicator function
 such that $\chi_{AX = 0}$ is~$1$ iff $A X = 0$.
 %Note that the dot product $\langle \mathbf{v}_j, x \rangle$ is calculated over $\mathbb{Z}_2$,
 %while the summation $\sum_{j=1}^n$ on the exponent of $i = \sqrt{-1}$ is evaluated as a sum mod~$4$ of 0-1 terms.
 We use $\mathscr{A}$ to denote the set of all affine signatures.
\end{definition}

\iffalse
\paragraph{Product-Type Signatures $\mathscr{P}$}
\begin{definition}%[Definition~3.3 in~\cite{CLX14}]
\label{definition-product-2}
 A signature on a set of variables $X$
 is of \emph{product type} if it can be expressed as a
product of unary functions,
 binary equality functions $([1,0,1])$,
and binary disequality functions $([0,1,0])$, each on one or two
variables of $X$.
 We use $\mathscr{P}$ to denote the set of product-type functions.
\end{definition}
\fi
\paragraph{Product-Type Signatures $\mathscr{P}$}
%\paragraph{Product-Type Signatures}
\begin{definition}%[Definition~3.3 in~\cite{CLX14}]
\label{definition-product-2}
 A signature on a set of variables $X$
 is of \emph{product type} if it can be expressed as a
product of unary functions,
 binary equality functions $([1,0,1])$,
and binary disequality functions $([0,1,0])$, each on one or two
variables of $X$.
 We use $\mathscr{P}$ to denote the set of product-type functions.
\end{definition}

 %A symmetric signature of the form $[a, 0, \ldots, 0, b]$ is called a
%{\sc Generalized Equality}.
%\begin{proposition}(cf.~Lemma~A.1 in the full version of~\cite{HL12})\label{symmetric:prod:type:signatures}
%%% JYC Cite their journal Computatiuonal Complexity paper.
%%% really need to cite that?
%i am pretty sure thsiw a done before in my earlier papers...
% If $f$ is a symmetric signature in $\mathscr{P}$,
%then $f$ is either degenerate,
%binary \textsc{Disequality} $(\neq_2) = [0,1,0]$,
%or $[a,0,\dotsc,0,b]$ for some $a, b \in \mathbb{C}$.
%\end{proposition}
%
%\begin{corollary}\label{[1,0,1,0]:not:prod}
%$[1, 0, 1, 0]\notin\mathscr{P}$.
%\end{corollary}%
%
%We will use Corollary~\ref{[1,0,1,0]:not:prod} in the proof of Theorem~\ref{dichotomy-csp-2}.

Definition~\ref{definition-product-2} is succinct.
%and is from \cite{CLX14}.
 But to deal with asymmetric signatures,
an alternative definition of $\mathscr{P}$ is
useful. This is given below in Definition~\ref{definition-product-1}.
To state it we need some notations.

Suppose $f$ is a signature of arity $n$ and $\mathcal{I}=\{I_1, I_2, \ldots, I_k\}$ is a partition of $[n]$.
If $f(X)=\displaystyle\prod_{j=1}^k f_j(X|_{I_j})$ for some signatures $f_1, f_2, \ldots, f_k$, where $X=\{x_1, x_2, \ldots, x_n\}$
and $X|_{I_j}=\{x_s|s\in I_j\}$ (we also denote it by $X_j$), then we
%say $f$ can be decomposed into type $\mathcal{I}$.
%%% JYC i changed this to below. I am not sure you ever use this term?
say $f$ can be decomposed as a tensor product of $f_1, f_2, \ldots, f_k$.
%according to $\mathcal{I}$.
We denote such a function
 by $f=\bigotimes_{\mathcal{I}}(f_1, f_2, \ldots, f_k)$.
If each $f_j$ is the signature of some
$\mathcal{F}$-gate, then $\bigotimes_{\mathcal{I}}(f_1, f_2, \ldots, f_k)$ is
the signature
of the $\mathcal{F}$-gate which is the disjoint union of
 the $\mathcal{F}$-gates for $f_j$, with variables
renamed and ordered according to $\mathcal{I}$.
%(In general this is not a planar $\mathcal{F}$-gate
%even when the $\mathcal{F}$-gates for all $f_j$ are planar,
%unless the sets $I_1, I_2, \ldots, I_k$ partition $[n]$ in order.)
 When
the indexing is clear, we also use
the notation $f_1\otimes f_2\otimes\cdots\otimes f_k$.
Note that this tensor product notation
$\otimes$ is consistent with tensor product of matrices.
We say a signature set $\mathcal{F}$ is closed under tensor product,
if for any partition  $\mathcal{I}=\{I_1, I_2\}$, and any  $f, g\in\mathcal{F}$
on $X_1$ and $X_2$ respectively,
we have $\bigotimes_{\mathcal{I}}(f, g)\in\mathcal{F}$.
The
tensor closure $\langle \mathcal{F}\rangle$
of $\mathcal{F}$
is the minimum set containing $\mathcal{F}$, closed under tensor product.

\begin{definition}\label{definition-product-1}
Let $\mathcal{E}$ be the set of all signatures $f$ such that
the support set
$\operatorname{supp}(f)$ is contained in two antipodal points, i.e.,
if $f$ has arity $n$,
then $f$ is zero except on (possibly) two inputs $\alpha
= (a_1, a_2, \ldots, a_n)$
and $\overline{\alpha} =
(\overline{a}_1, \overline{a}_2, \ldots, \overline{a}_n)=(1- a_1, 1- a_2,
 \ldots, 1 - a_n)$.
Then $\mathscr{P}=\langle \mathcal{E}\rangle$.
\end{definition}

By Definition~\ref{definition-product-1}, if a signature $f\in\mathscr{P}$, then the support of $f$ is affine.
Thus the number of nonzero entries of $f$ is a power of 2.

\begin{lemma}\label{lm:zhiguo}
Let $f$ be a 4-ary signature with the signature matrix
$M(f)=\left[\begin{smallmatrix}
0 & 0 & 0 & 0\\
0 & c & d & 0\\
0 & w & z & 0\\
0 & 0 & 0 & 0
\end{smallmatrix}\right]$ with $cdwz\neq 0$,
then $f\in\mathscr{P}$ iff $cz=dw$.
%
% $\left[\begin{smallmatrix}
% c & d \\
%  w & z
%\end{smallmatrix}\right]$ is degenerate.
%
Similarly, let $f$ be a 4-ary signature with the signature matrix
$M(f)=\left[\begin{smallmatrix}
1 & 0 & 0 & 0\\
0 & c & 0 & 0\\
0 & 0 & z & 0\\
0 & 0 & 0 & 1
\end{smallmatrix}\right]$ with $cz\neq 0$,
then $f\in\mathscr{P}$ iff $cz=1$.

\end{lemma}
\begin{proof}
We prove the lemma for $M(f)=\left[\begin{smallmatrix}
0 & 0 & 0 & 0\\
0 & c & d & 0\\
0 & w & z & 0\\
0 & 0 & 0 & 0
\end{smallmatrix}\right]$. The proof for $M(f)=\left[\begin{smallmatrix}
1 & 0 & 0 & 0\\
0 & c & 0 & 0\\
0 & 0 & z & 0\\
0 & 0 & 0 & 1
\end{smallmatrix}\right]$ is similar and we omit it here.

%%% for second matrix, the only partition is 13, 24
%%% then matrix has 2 by 2 submatrix 1 c \\ z 1

If $cz=dw$,
%$\left[\begin{smallmatrix}
% c & d \\
%  w & z
%\end{smallmatrix}\right]$ is degenerate, i.e.,
then there exists $k\neq 0$ such that
$\left[\begin{smallmatrix}
 c & d \\
  w & z
\end{smallmatrix}\right]=\left[\begin{smallmatrix}
 c & d \\
  kc & kd
\end{smallmatrix}\right]$.
Thus $f(x_1, x_2, x_3, x_4)=g_1(x_1, x_2)g_2(x_3, x_4)$, where $g_1=(0, 1, k, 0)^T$ and $g_2=(0, c, d, 0)^T$,
i.e., $f=g_1\otimes g_2$. Note that $g_1, g_2\in\mathcal{E}$,
so $f\in\mathscr{P}$ by Definition~\ref{definition-product-1}.

Conversely, suppose $M(f)$ has the given form
and  $f\in\mathscr{P}$.  Firstly we claim that
there are two binary signatures $g_1$ and $g_2$
such that $f(x_1, x_2, x_3, x_4)=g_1(x_1, x_2)g_2(x_3, x_4)$.
By Definition~\ref{definition-product-1}, $f=\otimes_{i=1}^kg_i$,
where $g_i\in\mathcal{E}$,
according to some partition $\mathcal{I}$.
Note that ${\rm supp}(f)$, the support of $f$,
 %is $\{\sigma_1=0101, \sigma_2=0110, \sigma_3=1001, \sigma_4=1010\}$.
is $\{0101, 0110, 1001, 1010\}$.
%We consider the variable $x_1$.

We first show that no $g_i$ can be a unary signature.
Suppose for a contradiction $g_i$ is a unary signature on $x_1$.
If $g_i(0) =0$ or $g_i(1) =0$,
then the first, respectively the last, two rows of $M(f)$ would be 0.
So $g_i(0)g_i(1) \not =0$.
Then  ${\rm supp}(f^{x_1 =0}) = {\rm supp}(f^{x_1 =1})$, clearly not true.
The same argument shows that no  $g_i$ is a unary signature on
any $x_j$.
%So no  $g_i$ can be a unary signature.

Since $|{\rm supp}(f)| = 4$, there must be at least two tensor factors
in  $f=\otimes_{i=1}^kg_i$, i.e., $k \ge 2$.
%ow |{\rm supp}(f)| <=2.
Having no unary $g_i$,
the only possibility is $k=2$ and both $g_1$ and $g_2$ are
binary signatures.
% if k>=3, some unary exists.

Suppose $\mathcal{I} = \{I_1, I_2\}$.
By $f=\otimes_{i=1}^kg_i$, ${\rm supp}(f)$ is a direct product of
the restrictions of ${\rm supp}(f)$ on $I_1$ and $I_2$.
If the  partition $\mathcal{I} \not = \{ \{1,2\}, \{3,4\} \}$,
then projecting ${\rm supp}(f)$ to $I_1$ and $I_2$ would give
$\{00,01,10,11\}$, a contradiction.

Hence  $\mathcal{I} = \{ \{1,2\}, \{3,4\} \}$,
and we can assume $f(x_1,x_2,x_3,x_4) = g_1(x_1,x_2) g_2(x_3,x_4)$.
Therefore  $M(f)$ has rank one, i.e.,
$\det
\left[\begin{smallmatrix}
 c & d \\
  w & z
\end{smallmatrix}\right]= 0$.

\end{proof}

 %Valiant
%\cite{val02a, val02b} introduced matchgates, which we denote by $\mathscr{M}$. They can be locally expressed by weighted perfect
%matchings, so problems defined by them are tractable by the FKT algorithm over planar graphs.

\paragraph{Local Affine Signatures $\mathscr{L}$}
\begin{definition}
A function $f$ of arity  $n$ is in $\mathscr{L}$, if for any
$\sigma=s_1s_2\dots s_n$
in the support of $f$,
the transformed function
\[R_{\sigma}f: (x_1, x_2, \cdots, x_n)\mapsto \alpha^{\sum_{i=1}^ns_ix_i}f(x_1, x_2, \cdots, x_n)\]
is in $\mathscr{A}$. Here each $s_i$
is a 0-1 valued integer, and the sum
$\sum_{i=1}^ns_ix_i$
is evaluated as an integer (or an integer mod 8).
\end{definition}

{Note that all {\sc Equalities}  $(=_{2k})$  of  even  arity
 are in $\mathscr{L}$.
But $(=_{2k+1})\notin\mathscr{L}$ and $(\neq_2)\notin\mathscr{L}$.
For an $n$-ary signature $f$, if $f(0, 0, \ldots, 0)\neq 0$, then $f\in\mathscr{L}$ implies that $f\in\mathscr{A}$.}

\paragraph{Transformable}

%An important definition involving a holographic transformation is the notion of a signature set being transformable.

\begin{definition} \label{def:prelim:trans}
 We say a pair of signature sets $(\mathcal{G}, \mathcal{F})$
is $\mathscr{C}$-transformable for  $\holant{\mathcal{G}}{\mathcal{F}}$
%and $\plholant{\mathcal{G}}{\mathcal{F}}$
 if there exists $T \in \mathbf{GL}_2(\mathbb{C})$ such that
$\mathcal{G}T \subseteq  \mathscr{C}$ and $T^{-1} {\mathcal{F}}
\subseteq \mathscr{C}$.
For $\mathcal{G} = \{(\neq_2)\}$,
 we say simply that $\mathcal{F}$ is
$\mathscr{C}$-transformable if $(\neq_2)T \in  \mathscr{C}$ and $T^{-1} {\mathcal{F}}
\subseteq \mathscr{C}$.
%For  $\mathcal{G} = \mathcal{EQ}_d$,
%$ \holant{\mathcal{EQ}_d}{\mathcal{F}}  \equiv^p_T \CSP^d(\mathcal{F})$,
%we say that $\mathcal{F}$ is
%$\mathscr{C}$-transformable for \#$\operatorname{CSP}^d$
%if $(\mathcal{EQ}_d)T \in  \mathscr{C}$.
%%We define similarly for \#$\operatorname{CSP}^d$ when   $\mathcal{G} = \mathcal{EQ}_d$.
\end{definition}

Notice that if
$\operatorname{Holant}(\mathscr{C})$ is tractable, and
$(\mathcal{G}, \mathcal{F})$
is $\mathscr{C}$-transformable,
then $\operatorname{Holant}(\mathcal{G} | \mathcal{F})$ is tractable
by a holographic transformation.
For example, let $T=
\left[\begin{smallmatrix} 1 & 0 \\
0 & \alpha \end{smallmatrix}\right]$.
Note that $(\mathcal{EQ}_2)T \subset
\mathscr{A}$. If $T^{-1}\mathcal{F}\subseteq\mathscr{A}$,
then $\#\operatorname{CSP}^2(\mathcal{F})$
is tractable.
We {\emph define} the signature set
\[\alpha\mathscr{A}=\left\{ \left[\begin{smallmatrix} 1 & 0 \\
0 & \alpha \end{smallmatrix}\right]^{\otimes \operatorname{ary}(f)} f
\mid f\in\mathscr{A}
\right\}.\]

The classes
$\mathscr{A}$ and $\mathscr{P}$ are
known to be the tractable classes for \#CSP.% \cite{cailuxia-2014}.
\begin{theorem}\cite{cailuxia-2014}\label{csp:dichotomy}
Let $\mathcal{F}$ be any set of complex-valued signatures in Boolean variables. Then $\operatorname{\#CSP}(\mathcal{F})$
is \#$\operatorname{P}$-hard unless
$\mathcal{F}\subseteq\mathscr{A}$ or
$\mathcal{F}\subseteq\mathscr{P}$,
  in which case the problem is computable in polynomial time.
\end{theorem}

Lin and Wang proved the following lemma (Lemma 3.4
in~\cite{wang-lin-2016}),
which says that one can always reduce a signature to its tensor power
in Holant problems.
We will only need a special case; for the  convenience of readers we
state it below with a short proof.

\begin{lemma}[Lin-Wang]\label{lm:Lin-Wang}
For any set of signatures $\mathcal{F}$,
and a signature $f$,
\[\operatorname{Holant}(\mathcal{F}, f)
\leq_{\rm T}^p
\operatorname{Holant}(\mathcal{F}, f^{\otimes 2}).\]
\end{lemma}
\begin{proof}
We ask the question: Is there a signature
grid $\Omega$ for $\operatorname{Holant}(\mathcal{F}, f)$
in which $f$ appears an odd number of times,
and the value
$\operatorname{Holant}_{\Omega}(\mathcal{F}, f)$ is nonzero?
If the answer is no, then here is a simple reduction:
For any input signature
grid $\Omega$ for $\operatorname{Holant}(\mathcal{F} ,f)$,
if  $f$ appears an odd number of times,
then $\operatorname{Holant}_{\Omega}(\mathcal{F} ,f) =0$,
otherwise, pair up occurrences of $f$ two at a time and
replace them by one copy of $f^{\otimes 2}$.

Now suppose the answer is yes, and let
$c = \operatorname{Holant}_{\Omega_0}
(\mathcal{F}, f) \not =0$, where $\Omega_0$ is
a signature grid in which $f$ appears $2k+1$ times.
Replace $2k$ occurrences of $f$ in $\Omega_0$
by $k$ copies of $f^{\otimes 2}$.
Now use one more copy of  $f^{\otimes 2}$.
Suppose $f^{\otimes 2}(x_1, \ldots, x_s, y_1,  \ldots, y_s)
= f(x_1, \ldots, x_s) f(y_1,  \ldots, y_s)$, where $s$ is
the arity of $f$.
Replace the $(2k+1)$-th occurrence of $f$ in $\Omega_0$
by $f^{\otimes 2}$, using variables $y_1,  \ldots, y_s$ of  $f^{\otimes 2}$
to connect to the $s$ edges of the $(2k+1)$-th occurrence of $f$,
and leaving $x_1, \ldots, x_s$ as dangling edges.
This creates a $(\mathcal{F} \cup \{f^{\otimes 2}\})$-gate
with singature $c f$.
Hence
\[\operatorname{Holant}(\mathcal{F}, f)
\leq_{\rm T}^p
\operatorname{Holant}(\mathcal{F}, f^{\otimes 2}).\]
\end{proof}

\begin{theorem}\label{construct:pin}\cite{huang-lu-2016}
For any signature set $\mathcal{F}$,
\[
\#\operatorname{CSP}^2([1, 0]^{\otimes 2}, [0, 1]^{\otimes 2}, \mathcal{F})\leq_T^p\#\operatorname{CSP}^2(\mathcal{F}).
\]
\end{theorem}

Using Lemma~\ref{lm:Lin-Wang} and Theorem~\ref{construct:pin}, we can state the
following  dichotomy theorem from \cite{cailuxia-2017}
for  \#CSP$^2$.
\begin{theorem}\label{CSP2}
Let $\mathcal{F}$ be any set of complex-valued signatures in Boolean variables. Then $\#\operatorname{CSP}^2(\mathcal{F})$
is $\#\operatorname{P}$-hard unless
$\mathcal{F}\subseteq\mathscr{A}$ or
$\mathcal{F}\subseteq\alpha\mathscr{A}$ or
$\mathcal{F}\subseteq\mathscr{P}$ or
$\mathcal{F}\subseteq\mathscr{L}$,
  in which cases the problem is computable in polynomial time.
\end{theorem}

The six-vertex model is a special case of the eight-vertex model with $a=x=0$ in $M(f)$.

\begin{theorem}\cite{cai:fu:xia}\label{dichotomy-six-vertex}
Let $f$ be a 4-ary signature with the signature matrix
$M(f)=\left[\begin{smallmatrix}
0 & 0 & 0 & b\\
0 & c & d & 0\\
0 & w & z & 0\\
y & 0 & 0 & 0
\end{smallmatrix}\right]$, then
$\operatorname{Holant}(\neq_2\mid f)$ is $\#\operatorname{P}$-hard except for the following cases:
\begin{itemize}
\item $f\in\mathscr{P}$;
\item $f\in\mathscr{A}$;
\item there is a zero in each pair $(b, y), (c, z), (d, w)$;
\end{itemize}
in which cases $\operatorname{Holant}(\neq_2\mid f)$ is computable in polynomial time.
\end{theorem}
Let $f$ be a 4-ary signature with the signature matrix
$M(f)=\left[\begin{smallmatrix}
0 & 0 & 0 & b\\
0 & c & d & 0\\
0 & w & z & 0\\
y & 0 & 0 & 0
\end{smallmatrix}\right]$.
If there are exactly three nonzero entries in the inner matrix $\left[\begin{smallmatrix}
 c & d \\
  w & z
\end{smallmatrix}\right]$, then the support of $f$ is not affine.
This is because $0101 \oplus 0110 \oplus 1001 = 1010$,
and if ${\rm supp}(f)$ were affine, then three entries in ${\rm supp}(f)$
would imply the fourth entry also belongs to ${\rm supp}(f)$.
Thus $f\notin\mathscr{P}\cup\mathscr{A}$.
Moreover, there is one pair in $\{(c, z), (d, w)\}$
that has two nonzero entries.
By Theorem~\ref{dichotomy-six-vertex},
$\operatorname{Holant}(\neq_2\mid f)$ is $\#\operatorname{P}$-hard.
Thus we have the following corollary.

\begin{corollary}\label{six:exact:3nonzero:inner:matrix}
Let $f$ be a 4-ary signature with the signature matrix
$M(f)=\left[\begin{smallmatrix}
0 & 0 & 0 & b\\
0 & c & d & 0\\
0 & w & z & 0\\
y & 0 & 0 & 0
\end{smallmatrix}\right]$.
If there are exactly three nonzero entries in
$\left[\begin{smallmatrix}
 c & d \\
  w & z
\end{smallmatrix}\right]$
 then
$\operatorname{Holant}(\neq_2\mid f)$ is $\#\operatorname{P}$-hard.
\end{corollary}

\begin{lemma}\label{normalize:a=x}\label{ax=0}
Let $f$ be a 4-ary signature with the signature matrix $M(f)=\left[\begin{smallmatrix}
 a & 0 & 0 & b\\
 0 & c & d & 0\\
 0 & w & z & 0\\
 y & 0 & 0 & x\\
\end{smallmatrix}\right]$,
then
\[
\operatorname{Holant}(\neq_2|f)\equiv^p_{T} \operatorname{Holant}(\neq_2|\tilde{f}),
\]
where $M(\tilde{f})=\left[\begin{smallmatrix}
 \tilde{a} & 0 & 0 & b\\
 0 & c & d & 0\\
 0 & w & z & 0\\
 y & 0 & 0 & \tilde{a}\\
\end{smallmatrix}\right]$ with $\tilde{a}=\sqrt{ax}$.
\end{lemma}
\begin{proof}
For any signature grid $\Omega$
with 4-regular graph $G$,
 any valid orientation
on $G$
for  Holant$_{\Omega}(\neq_2|f)$
must have an equal number of sources and sinks.
Hence the value  Holant$_{\Omega}(\neq_2|f)$
as a polynomial in $a$ and $x$ is in fact a polynomial in the product
$ax$.
So we can replace $(a,x)$ by any $(\tilde{a}, \tilde{x})$
such that $\tilde{a}\tilde{x} = ax$.
In particular,
let $\tilde{f}$  be a 4-ary signature with the signature matrix
$M(\tilde{f})=\left[\begin{smallmatrix}
\tilde{a} & 0 & 0 & b\\
0 & c & d & 0\\
0 & w & z & 0\\
y & 0 & 0 & \tilde{a}
\end{smallmatrix}\right]$ where $\tilde{a}=\sqrt{ax}$,
then
Holant$_{\Omega}(\neq_2|f)=$Holant$_{\Omega}(\neq_2|\tilde{f})$.
Thus \[
\operatorname{Holant}(\neq_2|f)\equiv^p_{T} \operatorname{Holant}(\neq_2|\tilde{f}).
\]
%
%********************
%
%In an instance of Holant$_{\Omega}(\neq_2|f)$ on  a bipartite graph $G=(V_1, V_2, E)$, where all the vertices in $V_1$ are labeled $\neq_2$
%and all the vertices in $V_2$ are labeled $f$,
%  if an assignment $\sigma$
%such that the evaluation $\prod_{v\in(V_1\cup V_2)}f_v(\sigma|E(v))$ is nonzero, then the number of the edges that are assigned $0$ has to be equal to
%the number of the edges that are assigned $1$ since all the vertices on the left side are labeled $\neq_2$.
%On the right side, let $V_a$ be the set of the vertices whose incident edges are all assigned $0$ and
%$V_x$ be the set of the vertices whose incident edges are all assigned $1$.
%For each vertex in $V_2\setminus(V_a\cup V_x)$, its four incident edges are
% assigned $0011, 0101, 0110, 1001, 1010$ or $ 1100$, i.e.,
%exact half of the incident edges at  the vertices in $V_2\setminus(V_a\cup V_x)$ are assigned $0$.
%This implies that $|V_a|=|V_x|$.
%So  Holant$_{\Omega}(\neq_2|f)$ can be expressed as a polynomial $\sum_{i=1}^{\lfloor\frac{|V_2|}{2}\rfloor}c_ia^ix^i$.
%Note that $\sum_{i=1}^{\lfloor\frac{|V_2|}{2}\rfloor}c_ia^ix^i=\sum_{i=1}^{\lfloor\frac{|V_2|}{2}\rfloor}c_i(\sqrt{ax})^i(\sqrt{ax})^i$.
%Thus Holant$_{\Omega}(\neq_2|f)=$Holant$_{\Omega}(\neq_2|f')$, where $f'$ has the signature matrix
%$M(f')=\left[\begin{smallmatrix}
% \sqrt{ax} & 0 & 0 & b\\
% 0 & c & d & 0\\
% 0 & w & z & 0\\
% y & 0 & 0 & \sqrt{ax}\\
%\end{smallmatrix}\right]$.
\end{proof}
By Lemma~\ref{normalize:a=x}, we may assume that $a=x$.
%And we can also switch the sign of both entries $(a,x)$.
%In particular,
%\begin{lemma}\label{ax=0}
Let $f$ be a signature  with the signature matrix
% and $f'$ be two 4-ary signatures with signature matrices
$M(f)=\left[\begin{smallmatrix}
a & 0 & 0 & b\\
0 & c & d & 0\\
0 & w & z & 0\\
y & 0 & 0 & x
\end{smallmatrix}\right]$. Suppose $a=0$ or $x=0$,
then let  $f'$ be a signature with $M(f')=\left[\begin{smallmatrix}
0 & 0 & 0 & b\\
0 & c & d & 0\\
0 & w & z & 0\\
y & 0 & 0 & 0
\end{smallmatrix}\right]$,  by Lemma~\ref{ax=0},  we have
\[
\operatorname{Holant}(\neq_2|f)\equiv^p_{T}\operatorname{Holant}(\neq_2|f').
\]
%\end{lemma}
%\begin{proof}
%We prove the lemma for the case that $a=0$. The proof for the case that $x=0$ is similar and we omit it here.%
%
%In an instance of Holant$_{\Omega}(\neq_2|f)$ on  a bipartite graph $G=(V_1, V_2, E)$, where all the vertices in $V_1$ are labeled $(\neq_2)$
%and all the vertices in $V_2$ are labeled $f$,
%  if an assignment $\sigma$
%such that the evaluation $\prod_{v\in(V_1\cup V_2)}f_v(\sigma|E(v))$ is nonzero, then the number of the edges that are assigned $0$ has to be equal to
%the number of the edges that are assigned $1$ since all the vertices on the left side are labeled $(\neq_2)$.
%%
%%We claim that for each vertex in $V_2$, its four incident edges are
%% assigned $0011, 0101, 0110, 1001, 1010$ or $ 1100$.
%Thus if there is a vertex in $V_2$ whose incident edges are assigned $0000$, then there is at least one vertex $v_a$ in $V_2$
% such that there are more than 2 incident edges of $v_a$ that are assigned $1$.
% Then the signature labeled on $v_a$ gives 0. This contradicts $\prod_{v\in(V_1\cup V_2)}f_v(\sigma|E(v))\neq 0$.
%ence for each vertex in $V_2$, its four incident edges are
% assigned $0011, 0101, 0110, 1001, 1010$ or $ 1100$.
%This implies that Holant$_{\Omega}(\neq_2|f)=$Holant$_{\Omega}(\neq_2|g)$.
%Thus \[
%\operatorname{Holant}(\neq_2|f)\equiv^p_{T}\operatorname{Holant}(\neq_2|g).
%\]
%\end{proof}
%This gives some new tractable class.% for eight vertex model.
The following lemma follows.
\begin{lemma}\label{aneq0}
Let
the signature matrix of $f$ be $M(f)=\left[\begin{smallmatrix}
 a & 0 & 0 & b\\
 0 & c & d & 0\\
 0 & w & z & 0\\
 y & 0 & 0 & x\\
\end{smallmatrix}\right]$, where $ax=0$, and
the signature matrix of $f'$ be $M(f')=\left[\begin{smallmatrix}
 0 & 0 & 0 & b\\
 0 & c & d & 0\\
 0 & w & z & 0\\
 y & 0 & 0 & 0\\
\end{smallmatrix}\right]$,
then $\operatorname{Holant}(\neq_2\mid f)$ is computable in polynomial time
iff $\operatorname{Holant}(\neq_2\mid f')$ is computable in polynomial time,
and one is \#P-hard iff the other one is \#P-hard.
\end{lemma}

\subsection{Interpolation}
Polynomial interpolation is a powerful tool in the study of counting problems. In this subsection,
we give some lemmas by polynomial interpolation.

\begin{lemma}\label{eigenvalue-interpolation}\label{eigenvalue-interpolation-2}
Let $f$ be a 4-ary signature with the signature matrix
$M(f)=\left[\begin{smallmatrix}
1 & 0 & 0 & t\\
0 & \frak i^r & \epsilon \frak i^r t & 0\\
0 & \epsilon \frak i^r t & \frak i^r  & 0\\
t & 0 & 0 & 1\\
\end{smallmatrix}\right]$,
% or $M(f)=\left[\begin{smallmatrix}
%1 & 0 & 0 & t\\
%0 & i & it & 0\\
%0 & it & i & 0\\
%t & 0 & 0 & 1\\
%\end{smallmatrix}\right],$
where $r \in \{0, 1, 2, 3\}$,
$\epsilon=\pm 1$,
%$t^4\neq 0, 1$  and $t$ is not a pure imagine number,
$t \not = \pm 1$ and $t \not \in \frak i \mathbb{R}$.
 then for any $\lambda\in\mathbb{C}$,
\[\operatorname{Holant}(\neq_2|f, g_{\lambda})\leq^p_{T}\operatorname{Holant}(\neq_2|f),\]
where $g_{\lambda}$ is a 4-ary signature with the signature matrix
$M(g_{\lambda})=\left[\begin{smallmatrix}
1+\epsilon\lambda & 0 & 0 & 1-\epsilon\lambda\\
0 & 1+\lambda & \epsilon(1-\lambda) & 0\\
0 & \epsilon(1-\lambda) & 1+\lambda & 0\\
1-\epsilon\lambda & 0 & 0 & 1+\epsilon\lambda\\
\end{smallmatrix}\right].$
%
%Moreover, \[\operatorname{Holant}(\neq_2|f, [1, 0, \epsilon, 0, 1])\leq^p_{T}\operatorname{Holant}(\neq_2|f).\]
%Note that $M(g_{0})=\left[\begin{smallmatrix}
%1 & 0 & 0 & 1\\
%%0 & \epsilon & 1 & 0\\
%1 & 0 & 0 & 1\\
%\end{smallmatrix}\right].$
\end{lemma}
\begin{proof}
Firstly, we consider the case $\epsilon=1$,
i.e.,  $M = M(f)=\left[\begin{smallmatrix}
1 & 0 & 0 & t\\
0 & \frak i^r & \frak i^r t & 0\\
0 & \frak i^r t & \frak i^r & 0\\
t & 0 & 0 & 1\\
\end{smallmatrix}\right]$.
%Let $N=\left [ \begin{smallmatrix}
%0 & 1 \\
%1 & 0
%\end{smallmatrix} \right ]^{\otimes 2}$.
We construct a series of
   gadgets by a chain of $4s$ copies of $f$
linked by two $(\neq_2)$'s
in between. We connect the variables $x_3$ and $x_4$ of
a preceding $f$
with the  variables $x_1$ and $x_2$ of a succeeding $f$ respectively.
See Figure~\ref{interpolation}.
%with  one leading copy of $f$ and
%a sequence of twisted $4s-1$ copies of $f$ linked by two $(\neq_2)$'s
%in between.
It has the signature matrix $D_{4s}= M (N M)^{4s-1} = N (N M)^{4s}$
for $s \ge 1$,
where   $N$ is the double {\sc Disequality}.
This is on the right side of
 $\operatorname{Holant}(\neq_2|f)$.

\begin{figure}
\centering
\subfloat[${\scriptstyle M(f)NM(f)}$]{
\begin{tikzpicture}[scale=0.3]
\node [external, scale=0.5] (-0) at (-5, 2) {};
\node [external] (+0) at (9.5, 5) {};
\node [internal, scale=0.5] (0) at (0, 2) {};
\node at (-0.5, 3) {\tiny{$x_1$}};
\node at (-0.5, 1) {\tiny{$x_2$}};
\node at (0.5, 3) {\tiny{$x_3$}};
\node at (0.5, 1) {\tiny{$x_4$}};
\node [external] (1) at (3, 2) {};
\node [external] (0a) at (-2, 2.7) {};
\node [external] (0b) at (-2, 1.3) {};
\node [internal, scale=0.5] (2) at (4, 2) {};
\node at (3.5, 3) {\tiny{$x_1$}};
\node at (3.5, 1) {\tiny{$x_2$}};
\node at (4.5, 3) {\tiny{$x_3$}};
\node at (4.5, 1) {\tiny{$x_4$}};
\node [external] (2a) at (5.5, 3) {};
\node [external] (+0) at (7.5, 3.2) {};
\node [external] (2b) at (5.5, 1) {};
\draw (0)  to [bend right=30] (2)[postaction={decorate, decoration={
                                        markings,
                                        mark=at position 0.5 with {\arrow[>=square,white, scale=0.7] {>}; },
                                        mark=at position 0.5 with {\arrow[>=open square, scale=0.7]  {>}; } } }];
\draw (0)  to [bend left=30] (2)[postaction={decorate, decoration={
                                        markings,
                                        mark=at position 0.5 with {\arrow[>=square,white, scale=0.7] {>}; },
                                        mark=at position 0.5 with {\arrow[>=open square, scale=0.7]  {>}; } } }];
\draw (0) to [bend right=15] (0a);
\draw (0) to [bend left=15] (0b);
\draw (2) to [bend right=15] (2b);
\draw (2) to [bend left=15] (2a);
\end{tikzpicture}
}
\subfloat[${\scriptstyle M(f)(NM(f))^2}$]{
\begin{tikzpicture}[scale=0.3]
\node [external, scale=0.5] (-0) at (-5, 2) {};
\node [external] (+0) at (10.5, 5) {};
\node [internal, scale=0.5] (0) at (0, 2) {};
\node at (-0.5, 3) {\tiny{$x_1$}};
\node at (-0.5, 1) {\tiny{$x_2$}};
\node at (0.5, 3) {\tiny{$x_3$}};
\node at (0.5, 1) {\tiny{$x_4$}};
\node [external] (1) at (3, 2) {};
\node [external] (0a) at (-2, 2.7) {};
\node [external] (0b) at (-2, 1.3) {};
\node [internal, scale=0.5] (2) at (4, 2) {};
\node at (3.5, 3) {\tiny{$x_1$}};
\node at (3.5, 1) {\tiny{$x_2$}};
\node at (4.5, 3) {\tiny{$x_3$}};
\node at (4.5, 1) {\tiny{$x_4$}};
\node [external, scale=0.5] (3) at (9, 2) {};
\node [internal, scale=0.5] (4) at (8, 2) {};
\node at (7.5, 3) {\tiny{$x_1$}};
\node at (7.5, 1) {\tiny{$x_2$}};
\node at (8.5, 3) {\tiny{$x_3$}};
\node at (8.5, 1) {\tiny{$x_4$}};
\node [external] (4a) at (9.5, 3) {};
\node [external] (4b) at (9.5, 1) {};
\draw (0)  to [bend right=30] (2)[postaction={decorate, decoration={
                                        markings,
                                        mark=at position 0.5 with {\arrow[>=square,white, scale=0.7] {>}; },
                                        mark=at position 0.5 with {\arrow[>=open square, scale=0.7]  {>}; } } }];
\draw (0) to [bend left=30] (2)[postaction={decorate, decoration={
                                        markings,
                                        mark=at position 0.5 with {\arrow[>=square,white, scale=0.7] {>}; },
                                        mark=at position 0.5 with {\arrow[>=open square, scale=0.7]  {>}; } } }];
\draw (2)  to [bend right=30] (4)[postaction={decorate, decoration={
                                        markings,
                                        mark=at position 0.5 with {\arrow[>=square,white, scale=0.7] {>}; },
                                        mark=at position 0.5 with {\arrow[>=open square, scale=0.7]  {>}; } } }];
\draw (2) to [bend left=30] (4)[postaction={decorate, decoration={
                                        markings,
                                        mark=at position 0.5 with {\arrow[>=square,white, scale=0.7] {>}; },
                                        mark=at position 0.5 with {\arrow[>=open square, scale=0.7]  {>}; } } }];
\draw (0) to [bend right=15] (0a);
\draw (0) to [bend left=15] (0b);
\draw (4) to [bend right=15] (4b);
\draw (4) to [bend left=15] (4a);
\end{tikzpicture}
}
\subfloat[${\scriptstyle M(f)(NM(f))^{4s-1}}$]{
\begin{tikzpicture}[scale=0.3]
\node [external, scale=0.5] (-0) at (-2.5, 2) {};
\node [external] (+0) at (12.5, 5) {};
\node [internal, scale=0.5] (0) at (0, 2) {};
\node at (-0.5, 3) {\tiny{$x_1$}};
\node at (-0.5, 1) {\tiny{$x_2$}};
\node at (0.5, 3) {\tiny{$x_3$}};
\node at (0.5, 1) {\tiny{$x_4$}};
\node [external] (1) at (3, 2) {};
\node [external] (0a) at (-2, 2.7) {};
\node [external] (0b) at (-2, 1.3) {};
\node [external] (2) at (4, 2) {};
\node at (4.4, 2) {$\cdots$};
\node [external] (3) at (5, 2) {};
\node [internal, scale=0.5] (4) at (9, 2) {};
\node at (8.5, 3) {\tiny{$x_1$}};
\node at (8.5, 1) {\tiny{$x_2$}};
\node at (9.5, 3) {\tiny{$x_3$}};
\node at (9.5, 1) {\tiny{$x_4$}};
\node [external, scale=0.5] (4a) at (10.5, 2.7) {};
\node [external, scale=0.5] (4b) at (10.5, 1.3) {};
\draw (0)  to [bend right=30] (2)[postaction={decorate, decoration={
                                        markings,
                                        mark=at position 0.5 with {\arrow[>=square,white, scale=0.7] {>}; },
                                        mark=at position 0.5 with {\arrow[>=open square, scale=0.7]  {>}; } } }];
\draw (0)  to [bend left=30] (2)[postaction={decorate, decoration={
                                        markings,
                                        mark=at position 0.5 with {\arrow[>=square,white, scale=0.7] {>}; },
                                        mark=at position 0.5 with {\arrow[>=open square, scale=0.7]  {>}; } } }];
 \draw (3)  to [bend right=30] (4)[postaction={decorate, decoration={
                                        markings,
                                        mark=at position 0.5 with {\arrow[>=square,white, scale=0.7] {>}; },
                                        mark=at position 0.5 with {\arrow[>=open square, scale=0.7]  {>}; } } }];
\draw (3)  to [bend left=30] (4)[postaction={decorate, decoration={
                                        markings,
                                        mark=at position 0.5 with {\arrow[>=square,white, scale=0.7] {>}; },
                                        mark=at position 0.5 with {\arrow[>=open square, scale=0.7]  {>}; } } }];
\draw (0)  to [bend right=15] (0a);
\draw (0)  to [bend left=15] (0b);
\draw (4)  to [bend left=25] (4a);
\draw (4)  to [bend right=25] (4b);
\end{tikzpicture}
}
\caption{{A recursive construction for the interpolation in Lemma~\ref{eigenvalue-interpolation} for the case $\epsilon=1$.
The circles are assigned $f$ and the squares are assigned $(\neq_2)$.}}
\label{interpolation}
\end{figure}
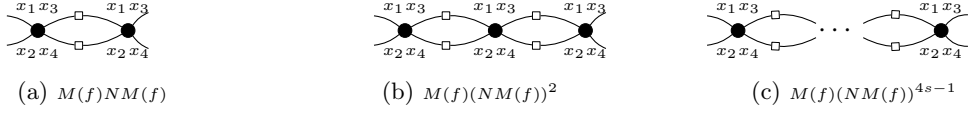
\begin{figure}
\centering
\subfloat[${\scriptstyle M_{x_1x_2, x_4x_3}(f)NM(f)}$]{
\begin{tikzpicture}[scale=0.3]
\node [external, scale=0.5] (-0) at (-6, 2) {};
\node [external] (+0) at (10.5, 5) {};
\node [internal, scale=0.5] (0) at (0, 2) {};
\node at (-0.5, 3) {\tiny{$x_1$}};
\node at (-0.5, 1) {\tiny{$x_2$}};
\node at (0.5, 3) {\tiny{$x_3$}};
\node at (0.5, 1) {\tiny{$x_4$}};
\node [external, scale=0.01] (1) at (3, 2) {};
\node [external] (0a) at (-2, 2.7) {};
\node [external] (0b) at (-2, 1.3) {};
\node [internal, scale=0.5] (2) at (6, 2) {};
\node at (5.5, 3) {\tiny{$x_1$}};
\node at (5.5, 1) {\tiny{$x_2$}};
\node at (6.5, 3) {\tiny{$x_3$}};
\node at (6.5, 1) {\tiny{$x_4$}};
\node [external] (2a) at (7.5, 3) {};
\node [external] (+0) at (7.5, 3.2) {};
\node [external] (2b) at (7.5, 1) {};
\draw (0) to [bend left=45] (1) to [bend right=45] (2)[postaction={decorate, decoration={
                                        markings,
                                        mark=at position 0.2 with {\arrow[>=square,white, scale=0.7] {>}; },
                                        mark=at position 0.2 with {\arrow[>=open square, scale=0.7]  {>}; } } }];
\draw (0) to [bend right=45] (1) to [bend left=45] (2)[postaction={decorate, decoration={
                                        markings,
                                        mark=at position 0.2 with {\arrow[>=square,white, scale=0.7] {>}; },
                                        mark=at position 0.2 with {\arrow[>=open square, scale=0.7]  {>}; } } }];
\draw (0) to [bend right=15] (0a);
\draw (0) to [bend left=15] (0b);
\draw (2) to [bend right=15] (2b);
\draw (2) to [bend left=15] (2a);
\end{tikzpicture}
}
\subfloat[${\scriptstyle M_{x_1x_2, x_4x_3}(f)(NM(f))^{4s-1}}$]{
\begin{tikzpicture}[scale=0.3]
\node [internal, scale=0.5] (0) at (0, 2) {};
\node at (-0.5, 3) {\tiny{$x_1$}};
\node at (-0.5, 1) {\tiny{$x_2$}};
\node at (0.5, 3) {\tiny{$x_3$}};
\node at (0.5, 1) {\tiny{$x_4$}};
\node [external, scale=0.01] (1) at (3, 2) {};
\node [external] (0a) at (-2, 2.7) {};
\node [external] (0b) at (-2, 1.3) {};
\node [internal, scale=0.5] (2) at (6, 2) {};
\node at (5.5, 3) {\tiny{$x_1$}};
\node at (5.5, 1) {\tiny{$x_2$}};
\node at (6.5, 3) {\tiny{$x_3$}};
\node at (6.5, 1) {\tiny{$x_4$}};
\node [external, scale=0.5] (3) at (9, 2) {};
\node [internal, scale=0.5] (4) at (10, 2) {};
\node [internal, scale=0.5] (6) at (14, 2) {};
\node at (9.5, 3) {\tiny{$x_1$}};
\node at (9.5, 1) {\tiny{$x_2$}};
\node at (10.5, 3) {\tiny{$x_3$}};
\node at (10.5, 1) {\tiny{$x_4$}};
%%%%%%%%%%%%%%%%%%%%%%%%%%%%%%%%%%%%%%%%%%%%%%%%%%%%%%%%%%%%%%%%
\node at (13.5, 3) {\tiny{$x_1$}};
\node at (13.5, 1) {\tiny{$x_2$}};
\node at (14.5, 3) {\tiny{$x_3$}};
\node at (14.5, 1) {\tiny{$x_4$}};
%%%%%%%%%%%%%%%%%%%%%%%%%%%%%%%%%%%%%%%%%%%%%%%%%%%%%%%%%%%
\node at (22.5, 3) {\tiny{$x_1$}};
\node at (22.5, 1) {\tiny{$x_2$}};
\node at (23.5, 3) {\tiny{$x_3$}};
\node at (23.5, 1) {\tiny{$x_4$}};
\node [external] (7) at (18, 2) {};
\node at (18.5, 2) {$\cdots$};
\node [external] (8) at (19, 2) {};
\node [internal, scale=0.5] (9) at (23, 2) {};
\node [external, scale=0.5] (9a) at (24.5, 2.7) {};
\node [external, scale=0.5] (9b) at (24.5, 1.3) {};
\draw (6) to [bend right=30] (7) [postaction={decorate, decoration={
                                        markings,
                                        mark=at position 0.5 with {\arrow[>=square,white, scale=0.7] {>}; },
                                        mark=at position 0.5 with {\arrow[>=open square, scale=0.7]  {>}; } } }];
\draw (6) to [bend left=30] (7) [postaction={decorate, decoration={
                                        markings,
                                        mark=at position 0.5 with {\arrow[>=square,white, scale=0.7] {>}; },
                                        mark=at position 0.5 with {\arrow[>=open square, scale=0.7]  {>}; } } }];
\draw (8) to [bend right=30] (9) [postaction={decorate, decoration={
                                        markings,
                                        mark=at position 0.5 with {\arrow[>=square,white, scale=0.7] {>}; },
                                        mark=at position 0.5 with {\arrow[>=open square, scale=0.7]  {>}; } } }];
\draw (8) to [bend left=30] (9) [postaction={decorate, decoration={
                                        markings,
                                        mark=at position 0.5 with {\arrow[>=square,white, scale=0.7] {>}; },
                                        mark=at position 0.5 with {\arrow[>=open square, scale=0.7]  {>}; } } }];
\draw (9) to [bend left=25] (9a);
\draw (9) to [bend right=25] (9b);
%\node [external] (4a) at (13.5, 3) {};
%\node [external] (4b) at (13.5, 1) {};
\draw (0) to [bend left=45] (1) to [bend right=45] (2)[postaction={decorate, decoration={
                                        markings,
                                        mark=at position 0.2 with {\arrow[>=square,white, scale=0.7] {>}; },
                                        mark=at position 0.2 with {\arrow[>=open square, scale=0.7]  {>}; } } }];
\draw (0) to [bend right=45] (1) to [bend left=45] (2)[postaction={decorate, decoration={
                                        markings,
                                        mark=at position 0.2 with {\arrow[>=square,white, scale=0.7] {>}; },
                                        mark=at position 0.2 with {\arrow[>=open square, scale=0.7]  {>}; } } }];
\draw (2) to  [bend right=30] (4)[postaction={decorate, decoration={
                                        markings,
                                        mark=at position 0.5 with {\arrow[>=square,white, scale=0.7] {>}; },
                                        mark=at position 0.5 with {\arrow[>=open square, scale=0.7]  {>}; } } }];
\draw (2) to  [bend left=30] (4)[postaction={decorate, decoration={
                                        markings,
                                        mark=at position 0.5 with {\arrow[>=square,white, scale=0.7] {>}; },
                                        mark=at position 0.5 with {\arrow[>=open square, scale=0.7]  {>}; } } }];
\draw (4) to  [bend left=30] (6)[postaction={decorate, decoration={
                                        markings,
                                        mark=at position 0.5 with {\arrow[>=square,white, scale=0.7] {>}; },
                                        mark=at position 0.5 with {\arrow[>=open square, scale=0.7]  {>}; } } }];
                                        \draw (2) to  [bend right=30] (4)[postaction={decorate, decoration={
                                        markings,
                                        mark=at position 0.5 with {\arrow[>=square,white, scale=0.7] {>}; },
                                        mark=at position 0.5 with {\arrow[>=open square, scale=0.7]  {>}; } } }];
\draw (4) to  [bend right=30] (6)[postaction={decorate, decoration={
                                        markings,
                                        mark=at position 0.5 with {\arrow[>=square,white, scale=0.7] {>}; },
                                        mark=at position 0.5 with {\arrow[>=open square, scale=0.7]  {>}; } } }];
\draw (0) to [bend right=15] (0a);
\draw (0) to [bend left=15] (0b);
%\draw (4) to [bend right=15] (4b);
%\draw (4) to [bend left=15] (4a);
\end{tikzpicture}
}
\caption{{Another recursive construction for the interpolation in Lemma~\ref{eigenvalue-interpolation} when $\epsilon=-1$, with a twist
at the beginning.
The variables  $x_3$ and $x_4$ of the first
copy of $f$ are connected to $x_2$ and $x_1$ of the
second copy of $f$ respectively.
The circles are assigned $f$ and the squares are $(\neq_2)$.}}
\label{interpolation1}
\end{figure}
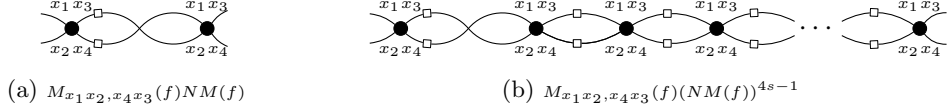

The signature matrix  of this gadget is
 given as a product of matrices. Each matrix is a function of arity $4$.
Notice that
$N M=
\left [ \begin{smallmatrix}
t & 0 & 0 & 1 \\
0 & \frak i^r t & \frak i^r & 0 \\
0 & \frak i^r & \frak i^r t & 0 \\
1 & 0 & 0 & t
\end{smallmatrix} \right ]$, and thus
\[ D_{4s}
=N P \cdot
\left [ \begin{smallmatrix}
t+1 & 0 & 0 & 0 \\
0 & \frak i^r (t+1) & 0 & 0 \\
0 & 0 & \frak i^r (t-1) & 0 \\
0 & 0 & 0 & t-1
\end{smallmatrix} \right ]^{4s}
\cdot P
= P
\left [ \begin{smallmatrix}
1 & 0 & 0 & 0 \\
0 & 1 & 0 & 0 \\
0 & 0 & -1 & 0 \\
0 & 0 & 0 & -1
\end{smallmatrix} \right ]
\cdot
\left [ \begin{smallmatrix}
t+1 & 0 & 0 & 0 \\
0 & t+1 & 0 & 0 \\
0 & 0 & t-1 & 0 \\
0 & 0 & 0 & t-1
\end{smallmatrix} \right ]^{4s}
\cdot P
=(t+1)^{4s}
P\Lambda_{4s} P,\]
where
$P= P^{-1} =\frac{1}{\sqrt{2}}\left [ \begin{smallmatrix}
1 & 0 & 0 & 1 \\
0 & 1 & 1 & 0\\
0 & 1 & -1 & 0\\
1 & 0 & 0 & -1\\
\end{smallmatrix} \right ]$,
and $\Lambda_{4s}=\left [ \begin{smallmatrix}
1 & 0 & 0 & 0 \\
0 & 1 & 0 & 0\\
0 & 0 & -\rho^{4s} & 0\\
0 & 0 & 0 & - \rho^{4s}\\
\end{smallmatrix} \right ]$,
with $\rho = \frac{t-1}{t+1}$.
%where $T=\left [ \begin{array}{cc}  b & c \\ c & b \end{array} \right ]^s=H \left [ \begin{array}{cc}  (b+c)^s & 0 \\ 0 & (b-c)^s \end{array} \right ]  H$.
By assumption, $t \not = \pm 1$ and $t \not \in \frak i \mathbb{R}$,
then $\rho$ is well defined, $\rho \not =0$,  and $|\rho| \not =1$.

This matrix $\Lambda_{4s}$ has a good form for polynomial interpolation.
Suppose  $\Omega$ is
a signature grid of  $\operatorname{Holant}(\neq_2\mid f, g_{\lambda})$,
 %to be reduced to
 %$\operatorname{Holant}(\neq_2\mid f)$,
in which
 $g_{\lambda}$ appears $m$ times.
We have
\[M(g_{\lambda}) =
\left[\begin{smallmatrix}
1+\lambda & 0 & 0 & 1-\lambda\\
0 & 1+\lambda & 1-\lambda & 0\\
0 & 1-\lambda & 1+\lambda & 0\\
1-\lambda & 0 & 0 & 1+\lambda\\
\end{smallmatrix}\right]
=
2 P \cdot
\left [ \begin{smallmatrix}
1 & 0 & 0 & 0 \\
0 & 1 & 0 & 0\\
0 & 0 & \lambda & 0\\
0 & 0 & 0 & \lambda\\
\end{smallmatrix} \right ]  \cdot P.\]
Take out  a factor $2$,
we can treat each of the $m$ appearances of $M(g_{\lambda})$
 as being composed of three functions in sequence
$P$, $D_{\lambda} = {\rm diag}\{1,1,\lambda,\lambda\}$, and $P$.
Define a new signature grid $\Omega'$
consisting of two parts.
 One part is composed of $m$ functions $D_{\lambda}$.
The second part consists of
 the rest of the functions, including $2m$ occurrences
of $P$, and the signature of this second part
 is represented by $X$ (which is a tensor expressed
as a row vector).
 The Holant value  $\operatorname{Holant}_{\Omega'}
=\operatorname{Holant}_{\Omega}$
is the dot product $\langle X, D_{\lambda}^{\otimes m}  \rangle$,
 which is a summation over $4m$ bits, that is, the values of the $4m$ edges connecting the two parts. We can  stratify
all 0-1 assignments of these  $4m$ bits having a nonzero
evaluation of $\operatorname{Holant}_{\Omega'}$ into the following categories:
\begin{itemize}
\item
There are $i$ many copies of $D_{\lambda}$ receiving inputs $0000$ or $0101$, and
\item
There are $j$ many copies of $D_{\lambda}$ receiving inputs $1010$ or $1111$,
\end{itemize}
such that $i+j= m$.
 Therefore
 the value
$\operatorname{Holant}_{\Omega}$
for the problem
 $\operatorname{Holant}(\neq_2\mid f, g_{\lambda})$
is
\begin{equation}  \label{eqn: holant-value-for-g_lambda}
\operatorname{Holant}_{\Omega}
=2^m \sum_{i+j=m} \lambda^{j} x_{i,j},
\end{equation}
where $x_{i,j}$ is the sum of values of
the second part $X$ over all assignments in the category $(i,j)$.

 Now we can replace each appearance of $g_{\lambda}$ by a copy of
the gadget in  Figure~\ref{interpolation} with the signature $D_{4s}$, to get an instance $\Omega_{4s}$ of $\operatorname{Holant}(\neq_2\mid f)$.
Ignoring the nonzero scalar $(t+1)^{4s}$,
we can treat each of the $m$ appearances of $D_{4s}$ as
being composed of three functions in sequence
$P$, $\Lambda_{4s}$ and $P$, and denote this new instance by $\Omega'_{4s}$. We divide  $\Omega'_{4s}$ into two parts. One part is composed of $m$ functions $\Lambda_{4s}$. The second part is the rest of the functions, including
the  $2m$ occurrences
of $P$, and its signature $X$ is the same
as in (\ref{eqn: holant-value-for-g_lambda}) for $\operatorname{Holant}_{\Omega}$.
 The Holant value of  $\Omega'_{4s}$ is the dot product $\langle X, \Lambda_{4s}^{\otimes m}  \rangle$, which is a summation over $4m$ bits, that is, the values of the $4m$ edges connecting the two parts. With the same stratification of
all 0-1 assignments of these  $4m$ bits having a nonzero
evaluation of $\operatorname{Holant}_{\Omega'_{4s}}$, we have
\begin{equation}  \label{eqa: interpo1}
 \operatorname{Holant}_{\Omega_{4s}} =
 \operatorname{Holant}_{\Omega'_{4s}} =
%{{(G_s)=\#(G'_s)=\langle X, \Lambda_s^{\otimes m}  \rangle
\langle X, \Lambda_{4s}^{\otimes m}  \rangle
= (t+1)^{4sm} \sum_{i+j=m}  (-1)^j \rho^{4sj}x_{i, j},
\end{equation}
where $x_{i,j}$ are the same as in (\ref{eqn: holant-value-for-g_lambda}).
%is the sum of values of
%the second part $X$ over all assignments in the category $(i,j)$.
%Let $y_{i, j}=(-1)^jx_{i, j}$, then
%\begin{equation}  \label{eqa: interpo}
% \operatorname{Holant}_{\Omega_{4s}} =
% \operatorname{Holant}_{\Omega'_{4s}} =
%%{{(G_s)=\#(G'_s)=\langle X, \Lambda_s^{\otimes m}  \rangle
%\langle X, \Lambda_{4s}^{\otimes m}  \rangle
%=\sum_{i+j=m}  y_{i, j}(\frac{t-1}{t+1})^{4sj}.
%\end{equation}

We pick $s= 1, 2, \ldots, m+1$, and get a system of linear equations
in $x_{i,j}$.
By $\rho \not =0$ and $|\rho| \not= 1$,
the Vandermonde coefficient matrix
of the linear system has full rank, and then we can solve for
each $(-1)^j x_{i,j}$, and therefore $x_{i,j}$.
And so we can compute
$\operatorname{Holant}_{\Omega}$ in (\ref{eqn: holant-value-for-g_lambda}).
Hence
\[\operatorname{Holant}(\neq_2|f, g_{\lambda})\leq_T^p\operatorname{Holant}(\neq_2|f).\]
%where $g_{\lambda}$ is a 4-ary signature whose signature matrix
%\[M(g_{\lambda})=P\left[\begin{smallmatrix}
%1 & 0 & 0 & 0\\
%0 & 1 & 0 & 0\\
%0 & 0 & \lambda & 0\\
%0 & 0 & 0 & \lambda\\
%\end{smallmatrix}\right]P=\frac{1}{2}\left[\begin{smallmatrix}
%1+\lambda & 0 & 0 & 1-\lambda\\
%0 & 1+\lambda & 1-\lambda & 0\\
%0 & 1-\lambda & 1+\lambda & 0\\
%1-\lambda & 0 & 0 & 1+\lambda\\
%\end{smallmatrix}\right].\]
%

Now we consider the case $\epsilon=-1$, i.e.,
 $M(f)=\left[\begin{smallmatrix}
1 & 0 & 0 & t\\
0 & \frak i^r  & -\frak i^r t & 0\\
0 & -\frak i^r t & \frak i^r & 0\\
t & 0 & 0 & 1\\
\end{smallmatrix}\right]$.
Note that $M_{x_1x_2, x_4x_3}(f)=M^{\sf R_{(34)}}(f)=\left[\begin{smallmatrix}
1 & 0 & 0 & t\\
0 &   -\frak i^r t & \frak i^r  & 0\\
0 &  \frak i^r &-\frak i^r t  & 0\\
t & 0 & 0 & 1\\
\end{smallmatrix}\right]$.
%=\left[\begin{smallmatrix}
%1 &  &  & 0\\
% &  0 & 1 & \\
% &  1 & 0  & \\
%0 &  &  & 1\\
%\end{smallmatrix}\right]
%M(f)$.
We construct a new series of
   gadgets as in Figure~\ref{interpolation1}.
This is also a  chain of  $4s$ copies of $f$,
but we introduce a twist in the connection
between the first and second copy of $f$.
Specifically we connect respectively
the variables $x_4$ and $x_3$ of the first copy of $f$
with the  variables $x_1$ and $x_2$ of the second  copy of $f$ via $N$.
% (See Figure~\ref{interpolation1}).
Thus
 the signature matrix is
\begin{eqnarray*}
D'_{4s} &=&
 M_{x_1x_2, x_4x_3}(f) (N M(f))^{4s-1}=
\left[\begin{smallmatrix}
0 &  &  & 1\\
 &  1 & 0 & \\
 &  0 & 1  & \\
1 &  &  & 0\\
\end{smallmatrix}\right]
(N M(f))^{4s} \\
&=&
P
\left[\begin{smallmatrix}
1 &  &  & \\
 &  1 &  & \\
 &   & 1  & \\
 &  &  &  -1\\
\end{smallmatrix}\right]  P^{-1} P
\left[\begin{smallmatrix}
1+t &  &  & \\
 &  {\frak i}^r (1-t) &  & \\
 &   & - {\frak i}^r (1+t)  & \\
 &  &  &  t-1\\
\end{smallmatrix}\right] ^{4s} P^{-1}\\
&=&P \left[\begin{smallmatrix}
1 &  &  & \\
 &  1 &  & \\
 &   & 1  & \\
 &  &  &  -1\\
\end{smallmatrix}\right]
\left[\begin{smallmatrix}
t+1 &  &  & \\
 &  t-1 &  & \\
 &   & t+1  & \\
 &  &  &  t-1\\
\end{smallmatrix}\right] ^{4s} P
=
(t+1)^{4s} P
\left[\begin{smallmatrix}
1 &  &  & \\
 &  \rho^{4s} &  & \\
 &   & 1  & \\
 &  &  &  -\rho^{4s}\\
\end{smallmatrix}\right]  P,
\end{eqnarray*}
%(1+t)^{4s}P\left [ \begin{smallmatrix}
%1 & 0 & 0 & 0 \\
%0 & (\frac{t-1}{t+1})^{4s} & 0 & 0\\
%0 & 0 & 1 & 0\\
%0 & 0 & 0 & -(\frac{t-1}{t+1})^{4s}\\
%\end{smallmatrix} \right ]P.\]
where $P= P^{-1} =\frac{1}{\sqrt{2}}\left [ \begin{smallmatrix}
1 & 0 & 0 & 1 \\
0 & 1 & 1 & 0\\
0 & 1 & -1 & 0\\
1 & 0 & 0 & -1\\
\end{smallmatrix} \right ]$, and  $\rho = \frac{t-1}{t+1}$ are as
above.

Suppose  $\Omega$ is an instance
of $\operatorname{Holant}(\neq_2| f, g_{\lambda})$ in which
 $g_{\lambda}$ appears $m$ times.
%We replace each appearance of $g_{\lambda}$ by a copy of
%the gadget $D_{4s}$, to get an instance $\Omega_{4s}$ of $\operatorname{Holant}(\neq_2\mid f)$.
%We can treat each of the $m$ appearances of $D_{4s}$ as a new gadget composed of three functions in sequence
%$P$, $\Lambda_{4s}$ and $P$, and denote
We similarly construct instances  $\Omega_{4s}$ of $\operatorname{Holant}(\neq_2\mid f)$
by replacing each occurrence of $g_{\lambda}$ with
$D'_{4s}$, for $s \ge 1$.
Decompose each $D'_{4s}$ as a product of three functions
in sequence, $P$, $\Lambda_{4s} = \diag\{1, \rho^{4s}, 1, -\rho^{4s}\}$, $P$,
and
%define
%a new signature grid $\Omega'_{4s}$
%of $\operatorname{Holant}(\neq_2\mid f)$
% by  dividing  $\Omega'_{4s}$
divide the signature grid  into two parts.
One part is composed of $m$ functions $\Lambda_{4s}$.
The second part is the rest of the functions, including the $2m$ occurrences
of $P$, and its signature is represented by $X$ (which is a tensor expressed
as a row vector).
Note that $X$ is independent of $s$.
 %The Holant value of  $\Omega'_{4s}$ is the dot product $\langle X, \Lambda_{4s}^{\otimes m}  \rangle$, which is a summation over $4m$ bits, that is, the values of the $4m$ edges connecting the two parts. We can
 Then we stratify
all 0-1 assignments of these  $4m$ bits having a nonzero
evaluation of the Holant value into the following categories:
% each of which corresponds to an ordered input array of $\Lambda_s$.
\begin{itemize}
\item
There are $i$ many copies of $\Lambda_{4s}$ receiving inputs $0000$ or $1010$;
\item
There are $j$ many copies of $\Lambda_{4s}$ receiving inputs $0101$;
\item
There are $k$ many copies of $\Lambda_{4s}$ receiving inputs $1111$
\end{itemize}
such that $i+j+k=m$.

For any assignment in the category with parameter $(i,j, k)$, the evaluation of
$\Lambda_{4s}^{\otimes m}$ is clearly $(-1)^k\rho^{4s(j+k)}$.
 We can rewrite the dot product  and  get
\begin{equation}  \label{eqa: interpo2}
 \operatorname{Holant}_{\Omega_{4s}} =
%{{(G_s)=\#(G'_s)=\langle X, \Lambda_s^{\otimes m}  \rangle
\langle X, \Lambda_{4s}^{\otimes m}  \rangle
=\sum_{i+j+k=m}  (-1)^k\rho^{4s(j+k)}x_{i, j, k},
\end{equation}
where $x_{i,j, k}$ is the summation of values of
the second part $X$ over all assignments in the category $(i,j, k)$.
Let $y_i=\sum_{j+k=m-i}(-1)^kx_{i, j, k}$, for $0 \le i \le m$, then
\begin{equation}  \label{eqa: interpo3}
 \operatorname{Holant}_{\Omega_{4s}} =
%{{(G_s)=\#(G'_s)=\langle X, \Lambda_s^{\otimes m}  \rangle
%\langle X, \Lambda_{4s}^{\otimes m}  \rangle
\sum_{0 \le i \le m}  y_{i}\rho^{4s(m-i)}.
\end{equation}

Since  $|\rho| \not =0, 1$, by the same argument as for the case
$\epsilon=1$, we can compute $y_i$.
%Again, we pick polynomially many values of $s$, and get a system of linear equations in $y_{i, n}$.
%By $t^4\neq 0, 1$, we have $t+1\neq 0, t-1\neq 0$, and  $t$ is not a pure imagine number, $|\frac{t-1}{t+1}|\neq 1$. So the Vandermonde coefficient matrix
%of the linear equations has full rank, and then we can solve for
%each $y_{i,j}$.
Once we have $y_{i}$ we can compute
\begin{equation*}
% \operatorname{Holant}_{\Omega}
%{{(G_s)=\#(G'_s)=\langle X, \Lambda_s^{\otimes m}  \rangle
%\langle X, \Lambda_{4s}^{\otimes m}  \rangle
\sum_{0 \le i \le m}  y_{i}\lambda^{m-i}=
\sum_{i+j+k=m}  x_{i, j, k}(-1)^k\lambda^{j+k}
\end{equation*} for any $\lambda\in\mathbb{C}$.

Since  for $\epsilon =-1$,
\[
\frac{1}{2}M(g_{\lambda})=
\frac{1}{2}\left[\begin{smallmatrix}
1-\lambda & 0 & 0 & 1+\lambda\\
0 & 1+\lambda & \lambda-1 & 0\\
0 & \lambda-1 & 1+\lambda & 0\\
1+\lambda & 0 & 0 & 1-\lambda\\
\end{smallmatrix}\right]
=P
\left[\begin{smallmatrix}
1  &  &  & \\
 & \lambda &  & \\
 &  & 1 &  \\
 &  &  & -\lambda \\
\end{smallmatrix}\right]
P,\]
we have
$ \operatorname{Holant}_{\Omega} =
\sum_{i+j+k=m}  x_{i, j, k}(-1)^k\lambda^{j+k}$, and
\[\operatorname{Holant}(\neq_2|f, g_{\lambda})\leq_T^p\operatorname{Holant}(\neq_2|f).\]
%
%
%The signature matrix  of this gadget is
% given as a product of matrices. Each matrix is a function of arity $4$.
%Notice that
%%\[N M(f)=
%\left [ \begin{smallmatrix}
%t & 0 & 0 & 1 \\
%%0 & \frak{i}t & \frak{i} & 0 \\
%0 & \frak{i} & \frak{i}t & 0 \\
%1 & 0 & 0 & t
%\end{smallmatrix} \right ], ~~~~\mbox{and}~~~~ D_{4s}=(t+1)^{4s}
%P\Lambda_{4s} P^{-1},\]
%where $\Lambda_s=(1+t)^{4s}\left [ \begin{smallmatrix}
%1 & 0 & 0 & 0 \\
%0 & 1 & 0 & 0\\
%0 & 0 & -(\frac{t-1}{t+1})^{4s} & 0\\
%0 & 0 & 0 & -(\frac{t-1}{t+1})^{4s}\\
%\end{smallmatrix} \right ], P=\frac{1}{\sqrt{2}}\left [ \begin{smallmatrix}
%1 & 0 & 0 & 1 \\
%0 & 1 & 1 & 0\\
%0 & 1 & -1 & 0\\
%1 & 0 & 0 & -1\\
%\end{smallmatrix} \right ].$
%%where $T=\left [ \begin{array}{cc}  b & c \\ c & b \end{array} \right ]^s=H \left [ \begin{array}{cc}  (b+c)^s & 0 \\ 0 & (b-c)^s \end{array} \right ]  H$.
%The remainging proof is completely same as the first case and we omit it here.
\end{proof}

% By the similar proof, we have the following lemma.
%\begin{lemma}\label{eigenvalue-interpolation-2}
%Let $f$ be a 4-ary signature with the signature matrix
%$M(f)=\left[\begin{smallmatrix}
%1 & 0 & 0 & t\\
%%0 & 1 & -t & 0\\
%0 & -t & 1 & 0\\
%t & 0 & 0 & 1\\
%\end{smallmatrix}\right]$ or $M(f)=\left[\begin{smallmatrix}
%1 & 0 & 0 & t\\
%0 & \frak{i} & -\frak{i}t & 0\\
%0 & -\frak{i}t & \frak{i} & 0\\
%t & 0 & 0 & 1\\
%\end{smallmatrix}\right]$
%where $t^4\neq 0, 1$  and $t$ is not a pure imagine number, then for any $\lambda\in\mathbb{C}$
%\[\operatorname{Holant}(\neq_2|f)\geq_{T}\operatorname{Holant}(\neq_2|f, g_{\lambda}),\]
%where $g_t$ is a 4-ary signature whose signature matrix
%$M(g_{\lambda})=\left[\begin{smallmatrix}
%1+\lambda & 0 & 0 & 1-\lambda\\
%0 & \lambda+1 & \lambda-1 & 0\\
%0 & \lambda-1 & \lambda+1 & 0\\
%1-\lambda & 0 & 0 & 1+\lambda\\
%\end{smallmatrix}\right].$
%\end{lemma}

\begin{lemma}\cite{cai:fu:plcsp}\label{construct-equality-4}
Let $f$ be a 4-ary signature with the signature matrix
$M(f)=\left[\begin{smallmatrix}
a & 0 & 0 & b\\
0 & 0 & 0 & 0\\
0 & 0 & 0 & 0\\
y & 0 & 0 & x\\
\end{smallmatrix}\right],$
where $\left[\begin{smallmatrix}
a &  b\\
y &  x\\
\end{smallmatrix}\right]$ has full rank,  then for any signature set $\mathcal{F}$,
\[\operatorname{Holant}(=_2|\mathcal{F}, f, =_4)\leq^p_T \operatorname{Holant}(=_2|\mathcal{F}, f).\]
\end{lemma}

\begin{lemma}\label{inter:2[1,0,1]}
Let $g$ be  a 4-ary signature with the signature matrix
 $M(g)=\left[\begin{smallmatrix}
t & 0 & 0 & 1\\
0 & 0 & 0 & 0\\
0 & 0 & 0 & 0\\
1 & 0 & 0 & t\\
\end{smallmatrix}\right]$, where $t \not = 0$ and $t^4\neq 1$.
Then for any signature set $\mathcal{F}$,% is a signature set and all the signatures in $\mathcal{F}$ have arity $\equiv 0 \pmod 4$.
\[\operatorname{Holant}(\neq_2|\mathcal{EQ}_2, \mathcal{F})\leq^p_T\operatorname{Holant}(\neq_2|\mathcal{F}, g).\]
In particular, {\rm \#CSP}$^2(\mathcal{F})$ $\leq^p_T\operatorname{Holant}(\neq_2|\mathcal{F}, g)$.
\end{lemma}
\begin{proof}
Firstly, we claim that
\begin{equation}\label{eqn:sec2-get-double101}
 \operatorname{Holant}(\neq_2|\mathcal{F}, g, [1, 0, 1]^{\otimes 2})\leq^p_T\operatorname{Holant}(\neq_2|\mathcal{F}, g).
\end{equation}
Note that
$M_{x_1x_3, x_2x_4}(g)=M^{\sf R_{(23)}}(g)=\left[\begin{smallmatrix}
t & 0 & 0 & 0\\
0 & 1 & 0 & 0\\
0 & 0 & 1 & 0\\
0 & 0 & 0 & t\\
\end{smallmatrix}\right]$.
We denote by $M=M_{x_1x_3, x_2x_4}(g)$.
Note that $NMN = M$, thus $(NM)^2 = NMNM = M^2$.
By connecting $2s+1$ copies of $g$ in the form of $M$, via $N$,
we get a signature $g_{2s+1}$ with the signature matrix
\[
M(NM)^{2s} = M^{2s+1} =\left [ \begin{smallmatrix}
t^{2s+1} & 0 & 0 & 0 \\
0 & 1 & 0 & 0\\
0 & 0 & 1 & 0\\
0 & 0 & 0 & t^{2s+1}\\
\end{smallmatrix} \right ].\]
If $t$ is not a root of unity, %we can use the same method interpolate $h$ whose signature matrix is
then by interpolation, we have
\[
 \operatorname{Holant}(\neq_2|\mathcal{F}, g, h)\leq^p_T\operatorname{Holant}(\neq_2|\mathcal{F}, g),
 \]
 for the following signature $h$ with the signature matrix
$M(h)=\left [ \begin{smallmatrix}
1 & 0 & 0 & 0 \\
0 & 1 & 0 & 0\\
0 & 0 & 1 & 0\\
0 & 0 & 0 & 1\\
\end{smallmatrix} \right ]$, i.e., $h=[1, 0, 1]^{\otimes 2}$,
a double copy of the binary {\sc Equality} $(=_2)$.

If $t$ is a root of unity, by connecting two copies of $g$
via $N$, we get a signature $g'$ with the signature matrix
$M(g)NM(g)=\left [ \begin{smallmatrix}
t & 0 & 0 & 1 \\
0 & 0 & 0 & 0\\
0 & 0 & 0 & 0\\
1 & 0 & 0 & t\\
\end{smallmatrix} \right ]N\left [ \begin{smallmatrix}
t & 0 & 0 & 1 \\
0 & 0 & 0 & 0\\
0 & 0 & 0 & 0\\
1 & 0 & 0 & t\\
\end{smallmatrix} \right ]=2t\left [ \begin{smallmatrix}
1 & 0 & 0 & \frac{1+t^2}{2t} \\
0 & 0 & 0 & 0\\
0 & 0 & 0 & 0\\
\frac{1+t^2}{2t} & 0 & 0 & 1\\
\end{smallmatrix} \right ]$.
Since $t \neq 0$,  $t^4\neq  1$
and $|t|=1$, we have  $|\frac{1+t^2}{2t}|\neq 0, 1$.
Then replacing $g$ by $g'$, we can get $[1, 0, 1]^{\otimes 2}$ by $g'$ again.
This finishes the proof of the claim,
 i.e., (\ref{eqn:sec2-get-double101}) holds.
%\begin{equation*}\label{re}
% \operatorname{Holant}(\neq_2|\mathcal{F}, g, [1, 0, 1]^{\otimes 2})\leq^p_T\operatorname{Holant}(\neq_2|\mathcal{F}, g).
% \end{equation*}

Then by Lemma~\ref{lm:Lin-Wang} (we apply Lemma~\ref{lm:Lin-Wang}
after a holographic transformation $Z$, and note that $(=_2)Z^{\otimes 2}
= (\ne_2)$), we have
\begin{equation*}\label{re1}
 \operatorname{Holant}(\neq_2|\mathcal{F}, g, [1, 0, 1])\leq^p_T\operatorname{Holant}(\neq_2|\mathcal{F}, g).
 \end{equation*}

 By $\neq_2$, we can move $[1, 0, 1]$ to the left side of the Holant
problem. Thus we have
 %\begin{equation}\label{re2}
 %\operatorname{Holant}(\neq_2|\mathcal{F}, g, [1, 0, 1]^{\otimes 2})\geq_{T}\operatorname{Holant}([1, 0, 1]^{\otimes 2}|\mathcal{F}, g).
 %\end{equation}
 %n an instance of $\operatorname{Hol}([1, 0, 1]|\mathcal{F}, g)$, since all the signatures on RHS have arity $\equiv \pmod$4, the number of $[1, 0, 1]$ on LHS
 %has to  be even. Thus we have
 \begin{equation}\label{re2}
 \operatorname{Holant}(\neq_2, [1, 0, 1]|\mathcal{F}, g)\leq^p_T\operatorname{Holant}(\neq_2|\mathcal{F}, g).
 \end{equation}
 Since $\left [ \begin{smallmatrix}
t & 1 \\
1 & t\\
\end{smallmatrix} \right ]$ has full rank,  by Lemma~\ref{construct-equality-4}
  we have
 \begin{equation}\label{re3}
 \operatorname{Holant}(\neq_2, [1, 0, 1]|\mathcal{F}, =_4)\leq^p_T\operatorname{Holant}(\neq_2|\mathcal{F}, g).
 \end{equation}
 Then by Lemma~\ref{equality-4-csp2}, we have
 \begin{equation}\label{re4}
 \operatorname{Holant}(\neq_2|\mathcal{F}, \mathcal{EQ}_2)\leq^p_T\operatorname{Holant}(\neq_2|\mathcal{F}, g),
 \end{equation}
and
\begin{equation}\label{re5}
{\rm \#CSP}^2(\mathcal{F}) \leq^p_T\operatorname{Holant}(\neq_2|\mathcal{F}, g).
 \end{equation}

%By (\ref{re2}), (\ref{re3}), and (\ref{re4}), we have
%  \[
% \#\operatorname{CSP}^2(\mathcal{F})\leq^p_T\operatorname{Holant}(\neq_2|\mathcal{F}, g).
%\]
\end{proof}

\subsection{M\"{o}bius Transformation}
A M\"{o}bius transformation~\cite{ahlfors} is
a mapping of the form ${\mathfrak z}
\mapsto \frac{a\mathfrak z+b}{c\mathfrak z+d}$,
where $\det \left[\begin{smallmatrix}
a & b \\
c & d
\end{smallmatrix}\right] \not =0$.
It is a bijective
 conformal map of the extended complex plane $\widehat{\mathbb{C}}
= \mathbb{C} \cup \{ \infty \}$ to itself.
%where the coefficients $a, b, c, d$ are complex numbers satisfying $ad-bc\neq 0$.
%A M\"{o}bius transformation is a bijective conformal function from the extended complex plane to the extended complex plane.
%It is a  conformal map that maps a circle to a circle or a line.
%In particular,
A M\"{o}bius transformation maps the unit circle $S^1=\{\mathfrak{z}
\in \mathbb{C} \mid |\mathfrak{z}|=1\}$ to itself iff it is of the form
$\varphi(\mathfrak z)=e^{i\theta}\frac{\mathfrak{z}+\lambda}{1+\bar{\lambda} \mathfrak{z}}$ denoted by $\mathcal{M}(\lambda, e^{i\theta})$, where $|\lambda|\neq 1$. When $|\lambda|< 1$, it maps the interior of $S^1$ to the interior,
and when $|\lambda|> 1$, it maps the interior of $S^1$ to the exterior.
A M\"{o}bius transformation is determined by its values on any 3
distinct points. In particular if there are  5 distinct
points $\mathfrak{z}_i \in S^1$, such that $|\varphi({\mathfrak z}_i)|$
is either 0 or 1 or $\infty$, then it must map $S^1$ to $S^1$ in
a bijection (and thus in fact $\varphi({\mathfrak z}_i) \not = 0, \infty$).

The following lemma gives a sufficient condition to
getting an arbitrary binary signature of the form
$(0, 1, u, 0)^T$.
\begin{lemma}\label{binary-interpolation}
Suppose there is a gadget construction in
$\operatorname{Holant}(\neq_2|\mathcal{F})$ that can produce
a sequence of polynomially many distinct binary signatures
of the form  $g=(0, 1, t, 0)^T$ of polynomial size description,
then for any $u\in\mathbb{C}$,
\[
\operatorname{Holant}(\neq_2|\mathcal{F}, (0, 1, u, 0)^T)\leq_{T}\operatorname{Holant}(\neq_2|\mathcal{F}).\]
\end{lemma}

\begin{corollary}\label{cor:binary-interpolation}
The conclusion of Lemma~\ref{binary-interpolation}
holds if $\mathcal{F}$
contains some  $(0, 1, t, 0)^T$,
where $t \neq 0$ and is not a root of unity.
%Also, the conclusion holds if $\mathcal{F}$
%contains three distinct $(0, 1, t, 0)^T$ with $|t| =1$,
% and some $f$ with $M(f) = \left [ \begin{smallmatrix}
%a & 0 & 0 & b \\
%0 & c & d & 0\\
%0 & w & z & 0\\
%y & 0 & 0 & a\\
%\end{smallmatrix} \right ]$
%such that
%$\varphi(\mathfrak z)=\frac{c\mathfrak{z}+d}{w\mathfrak{z}+
%z}$  has infinite order.
\end{corollary}

%\begin{proof}
%Forming a chain of $(0, 1, t, 0)^T$ connected by $\neq_2$
 %gives $(0, 1, t^i, 0)^T$, as
 %\[\left [ \begin{smallmatrix}
 %0 & 1\\
 %t^{i-1} & 0
 %\end{smallmatrix} \right ]
 %\left [ \begin{smallmatrix}
 %0 & 1\\
 %1 & 0
 %\end{smallmatrix} \right ]
 %\left [ \begin{smallmatrix}
 %0 & 1\\
 %t & 0
 %\end{smallmatrix} \right ]
 %=
 %\left [ \begin{smallmatrix}
 %0 & 1\\
 %t^i & 0
 %\end{smallmatrix} \right ].\]
%
%If $g(x_1, x_2) = (0, 1, t, 0)^T$, then
%connecting $x_3, x_4$ of $f$ to $x_1, x_2$ of $g$  respectively,
 %by two edges labeled $\neq_2$,
 %gives
 %\[M(f) N g
 %= \left[\begin{smallmatrix}
 %a & 0& 0& b\\
 %0 & c& d& 0\\
 %0 & w& z& 0\\
 %y & 0 & 0 & a
 %\end{smallmatrix}\right]
 %\left[\begin{smallmatrix}
 %0\\
 %t\\
 %1\\
 %0\\
 %\end{smallmatrix}\right]
 %=\left[\begin{smallmatrix}
 %0\\
 %ct+d\\
 %wt+z\\
 %0\\
%\end{smallmatrix}\right].
%\]
%%%

\section{Main Theorem and Proof Outline}
\begin{theorem}\label{main:theorem}
Let $f$ be a 4-ary signature with the signature matrix
$M(f)=\left[\begin{smallmatrix}
a & 0 & 0 & b\\
0 & c & d & 0\\
0 & w & z & 0\\
y & 0 & 0 & x
\end{smallmatrix}\right]$.
If  $ax=0$, then  $\operatorname{Holant}(\neq_2|f)$ is
equivalent to the six-vertex model $\operatorname{Holant}(\neq_2|f')$
where $f'$ is obtained from $f$ by setting $a=x=0$,  i.e.,
$M(f')=\left[\begin{smallmatrix}
0 & 0 & 0 & b\\
0 & c & d & 0\\
0 & w & z & 0\\
y & 0 & 0 & 0
\end{smallmatrix}\right]$. Explicitly,
$\operatorname{Holant}(\neq_2|f)$
is $\#\operatorname{P}$-hard except in the following cases:
\begin{itemize}
\item $f'\in\mathscr{P}$,
\item $f'\in\mathscr{A}$,
\item there is at least one zero in each pair $(b, y), (c, z), (d, w)$.
\end{itemize}
If $ax\neq 0$, then $\operatorname{Holant}(\neq_2|f)$
is $\#\operatorname{P}$-hard except in the following cases:
\begin{itemize}
\item $f$ is $\mathscr{P}$-transformable;
\item $f$ is $\mathscr{A}$-transformable;
\item $f$ is $\mathscr{L}$-transformable.
\end{itemize}
In all listed cases, $\operatorname{Holant}(\neq_2|f)$ is computable in polynomial time.
\end{theorem}%%end of theorem

By Lemma~\ref{ax=0}, we may assume $a=x$.
%For any given $f$ in the eight-vertex model and any signature grid $\Omega$
%with 4-regular graph $G$,
 %any valid orientation
%on $G$
%for  Holant$_{\Omega}(\neq_2|f)$
%must have an equal number of sources and sinks.
%Hence the value  Holant$_{\Omega}(\neq_2|f)$
%as a polynomial in $a$ and $x$ is in fact a polynomial in the product
%$ax$.
%So we can replace $(a,x)$ by any $(\tilde{a}, \tilde{x})$
%such that $\tilde{a}\tilde{x} = ax$.
%In particular,
%let $\tilde{f}$  be a 4-ary signature with the signature matrix
%$M(\tilde{f})=\left[\begin{smallmatrix}
%\tilde{a} & 0 & 0 & b\\
%0 & c & d & 0\\
%0 & w & z & 0\\
%y & 0 & 0 & \tilde{a}
%\end{smallmatrix}\right]$ where $\tilde{a}=\sqrt{ax}$,
%then
%Holant$_{\Omega}(\neq_2|f)=$Holant$_{\Omega}(\neq_2|\tilde{f})$(See Lemma~\ref{ax=0}.).
%Note that one can also switch the sign of both entries $(a,x)$.
If $a=0$,
  this is the six-vertex model and has been solved in Theorem~\ref{dichotomy-six-vertex} \cite{cai:fu:xia}. See Lemma~\ref{aneq0}.
In the following we assume $a=x\neq 0$. 

Let $\textsf{N}$ be the number of zeros in $\{b, c, d, y, z, w\}$.
We define Case \uppercase\expandafter{\romannumeral1} to be
 when there are
(at least) two
 pairs in $\{(b, y), (c, z), (d, w)\}$ that are $(0, 0)$.
We define Case \uppercase\expandafter{\romannumeral2} to be
 $\textsf{N}\geq 1$
and there is at most one pair in $\{(b, y), (c, z), (d, w)\}$
that is $(0, 0)$.
 Note that Case \uppercase\expandafter{\romannumeral1}
and  \uppercase\expandafter{\romannumeral2} cover all cases when $\textsf{N}\neq 0$.
Finally we define Case \uppercase\expandafter{\romannumeral3} to be $\textsf{N}=0$.
Formally,
the three Cases are defined as follows:
%%%%%%%%%%%%%%%%%%%%%%%%%%%%%%%%%%%%%%%%%%%%%
\begin{description}
\item {Case \uppercase\expandafter{\romannumeral1}:}
There are (at least) two pairs in
$\{(b, y), (c, z), (d, w)\}$ that are $(0, 0)$. (See Lemma~\ref{spin}.)

By the symmetry of the three inner
 pairs (the group action of $S_3$ induced by $S_4$),
 we may assume that $b=y=d=w=0$.
%%%%%%%%%%%%%%%%%%%%%
As $a \not =0$ we can normalize it to $a =1$.
We define a binary signature $g$ with the matrix
 $M(g) =
\left[\begin{smallmatrix}
1 & c \\
z & 1
\end{smallmatrix}\right]$.
We will reduce  \#CSP$^2(g)$ to
Holant$(\neq_2|f)$.
This will be accomplished by replacing every variable
in an instance $\Omega$ of  \#CSP$^2(g)$
by a cycle of even length,
such that there is a 1-1 correspondence between assignments
in $\Omega$ and valid configurations of the eight-vertex model
Holant$(\neq_2|f)$, that preserves the product of the weights.

%%%%%%%%%%%%%%%%%%%%
%We can define a set of circuits, and
%define a \#CSP problem where the variables correspond to
%the circuits, and the valid assignments in the eight-vertex model
%Holant$(\neq_2|f)$ ensure that there are exactly two valid
%configurations on these circuits that correspond to
%the two cyclic orientations.
%We further show that this is in fact a \#CSP$^2$ problem, where
%each variable appears an even number of times.  This is because
%the support structure of $f$ ensures that these circuits all have even
%length.
%
%Then we reduce \#CSP$^2(g)$ to
%Holant$(\neq_2|f)$,
%%, where $g$ is a binary signature whose entries come from the
%%support of $f$.
%where $g$ is a binary signature
%with matrix $M(g) =
%\left[\begin{smallmatrix}
%a & c \\
%z & a
%\end{smallmatrix}\right]$.
%%=(a, c, z, a)^T$.
%%Note that in \cite{cai:fu:xia}, the connection is between Holant and \#CSP.
%%\#CSP$^2$ is necessary for this paper since if the evaluation of Holant$(\neq_2|f)$ is not zero,
%%then the number of vertices in each circuit has to be even.

By Theorem~\ref{CSP2}, if $g\notin\mathscr{P}\cup\mathscr{A}\cup\alpha\mathscr{A}\cup\mathscr{L}$,
then \#CSP$^2(g)$ is \#P-hard. In this case Holant$(\neq_2|f)$ is \#P-hard.
If $g\in\mathscr{P}\cup\mathscr{A}\cup\alpha\mathscr{A}\cup\mathscr{L}$,
we show that $f\in\mathscr{P}$ or $\mathscr{A}$-transformable.
Thus  Holant$(\neq_2|f)$ is tractable.
\item {Case \uppercase\expandafter{\romannumeral2}:}
%There is are least one zero in $\{b, c, d, y, z, w\}$
$\textsf{N} \ge 1$ and there is at most one pair in
$\{(b, y), (c, z), (d, w)\}$ that is $(0, 0)$. (See Lemma~\ref{4:nozero}
to Lemma~\ref{5:nonzero}.)

In this case we prove that Holant$(\neq_2|f)$ is \#P-hard.
We prove this by constructing a 4-ary signature $g$
in the six-vertex model, such that Holant$(\neq_2|g)$ is \#P-hard
by Theorem~\ref{dichotomy-six-vertex}, and prove that  Holant$(\neq_2|g)
\leq_T^p $ Holant$(\neq_2|f)$.
\item {Case \uppercase\expandafter{\romannumeral3}:}
$\textsf{N}=0$, i.e., all values in $\{b, c, d, y, z, w\}$ are nonzero.
%%%%%%%%%%%%%%%%%%%%%%%%%%%%%%%%%%%%%%%%%%%%%%%%%%%%%%%%%%%%%%%%%%%%%555%%%%%%%%%%%%%%%%%%%%%%%%%%%%%%%%%%%%%%%%%%%%%%%%%%%%%%%%%%%%%%%%%%%%%%%%%%%%%%%%%%%

In this case we prove that  Holant$(\neq_2|f)$ is tractable
in the listed cases, and  \#P-hard otherwise.
The main challenge here is when we prove \#P-hardness for some
signatures $f$, the interpolation needs certain quantities
not to repeat after iterations. If we can only produce a root of unity, then its powers will
repeat after only a bounded number of steps in an iteration.
Typically one satisfies such a requirement
by producing quantities of complex norm not equal to 1.
   But for some $f$,
    provably the only such quantities
that can be produced are all of complex norm 1.

Our main new idea is to use M\"{o}bius transformations.
But before getting to that, there are some settings where we cannot do so, either because we don't
have the initial signature to start the process, or the matrix that
would define the M\"{o}bius transformation is singular.
   So we first treat the following two special cases.
      \begin{itemize}
      \item (See Lemma~\ref{just:one:binary} and Lemma~\ref{even}.) If $b=\epsilon y$, $c=\epsilon z$ and $d=\epsilon w$, where $\epsilon = \pm 1$, by a rotational symmetric gadget,
      we get some redundant signatures (Definition~\ref{redu-defi}).
 If one of the compressed matrices of these redundant signatures has full rank,
      then we can prove \#P-hardness.
      If all of these compressed matrices are degenerate, then we get a system of equations of $\{a, b, c, d\}$.
      To satisfy these equations,  $f$ has a very special form.
      Then we show that either
$f$ is $\mathscr{A}$-transformable or we can construct
an arity 4 signature
$g = \left[\begin{smallmatrix}
t & 0 & 0 & 1\\
0 & 1 & 1 & 0\\
0 & 1 & 1 & 0\\
1 & 0 & 0 & \frac{1}{t}
   \end{smallmatrix}
      \right]$, which is actually
 a symmetric signature $g = [g_0, g_1, g_2, g_3, g_4]
=[t, 0, 1, 0, \frac{1}{t}]$, with $t\neq 0$.
Here $g_w$ is the value of $g$ on all inputs of Hamming weight $w$.
      By a holographic transformation using $T=\left[
   \begin{smallmatrix}
   1 & \sqrt{t}\\
   \frak{i} & -\frak{i}\sqrt{t}
   \end{smallmatrix}
      \right]
=\sqrt{2} Z
\left[
   \begin{smallmatrix}
   1 & 0\\
   0 & \sqrt{t}
   \end{smallmatrix}
      \right]
=\left[
   \begin{smallmatrix}
   1 & 1\\
   \frak{i} & -\frak{i}
   \end{smallmatrix}
      \right]
\left[
   \begin{smallmatrix}
   1 & 0\\
   0 & \sqrt{t}
   \end{smallmatrix}
      \right]
=\left[
   \begin{smallmatrix}
   1 & 0\\
   0 & \frak{i}
   \end{smallmatrix}
      \right]
\left[
   \begin{smallmatrix}
   1 & 1\\
   1 & -1
   \end{smallmatrix}
      \right]
\left[
   \begin{smallmatrix}
   1 & 0\\
   0 & \sqrt{t}
   \end{smallmatrix}
      \right]
$ on
Holant$(\neq_2|f, g)$,
because $(=_2)Z^{\otimes 2} = (\neq_2)$,
% and therefore
%$(\neq_2)(T^{-1})^{\otimes 2} = \frac{1}{2\sqrt{t}} (=_2)$,
the transformed signature
$(\neq_2) (T^{-1})^{\otimes 2}$ is  $(=_2)$ on the LHS up to a nonzero scalar.
Similarly $g$ is
% $[t, 0, 1, 0, \frac{1}{t}]$ is
 transformed to $T^{\otimes 4} g$ on the RHS,
which is $(=_4)$ up to a nonzero scalar.
 Therefore we have the reduction
 \#CSP$^2(T^{\otimes 4}f) \leq_T^p $ Holant$(\neq_2|f, g)$.
%the equivalence
%      Holant$(\neq_2|f, g)\equiv^p_T$ \#CSP$^2(T^{\otimes 4}f)$.
      This implies that
       Holant$(\neq_2 \mid f)$ is either \#P-hard, or
      $f$ is $\mathscr{P}$-transformable, or $\mathscr{A}$-transformable, or $\mathscr{L}$-transformable
      by Theorem~\ref{CSP2}.
         \item (See Lemma~\ref{by=cz=dw}) If $by=cz=dw$,
then either we can realize a non-singular redundant signature or
$f$ is $\mathscr{P}$-transformable, or $\mathscr{A}$-transformable, or $\mathscr{L}$-transformable.
    \end{itemize}
      If $f$ does not belong to the above two cases,
      by the symmetry of the pairs $\{(b, y), (c, z), (d, w)\}$, we may assume that
      $cz\neq dw$, i.e., the inner matrix
$\left[
   \begin{smallmatrix}
   c & d\\
   w & z
   \end{smallmatrix}
      \right]$
of $M(f)$ has full rank.
      Then
      we
want to realize
binary signatures of the form $(0, 1, t, 0)^T$,
      for arbitrary values of $t$. If this can be done,
      by carefully choosing the values of $t$,
      we will prove \#P-hardness by the following methods (See Lemma~\ref{with-any-binary}.)
      \begin{itemize}
      \item constructing a signature that belongs to Case \uppercase\expandafter{\romannumeral2}, or
      \item constructing a redundant signature with a 
compressed signature matrix of full rank, or
      \item
      constructing the symmetric signature $[1, 0, -1, 0, 1]$ by gadget construction. Then by
       the holographic transformation using
 $T=\left[
   \begin{smallmatrix}
   1 & \sqrt{s}\\
   \frak{i} & -\frak i\sqrt{s}
   \end{smallmatrix}
      \right]
=
\left[
   \begin{smallmatrix}
1 & 0\\
0 & \frak{i}
   \end{smallmatrix}
      \right]
\left[
   \begin{smallmatrix}
1 & 1\\
1 & -1
\end{smallmatrix}
      \right]
\left[
   \begin{smallmatrix}
1 & 0\\
0 & \sqrt{s}
   \end{smallmatrix}
      \right]
$, we have
%%%%%%%%%%%%%%%%%%%%%%%%%%%%%%%%%%%%%%%here
      \begin{center}
\#CSP$^2(T^{\otimes 4}f, T^{\otimes 2}(0, 1, t, 0)^T)$ $\leq_T^p$
      Holant$(\neq_2|f, [s, 0, 1, 0, \frac{1}{s}], (0, 1, t, 0)^T)$,
%\equiv^p_T$ \#CSP$^2(T^{\otimes 4}f, T^{\otimes 2}(0, 1, t, 0)^T)$,
      \end{center}
because $(\ne_2) (T^{-1})^{\otimes 2}$ is $(=_2)$,
and $T^{\otimes 4} [s, 0, 1, 0, \frac{1}{s}]$ is $(=_4)$, both up to  a nonzero scalar.
      Then we can prove \#P-hardness  by Theorem~\ref{CSP2}.
      \end{itemize}
      %Follows from the result of Case \Rmnum{3},
      %we get the \#P-hardness result.

\vspace{.1in}

      We realize binary signatures by connecting $f$ with $(\neq_2)$.
This corresponds naturally to a M\"{o}bius transformation.
      By discussing the following different forms of binary signatures we get,
      we can either realize arbitrary $(0, 1, t, 0)^T$, then  Holant$(\neq_2|f)$
      is \#P-hard, or $f$ is $\mathscr{A}$-transformable under some nontrivial holographic transformation.
\begin{itemize}
         \item If we can get a signature of the form $g = (0, 1, t, 0)^T$ where $t\neq0$ is not a root of unity, then by connecting a chain of $g$, we can
get polynomially many distinct binary signatures $g_i=(0, 1, t^i, 0)^T$.
         Then, by interpolation, we can realize
 arbitrary binary signatures of the form $(0, 1, s, 0)^T$. (See Corollary~\ref{cor:binary-interpolation}.)

         \item Suppose we can get a signature of the form $(0, 1, t, 0)^T$, where $t\neq0$ is an $n$-th primitive root of unity $(n\geq 5)$. Now, we only have $n$ many different signatures $g_i=(0, 1, t^i, 0)^T$.
         But we can relate $f$ to a M\"{o}bius transformation
$\varphi(\mathfrak{z}): \mathfrak{z} \mapsto
\dfrac{c \mathfrak{z} + d}{w \mathfrak{z} + z}$,
%%%%%%%%%%%%{{{{{{{{{{{{
due to
$\det\left[\begin{smallmatrix}
c & d\\
w & z
\end{smallmatrix}\right] \not =0$.
%$ax-cz \neq 0$ and $by-cz \neq 0$.
         For the M\"{o}bius transformation $\varphi$, we can realize the signatures $g= (0, 1, \varphi(t^i), 0)^T$.
If $|\varphi(t^i)|\neq 0, 1$ or $\infty$ for some $i$,
then this is treated above.
Otherwise, since $\varphi$ is a bijection on the extended complex plane
$\widehat{\mathbb{C}}$, it can map at most two points of  $S^1$ to
$0$ or $\infty$.
         Hence, $|\varphi(t^i)|= 1$ for at least three $t^i$.
But a M\"{o}bius transformation is determined by any three distinct points.
This implies that $\varphi$  maps  $S^1$ to itself.
 Such M\"{o}bius transformations  have a known special form
$e^{i\theta}\dfrac{\mathfrak{z}+\lambda}{1+\bar{\lambda}\mathfrak{z}}$.
By exploiting its property we can construct a signature $f'$ such that
its corresponding M\"{o}bius transformation $\varphi'$
%infinite projective order.
defines an infinite group. This implies
that $\varphi'^k(t)$ are all distinct. Then, we can get polynomially
 many distinct binary signatures $(0, 1, \varphi'^k(t), 0)$, and
realize arbitrary binary signatures of the form $(0, 1, s, 0)^T$. (See Lemma~\ref{construct-binary}.)

         \item Suppose we can get a signature of the form $(0, 1, t, 0)^T$ where $t\neq0$ is an $n$-th primitive root of unity (where $n=3$ or $4$). Then we can either relate it to two M\"{o}bius transformations mapping the unit circle to itself, or
realize the $(+, -)$-pinning $(0, 1, 0, 0)^T = (1,0) \otimes (0,1)$. (See Corollary~\ref{construct:(0,1,0,0)}.)
\item Suppose we can get $(0, 1, 0, 0)^T$. 
 By connecting $f$ with it, we can get new signatures of the form $(0, 1, t, 0)^T$. Similarly, by analyzing the value of $t$,
         we can either realize arbitrary binary signatures of the form $(0, 1, s, 0)^T$, or a redundant signature
         that its compressed signature matrix has full rank, or a signature in Case \uppercase\expandafter{\romannumeral1}, which is \#P-hard,
          or we prove that $f$ is $\mathscr{A}$-transformable under a
 holographic transformation
%such as
          $\left[\begin{smallmatrix}
1 & 0 \\
0 & \gamma
\end{smallmatrix}\right]$, where $\gamma^2= \alpha$  or $ \gamma^2= \frak{i}$
(See Theorem~\ref{with:(0,1,0,0)}).
         \item Suppose we can only get signatures of the form $(0, 1, \pm 1, 0)^{T}$. That implies $a=\epsilon x$, $b=\epsilon y$ and $c=\epsilon z$, where $\epsilon=\pm 1$.  This has been treated before.
    \end{itemize}
\end{description}

\section{Interpolation via M\"{o}bius Transformation}
Note that having $g=(0, 1, t, 0)^T$ is equivalent to having
$g'(x_1, x_2) = g(x_2, x_1) = (0, t, 1, 0)^T$.

\begin{lemma}\label{construct-binary}
Let $g=(0, 1, t, 0)^T$ be a binary signature where $t^i$ are distinct for $1\leq i\leq 5$,  and $f$
be a signature with the signature matrix
 $M(f)=\left[\begin{smallmatrix}
 a & 0 & 0 & b\\
 0 & c & d & 0\\
 0 & w & z & 0\\
 y & 0 & 0 & a\\
\end{smallmatrix}\right]$ with $abcdyzw\neq 0$, where
$\left[\begin{smallmatrix}
  c & d\\
  w & z
\end{smallmatrix}\right]$
has full rank, then for any $u\in\mathbb{C}$,
$\operatorname{Holant}(\neq_2 | f, (0, 1, u, 0)^T)\leq^p_{\operatorname{T}}\operatorname{Holant}(\neq_2 | f, g).$
\end{lemma}
%%%
%%% seems the only nonzero requirements for $abcdyzw\neq 0$
%%% is the inner matrix is not diagonal nor anti-diagonal.
%%%
\begin{proof}
By connecting two copies of $g$ using $(\neq_2)$, we get the signature $g_2$ with the signature matrix is
$
M(g_2)=
\left[\begin{smallmatrix}
 0& 1\\
 t& 0\\
\end{smallmatrix}\right]
\left[\begin{smallmatrix}
 0& 1\\
 1& 0\\
\end{smallmatrix}\right]
\left[\begin{smallmatrix}
 0& 1\\
 t& 0\\
\end{smallmatrix}\right]=
\left[\begin{smallmatrix}
 0& 1\\
 t^2& 0\\
\end{smallmatrix}\right]
.$
Similarly, we can construct
at least five distinct  $g_i=(0, 1, t^i, 0)^T$ for $1\leq i\leq 5$.
 %Since the order $n\geq 5$,
 % $g_i$ are distinct pairwise for $1\leq i\leq 5$.

%{\bf This part has been moved to preliminary. }
%\{For any binary signature $g$ in the form of %$g^{(x_1, x_2)}=(0, a, b, 0)$, notice that %$g^{(x_2, x_1)}=(0, b, a, 0)$. Then we have
%\begin{align*}
%& M_{x_s x_t, x_v x_u}Ng^{(x_1, x_2)}\\
%=& M_{x_s  x_t, x_v x_u}
%\begin{bmatrix}
%0 & 0 & 0 & 1 \\
%0 & 0 & 1 & 0 \\
%0 & 1 & 0 & 0 \\
%1 & 0 & 0 & 0 \\
%\end{bmatrix}
%\begin{bmatrix}
%0\\
%a\\
%b\\
%0\\
%\end{bmatrix} \\
%= & M_{x_s x_t, x_v x_u}
%\begin{bmatrix}
%0\\
%b\\
%a\\
%0\\
%\end{bmatrix}\\
%= & M_{x_s  x_t, x_v x_u}g^{(x_2, x_1)}.
%\end{align*}
%That means connecting the variables $x_v$, $x_u$ of the signature $f$ with the variables $x_1$, $x_2$ of the signature $g$ using $(\neq_{2})$ are equivalent to connecting the variables $x_v$, $x_u$ of the signature $f$ with the variables $x_2$, $x_1$ of the signature $g$ directly. Since $g$ is a binary signature, we can rotate it by $180$ degree. Then the variables $x_1$ and $x_2$ change their positions with each other, and that rotation do not destroy the planar graph. \}

By connecting the variables $x_3$ and $x_4$ of the signature $f$ with the variables $x_1$ and $x_2$ of $g_i$ using $\neq_2$
for $1 \leq i \leq 5$ respectively,  we get binary signatures
$$h_i=M(f)Ng_i=
\left[\begin{smallmatrix}
a & 0& 0& b\\
0 & c& d& 0\\
0 & w& z& 0\\
y & 0 & 0 & a
\end{smallmatrix}\right]
\left[\begin{smallmatrix}
0\\
t^i\\
1\\
0\\
\end{smallmatrix}\right]
=\left[\begin{smallmatrix}
0\\
ct^i+d\\
wt^i+z\\
0\\
\end{smallmatrix}\right].
$$
Let $\varphi(\mathfrak z)=\dfrac{c\mathfrak z+d}{w\mathfrak z+z}$.
Since $\det\left[\begin{smallmatrix}
c & d\\
w & z
\end{smallmatrix}\right]\neq 0$,
$\varphi(\mathfrak z)$ is a M\"{o}bius transformation  of the  extended  complex  plane $\widehat{\mathbb{C}}$.
We rewrite $h_i$ as $(wt^i+z)(0, \varphi(t^i), 1, 0)^T$, with the understanding that if $wt^i+z=0$, then $ \varphi(t^i)=\infty$, and we define
$(wt^i+z)(0, \varphi(t^i), 1, 0)^T$ to be $(0, ct^i+d, 0, 0)^T$.
Having $(0, \varphi(t^i), 1, 0)^T$ is equivalent to
having $(0, 1, \varphi(t^i), 0)^T$.
If there is a $t^i$ such that $\varphi(t^i)\neq 0, \infty$ or a root of unity for $1\leq i\leq 5$,
 then by
 %Lemma~\ref{binary-interpolation}
 Corollary~\ref{cor:binary-interpolation}
 $\operatorname{Holant}(\neq_2 | f, (0, 1, u, 0)^T)\leq^p_{\operatorname{T}}\operatorname{Holant}(\neq_2 | f, (0, 1, \varphi(t^i), 0)^T)$,
for all $u \in \mathbb{C}$.
 Otherwise, $\varphi(t^i)$ is $0, \infty$ or a root of unity for $1\leq i\leq 5$. Since
 $\varphi(\mathfrak z)$ is a bijection of $\widehat{\mathbb{C}}$, there is at most one $t^i$ such that $\varphi(t^i)=0$ and at most one $t^i$ such that $\varphi(t^i)=\infty$.
That means, there are at least three $t^i$ such that $|\varphi(t^i)|=1$.
 Since a M\"{o}bius transformation is determined by any 3 distinct points, mapping 3 distinct points from $S^1$ to $S^1$ implies that
 this $\varphi(\mathfrak z)$ maps $S^1$ homeomorphically onto $S^1$.

A M\"{o}bius transformation mapping 3 distinct points from $S^1$ to $S^1$ has a special form
$\mathcal{M}(\lambda, e^{\frak i\theta})$:
${\mathfrak z} \mapsto e^{\frak i\theta}\dfrac{\mathfrak z+\lambda}{1+\bar{\lambda}\mathfrak z}$,
where $|\lambda|\neq 1$.
%Since the inner matrix has full rank and is not diagonal or anti-diagonal matrix, by the symmetry of the three inner pairs, we can assume that $c\neq 0$.
By normalization in $f$, we may assume $z=1$, since $z \not =0$.
 Comparing coefficients with $\varphi(\mathfrak z)$ we have
 $c= e^{\frak i\theta}$,
$d = e^{\frak i\theta} \lambda$ and
$w = \bar{\lambda}$.
Since $w \not = 0$ we have $\lambda \not = 0$.
Thus $M(f)=\left[\begin{smallmatrix}
a & 0 & 0 & b\\
0 & e^{\frak i\theta} & e^{\frak i\theta}\lambda & 0\\
0 & \bar{\lambda} & 1 & 0\\
y & 0 & 0 & a\\
\end{smallmatrix}\right]$.
Note that
$M_{x_3x_4, x_2x_1}(f)=M^{\sf R_{(12)}\sf R_{T}}(f)=\left[\begin{smallmatrix}
a & 0 & 0 & y\\
0 & \bar{\lambda} & e^{\frak i\theta}  & 0\\
0 & 1 & e^{\frak i\theta}\lambda  & 0\\
b & 0 & 0 & a\\
\end{smallmatrix}\right]$.
% obtained from $M(f)=M_{x_1x_2,x_3x_4}(f)$
%by exchanging the two middle columns of $(M(f))^{T}$.
%first taking its transpose and then exchange two middle columns.
By taking two copies of $f$ and connecting
the variables $x_3, x_4$ of the first copy to the variables $x_3, x_4$ of the second copy using $(\neq_2)$,
we get a signature
 $f_1$ with the signature matrix
\[
M(f_1)=M(f)NM_{x_3x_4, x_2x_1}(f)=
\left[\begin{smallmatrix}
2ab & 0 & 0 & a^2+by\\
0 & e^{\frak i\theta}(1+|\lambda|^2) & 2e^{2\frak i\theta}\lambda & 0\\
0 &  2\bar{\lambda}  & e^{\frak i\theta}(1+|\lambda|^2) & 0\\
a^2+by & 0 & 0 & 2ay\\
\end{smallmatrix}\right].
\]
Then up to the nonzero scalar $s=e^{\frak i\theta}(1+|\lambda|^2)$,
and denote by  $\delta =
\dfrac{2e^{\frak i\theta}\lambda}{1+|\lambda|^2}$,
we have $\delta \not =0$ because $\lambda \not =0$, and $\bar{\delta} =
\dfrac{2e^{-\frak i\theta}\bar{\lambda}}{1+|\lambda|^2}$,
and
  the signature $f_1$ has the signature matrix
$M(f_1)=\left[\begin{smallmatrix}
\frac{2ab}{s} & 0 & 0 & \frac{a^2+by}{s}\\
0 & 1 & \delta & 0\\
0 & \bar{\delta} & 1 & 0\\
\frac{a^2+by}{s} & 0 & 0 & \frac{2ay}{s}\\
\end{smallmatrix}\right]$.
The inner matrix
$\left[\begin{smallmatrix}
1 & \delta\\
\bar{\delta} & 1
\end{smallmatrix}\right]$ of $M(f_1)$ is
the product of three nonsingular $2 \times 2$ matrices, thus
it is also nonsingular.
The two eigenvalues of
$\left[\begin{smallmatrix}
1 &  \delta \\
\bar{\delta} & 1
\end{smallmatrix}\right]$
 are $1 + |\delta|$ and $1- |\delta|$, both are real and must be nonzero.
In particular $|\delta| \not =1$.
Obviously $|{1+ |\delta|}| \neq  |{1- |\delta|}|$,
since $|\delta| >0$.
This implies that
 there are no integer $n> 0$ and complex number $\mu$ such that
$\left[\begin{smallmatrix}
1 &  {\delta} \\
  \bar{\delta} & 1
\end{smallmatrix}\right]^n=\mu I$,
i.e., $\left[\begin{smallmatrix}
1 &  \delta \\
\bar{\delta} & 1
\end{smallmatrix}\right]$ has infinite projective order.
Note that
$\left[\begin{smallmatrix}
1 &  \delta \\
\bar{\delta} & 1
\end{smallmatrix}\right]$
defines a M\"{o}bius transformation $\psi({\mathfrak z})$
of the form
$\mathcal{M}(\lambda, e^{{\frak i}\theta})$ with $\lambda = \delta$
and $\theta =0$:
${\mathfrak z} \mapsto \psi({\mathfrak z})=
\dfrac{{\mathfrak z}+\delta}{1+\bar{\delta}{\mathfrak z}}$,
mapping  $S^1$ to $S^1$.
Since $\left[\begin{smallmatrix}
1 &  \delta \\
\bar{\delta} & 1
\end{smallmatrix}\right]$ has infinite projective order,
the  M\"{o}bius transformation $\psi({\mathfrak z})$
defines an infinite group.
%
%Denote $\left[\begin{smallmatrix}
%\bar{\delta} & 1\\
%1 & \delta\\
%\end{smallmatrix}\right]^n$
%by $\left[\begin{smallmatrix}
%d_n & c_n\\
%z_n & w_n\\
%\end{smallmatrix}\right]$.
%Since $\left[\begin{smallmatrix}
%d_n & c_n\\
%z_n & w_n\\
%\end{smallmatrix}\right]$ has full rank,
%$\psi^{(n)}(\mathfrak z)=\frac{w_n\mathfrak z + z_n}{c_n\mathfrak z + d_n}$
%is a M\"{o}bius transformation mapping $S^1$ to $S^1$.

We can connect the binary signature
$g_i(x_1, x_2)$ via $N$ to $f_1$. This gives us
the binary signatures  (for $1 \leq i \leq 5$)
$$h^{(1)}_i=M(f_1)Ng_i=
\left[\begin{smallmatrix}
* & 0& 0& *\\
0 & 1 &  \delta & 0\\
0 & \bar{\delta}& 1 & 0\\
* & 0& 0& *\\
\end{smallmatrix}\right]
\left[\begin{smallmatrix}
0\\
t^i\\
1\\
0\\
\end{smallmatrix}\right]
=C_{(i, 0)}\left[\begin{smallmatrix}
0\\
\psi(t^i)\\
1\\
0\\
\end{smallmatrix}\right],
$$
where
$\psi({\mathfrak z})$ is the
 M\"{o}bius transformation
defined by
 the matrix
$\left[\begin{smallmatrix}
1 &  {\delta} \\
\bar{\delta}  & 1
\end{smallmatrix}\right]$,
and $C_{(i, 0)} = 1 + \bar{\delta} t^i$.
Since $\psi$ maps $S^1$ to $S^1$, and $|t^i| =1$,
clearly $C_{(i, 0)} \not =0$, and
$\psi(t^i) \in S^1$.

Now we can use $h^{(1)}_i(x_2, x_1)
= (0, 1, \psi(t^i), 0)^{T}$ in place of
$g_i(x_1, x_2) = (0, 1, t^i, 0)^{T}$
and repeat this construction. Then we get
\[
h^{(2)}_i=
\left[\begin{smallmatrix}
* & 0& 0& *\\
0 & 1 &  \delta & 0\\
0 & \bar{\delta}& 1 & 0\\
* & 0& 0& *\\
\end{smallmatrix}\right]
\left[\begin{smallmatrix}
0\\
\psi(t^i)\\
1\\
0\\
\end{smallmatrix}\right]
=C_{(i, 1)}\left[\begin{smallmatrix}
0\\
\psi^2(t^i)\\
1\\
0\\
\end{smallmatrix}\right],
\]
where $\psi^2$ is the composition $\psi \circ \psi$,
corresponding to
$\left[\begin{smallmatrix}
1 &  {\delta} \\
\bar{\delta}  & 1
\end{smallmatrix}\right]^2$,
and $C_{(i, 1)} = 1 + \bar{\delta} \psi(t^i) \not =0$.

We can iterate this process and get polynomially many
 $h^{(k)}_i = (0, \psi^k(t^i), 1, 0)^{T}$
for $1 \le i \le 5$ and $k \ge 1$.

If for each $i\in\{1, 2, 3\}$, there is some $n_i > 0$ such that
$\psi^{n_i}(t^i)=t^i$,  then $\psi^{n_0}(t^i)=t^i$,
 for $n_0=n_1n_2n_3 >0$,
and all $1\leq i\leq 3$, i.e., the M\"{o}bius transformation $\psi^{n_0}$
 fixes three distinct
complex numbers $t, t^2, t^3$.
So the M\"{o}bius transformation is the identity map, i.e.,
$\psi^{n_0}(\mathfrak z)=\mathfrak z$ for all $\mathfrak z\in\mathbb{C}$.
This implies that $\psi(\mathfrak z)$ defines a group of finite order,
a contradiction.
%$\left[\begin{smallmatrix}
%1 & \delta\\
%\bar{\delta} & 1
%\end{smallmatrix}\right]^{n_0}=C\left[\begin{smallmatrix}
%1 & 0\\
%0 & 1 &
%\end{smallmatrix}\right]$ for some constant $C$.
%This contradicts the fact that
%$\left[\begin{smallmatrix}
%1 & \delta\\
%\bar{\delta} & 1
%\end{smallmatrix}\right]$ does not have finite projective order.
Therefore, there is an $i \in \{1,2,3\}$
such that $\psi^{n}(t^i) \not = t^i$ for all  $n\in\mathbb{N}$.
This implies that
 $(1, \psi^{n}(t^i))$ are all distinct for $n\in\mathbb{N}$,
since $\psi$ maps $S^1$  1-1 onto $S^1$.
Then we can generate polynomially many distinct binary signatures
of the form $(0, 1, \psi^{n}(t^i), 0)^T$. By Lemma~\ref{binary-interpolation}, for any $u\in\mathbb{C}$ we have
$\operatorname{Holant}(\neq_2 | f, (0, 1, u, 0)^T)\leq^p_{\operatorname{T}}\operatorname{Holant}(\neq_2 | f, g).$
\end{proof}

Suppose we have the binary signature  $(0, 1, t, 0)^T$
where $t$ is
an $n$-th primitive root of unity, for $n=3$ or $n=4$.
Then $(0, 1, t^i, 0)^T$ are distinct for $i=1, 2, 3$.
In the proof of Lemma~\ref{construct-binary},
if one of $\varphi(t^i)=0$ or $\infty$, then we have $(0, 1, 0, 0)^T$
or $(0, 0, 1, 0)^T$, which means we have both.
If none of $\varphi(t^i)=0$ or $\infty$ for  $i=1, 2, 3$,
then the proof of  Lemma~\ref{construct-binary}
shows that we have $(0, 1, u, 0)^T$ for
any $u\in\mathbb{C}$. Thus we have $(0, 1, 0, 0)^T$ as well.
%Note that $(0, 1, 0, 0)^T$ and $(0, 0, 1, 0)^T$ are the same signature.
So we have the following corollary.
\begin{corollary}\label{construct:(0,1,0,0)}
Let $g=(0, 1, t, 0)^T$ be a binary signature
where $t$ is an $n$-th primitive root of unity for $n=3$ or 4,  and let $f$
be a signature with the signature matrix
 $M(f)=\left[\begin{smallmatrix}
 a & 0 & 0 & b\\
 0 & c & d & 0\\
 0 & w & z & 0\\
 y & 0 & 0 & a\\
\end{smallmatrix}\right]$ with $abcdyzw\neq 0$, where
$\left[\begin{smallmatrix}
  c & d\\
  w & z
\end{smallmatrix}\right]$
has full rank, then
$\operatorname{Holant}(\neq_2 | f, (0, 1, 0, 0)^T)\leq^p_{\operatorname{T}}\operatorname{Holant}(\neq_2 | f, g).$
\end{corollary}

%The next lemma shows when we can get $(0, 1, 0, 0)^T$ or $(0, 1, t, 0)^T$ with $t^2\neq 0, 1$.

\begin{lemma}\label{construct:basic:binary}
Let $f$ be a 4-ary signature with the signature matrix
$M(f)=\left[\begin{smallmatrix}
a & 0 & 0 & b\\
0 & c & d & 0\\
0 & w & z & 0\\
y & 0 & 0 & a
\end{smallmatrix}\right]$.
If the signature matrix is not of the form $\left[\begin{smallmatrix}
a & 0 & 0 & \epsilon x\\
0 & \epsilon z & \epsilon w & 0\\
0 & w & z & 0\\
x & 0 & 0 & a
\end{smallmatrix}\right]$, where $\epsilon=\pm 1$, then
%\begin{equation}\label{lm3-3-eqn1}
%\operatorname{Holant}(\neq_2|(0, 1, 0, 0), f)
%\leq^p_T\operatorname{Holant}(\neq_2|f),
%\end{equation}
%or
\begin{equation}\label{lm3-3-eqn2}
\operatorname{Holant}(\neq_2|(0, 1, t, 0), f)
\leq^p_T\operatorname{Holant}(\neq_2|f),
\end{equation}
where $t\neq  \pm  1$.
\end{lemma}
\begin{proof}
For any two pairs $(s,t)$ and $(u,v)$, we say
``${\sf S}
\left(\left[\begin{smallmatrix}
s & u\\
v & t
\end{smallmatrix}\right] \right)$ by $\epsilon$'',
if $s = \epsilon t$ and $u = \epsilon v$, for some $\epsilon = \pm 1$.
If
${\sf S}
\left(\left[\begin{smallmatrix}
c & d\\
w & z
\end{smallmatrix}\right] \right)$ by $\epsilon_1$,
${\sf S}
\left(\left[\begin{smallmatrix}
c & b\\
y & z
\end{smallmatrix}\right] \right)$ by $\epsilon_2$,
and
${\sf S}
\left(\left[\begin{smallmatrix}
b & d\\
w & y
\end{smallmatrix}\right] \right)$ by $\epsilon_3$,
then by assumption it is not the case that
$\epsilon_1 = \epsilon_2 = \epsilon_3$.
By the symmetry of three inner
pairs, we may assume $\epsilon_1 \not = \epsilon_2$.
Then from $\epsilon_1 z = c = \epsilon_2 z$, we get
$c=z=0$. But then $c = \epsilon_3 z$,
contradicting the assumption.
So by the symmetry of the three inner pairs,
we may assume
${\sf S}
\left(\left[\begin{smallmatrix}
c & d\\
w & z
\end{smallmatrix}\right] \right)$ \emph{does not} hold
for any $\epsilon = \pm 1$.
i.e., there is no $\epsilon = \pm 1$ such that
 $c = \epsilon z$ and $d = \epsilon w$.

We first deal with the case that $\det A=0$,
where $A =
\left[\begin{smallmatrix}
c & d\\
w & z
\end{smallmatrix}\right]$.
If so,
since ${\sf S}
(A)
$ does not hold
for any $\epsilon = \pm 1$, it is not the case that all
four entries of $A$ are 0.
 By the symmetry of the pairs, we may assume $c \not =0$
and normalize $c =1$. So
$A
=
\left[\begin{smallmatrix}
1 & d\\
w & dw
\end{smallmatrix}\right]$.
%By a loop on $x_1, x_2$ of $f$ via $(\neq_2)$,
%and a loop on $x_3, x_4$ of $f$ via $(\neq_2)$,
%we get both $(1+d)(0, 1, w, 0)^T$
%and $(1+w)(0, 1, d, 0)^T$.
%
%*
%%
By a loop on $x_1, x_2$ of $f$ via $(\neq_2)$,
and a loop on $x_3, x_4$ of $f$ via $(\neq_2)$,
we get both $(1+w)(0, 1, d, 0)^T$ and $(1+d)(0, 1, w, 0)^T$.
If $w = \pm 1$ and $d = \pm 1$,
then we have
${\sf S}
(A)$ by $\epsilon = \frac{1}{dw} = \frac{d}{w}$,
a contradiction.
So we have
$w  \not = \pm 1$  or $d \not = \pm 1$.
Assume $w  \not = \pm 1$; the other case is symmetric.
Since we have $(1+d) \cdot (0, 1, w, 0)^T$,
if $d  \not = - 1$, then we have $(0, 1, w, 0)^T$
satisfying (\ref{lm3-3-eqn2}).
Suppose $d = -1$, then from $(1+w) \cdot (0, 1, d, 0)^T$,
we get $(0, 1, -1, 0)^T$.
By a loop on $x_3, x_4$ of $f$ via $(0, 1, -1, 0)^T$,
we get $(1-d)  \cdot (0, 1, w, 0)^T = 2 (0, 1, w, 0)^T$
satisfying (\ref{lm3-3-eqn2}).
%
%*
%
%If $w  \not = \pm 1$  or $d \not = \pm 1$, then
% (\ref{lm3-3-eqn2}) is satisfied.
%Otherwise,
%$w = \pm 1$ and $d = \pm 1$,
%then we have
%${\sf S}
%(A)$ by $\epsilon = \frac{1}{dw} = \frac{d}{w}$,
%a contradiction.

In the following we have
 $\det A \not =0$.
The matrix $A$ and  its transpose $A^T$
define two
 M\"{o}bius transformations,
$\varphi(\mathfrak z)=
\dfrac{c\mathfrak z+d}{w\mathfrak z+z}$
and $\psi(\mathfrak z)=
\dfrac{c\mathfrak z+w}{d\mathfrak z+z}$.
By connecting $x_3$ and $x_4$ of $f$ via $(\neq_2)$
we get $(0, c+d, w+z, 0)^T$.
If  $w+z =0$ then $c+d \not = 0$,
and (\ref{lm3-3-eqn2}) is satisfied
with $t=0$. Otherwise we can write
 $(0, c+d, w+z, 0)^T = (w+z) \cdot (0, \varphi(1), 1, 0)^T$.
If $\varphi(1) \not = \pm 1$ then (\ref{lm3-3-eqn2}) is satisfied.
%%% already said about $\varphi(1) = \infty$, ie   $w+z =0$
%(Note that this is true even if $\varphi(1) = \infty$,
%in which case we have a nonzero $(0, c+d, 0, 0)^T
%= (c+d)  \cdot (0,1,0,0)^T$,
%which is equivalent to $(0,0,1,0)^T$ and (\ref{lm3-3-eqn2}) is satisfied.

Hence we may assume $\varphi(1) = \pm 1$.
Similarly we may assume $\psi(1) = \pm 1$,
by connecting $x_1$ and $x_2$  of $f$ via $(\neq_2)$
and getting $(0, c+w, d+z, 0)^T = (d+z) \cdot (0, \psi(1), 1, 0)^T$.
Since $\varphi(1), \psi(1) \not = \infty$, we have
$w+z \not = 0$ and $d+z \not = 0$,
and we have both $(0,1,\varphi(1),1)^T$ and $(0,1,\psi(1),0)^T$.

If either $\varphi(1) = -1$ or $\psi(1) = -1$, then we have
\begin{equation}\label{sum-cdwz-is-zero}
c+d + w +z  =  0,
\end{equation}
and the signature $(0,1,-1,0)^T$.
By connecting $x_3$ and $x_4$ of $f$  via $(0, 1, -1, 0)^T$
we get the binary signature $(0, c-d, w-z, 0)^T$.
If one of $c-d$ or $w-z=0$, then the other one is nonzero
because $\det A \not = 0$,
and we get $(0,1,0,0)^T$, satisfying  (\ref{lm3-3-eqn2}).
So we may assume $c-d \not = 0$, and $w-z \not = 0$,
and we have $(0,\varphi(-1), 1, 0)^T$,
where $\varphi(-1) = \frac{c-d}{w-z}$.
If $\varphi(-1) \not = \pm 1$, then   (\ref{lm3-3-eqn2}) is satisfied.
If  $\varphi(-1) = 1$, then $c-d = w-z$.
Combined with (\ref{sum-cdwz-is-zero}), we get $c= -z$ and $d= -w$,
and thus ${\sf S}
(A)$ by $\epsilon = -1$, a contradiction.
If  $\varphi(-1) = -1$, then $c-d = z-w$.
Combined with (\ref{sum-cdwz-is-zero}), we get $c= -w$ and $d = -z$.
This contradicts $\det A \not = 0$.

Thus we may assume
$\varphi(1) = \psi(1) = 1$.
Then from $c+d = w+z$ and $c+w = d+z$,
%%% consider c- z  &  d - w.
 we get $c= z$ and $d=  w$.
This gives ${\sf S}
(A)$ by $\epsilon = 1$, a contradiction.
\end{proof}

Note that $(0, 1, 0, 0)^T=(1, 0)^T\otimes (0, 1)^T$.
While doing a loop to the variables $x_i, x_j$ of $f$ using $(0, 1, 0, 0)^T$ via $N$, it is equivalent to
pin $x_i=1$ and $x_j=0$.
In the following, we will say pinning $x_i=1$ and $x_j=0$ of $f$ using $(0, 1, 0, 0)^T$
instead of, synonymously,
 doing a loop to the variables $x_i, x_j$ of $f$ using $(0, 1, 0, 0)^T$ via $N$.

\begin{corollary}\label{inner:matrix:three:nonzero}
Let $f$ be a 4-ary signature with the signature matrix $M(f)=\left[\begin{smallmatrix}
a & 0 & 0 & b\\
0 & c & d & 0\\
0 & w & z & 0\\
y & 0 & 0 & a\\
\end{smallmatrix}\right]$.
If there are exactly three nonzero entries in the inner matrix
$\left[\begin{smallmatrix}
 c & d \\
 w & z
\end{smallmatrix}\right]$
then for any $u\in\mathbb{C}$,
\[\operatorname{Holant}(\neq_2 | f, (0, 1, u, 0)^T)\leq_T^p\operatorname{Holant}(\neq_2 | f).\]
\end{corollary}
\begin{proof}
By the symmetry of the three inner pairs, we can assume that $z=0, cdw\neq 0$.
Then by a normalization in $f$, we can assume that $c=1$, i.e.,
 $M(f)=\left[\begin{smallmatrix}
a & 0 & 0 & b\\
0 & 1 & d & 0\\
0 & w & 0 & 0\\
y & 0 & 0 & a\\
\end{smallmatrix}\right]$.

By Lemma~\ref{construct:basic:binary} and $c\neq \pm z$, we 
have $(0, 1, t, 0)^T$ with $t\neq \pm 1$.
\begin{itemize}
\item If $t=0$, then we have $(0, 1, 0, 0)^T$, by pinning $x_1=0$ and $x_2=1$, $x_3=0$ and $x_4=1$ respectively,
we get the binary signatures $(0, 1, d, 0)^T, (0, 1, w, 0)^T$.
By Lemma~\ref{inverse}, we have the binary signatures  $(0, 1, d^{-1}, 0)^T,$ $(0, 1, w^{-1}, 0)^T$.
Then by doing binary modifications to the variables $x_1, x_3$ of $f$ using
$(0, 1, w^{-1}, 0)^T$, $(0, 1, d^{-1}, 0)^T$ respectively, we get a
 signature $f_1$ with the signature matrix
 $M(f_1)=\left[\begin{smallmatrix}
a & 0 & 0 & \frac{b}{d}\\
0 & 1 & 1 & 0\\
0 & 1 & 0 & 0\\
\frac{y}{w} & 0 & 0 & \frac{a}{dw}\\
\end{smallmatrix}\right]$.
By connecting $x_1, x_2$ of $f_1$ using $\neq_2$, we get the binary signature $2(0, 1, \frac{1}{2}, 0)$.
Then by Corollary~\ref{cor:binary-interpolation}, we have $(0, 1, u, 0)^T$ for any $u\in\mathbb{C}$.
\item  If we have $g=(0, 1, t, 0)^T$ with $t\neq 0$ and $t \neq \pm 1$, then by connecting $g$ using $\neq_2$, we can construct
$g_{i}=(0, 1, t^i, 0)^T$ for $1\leq i\leq 3$.
Since $t^2\neq 0, 1$, $g_{i}$ are distinct for $1\leq i\leq 3$.

By connecting the variables $x_3$ and $x_4$ of the signature $f$ with the variables $x_1$ and $x_2$ of $g_i$ using $\neq_2$
for $1 \leq i \leq 3$ respectively,  we get binary signatures
$$h_i=M(f)Ng_i=
\left[\begin{smallmatrix}
a & 0& 0& b\\
0 & 1& d& 0\\
0 & w& 0& 0\\
y & 0 & 0 & a
\end{smallmatrix}\right]
\left[\begin{smallmatrix}
0\\
t^i\\
1\\
0\\
\end{smallmatrix}\right]
=\left[\begin{smallmatrix}
0\\
t^i+d\\
wt^i\\
0\\
\end{smallmatrix}\right].
$$
Note that $\left[\begin{smallmatrix}
1 & d\\
w & 0&
\end{smallmatrix}\right]$ has full rank.
%Thus $(t^i+d, wt^i)\neq(0, 0)$  for each $i\in\{1, 2, 3\}$.
Let $\varphi(\mathfrak z)=\dfrac{\mathfrak z+d}{w\mathfrak z}$, then $\varphi(\mathfrak z)$
is a M\"{o}bius transformation.
If there exists $i$ such that
$\varphi(t^i)=0$, then we have $(0, 1, 0, 0)^T$ and we are done by the above case.

Otherwise, $h_i=wt^i(0, \varphi(t^i), 1, 0)^T$ and $\varphi(t^i)\neq 0$. Note that $wt^i\neq 0$.
Since $\varphi(\mathfrak z)$ is not a M\"{o}bius transformation mapping $S^1$ to $S^1$,
there is at least one $i$ ($1\leq i\leq 3$)
 such that the complex norm of $\varphi(t^i)$ is not 1.
Then by Corollary~\ref{cor:binary-interpolation}, we have $(0, 1, u, 0)^T$ for any $u\in\mathbb{C}$.
\end{itemize}
\end{proof}

\section{At Least One Entry Is Zero: ${\sf N} \ge 1$}\label{zero:entry}
Recall that ${\sf N}$ is the number of zero entries among the six
entries $\{b, c, d, y, z, w\}$ of the three inner pairs.
Let $f$ be a 4-ary signature with the signature matrix
$M(f)=\left[\begin{smallmatrix}
 a & 0 & 0 & b\\
 0 & c & d & 0\\
 0 & w & z & 0\\
 y & 0 & 0 & a\\
\end{smallmatrix}\right]$.
Recall that we have assumed that $a \not = 0$ from
Lemma~\ref{ax=0} and Lemma~\ref{aneq0},
%Section~\ref{sec3:main-and-outline},
or else we have already proved the classification
as a six-vertex model.
In this section, we assume ${\sf N} \ge 1$.
% that there is at least one of $\{b, c, d, y, z, w\}$ is zero.
Firstly,
if $b=y=c=z=d=w=0$, then $f\in\mathscr{P}$
and Holant$(\neq_2|f)$ can be computed in polynomial time.

\subsection{Two Inner Pairs Are $(0, 0)$}\label{two:(0,0)}
This subsection deals with the case where either ${\sf N} \ge 5$ or
 ${\sf N} =4$ and they appear as two $(0, 0)$ pairs.
Note that  ${\sf N} \ge 5$  implies that two pairs must be $(0, 0)$.
By the symmetry
of the three inner pairs $(b, y), (c, z), (d, w)$, we may assume that $b=y=d=w=0$.
Then we have the following lemma.
\begin{lemma}\label{spin}
Let $f$ be a 4-ary signature with the signature matrix $M(f)=\left[\begin{smallmatrix}
 a & 0 & 0 & 0\\
 0 & c & 0 & 0\\
 0 & 0 & z & 0\\
 0 & 0 & 0 & a\\
\end{smallmatrix}\right],$
%\begin{center}
%\begin{equation}\label{one-zero-pair-matrix-form}
%M_{x_sx_t, x_ux_v}(f)=\begin{bmatrix}
%\alpha & 0 & 0 & 0\\
%0 & \beta & 0 & 0\\
%0 & 0 & \gamma & 0\\
%0 & 0 & 0 & \delta\\
%\end{bmatrix},
%\end{equation}
%\end{center}
%where $\{s,t,u,v\}$ is a permutation of $\{1, 2, 3, 4\}$.
 then the problem $\operatorname{Holant}(\neq_2\mid f)$
is \#$\operatorname{P}$-hard unless
$f$ is $\mathscr{A}$-transformable, more precisely, $\left[\begin{smallmatrix}
 1 & 0\\
0 & \beta&
\end{smallmatrix}\right]^{\otimes 4}f\in\mathscr{A}$ with $\beta^{16}=1$, or
$f\in\mathscr{P}$,
  in which cases the problem is computable in polynomial time.
\end{lemma}
\begin{proof}
Tractability follows from Theorem~\ref{CSP2}.

As $a \not =0$ we can normalize it to $a=1$.
Let $g(x_1, x_2)$ be the binary signature
  $M(g) = \left[\begin{smallmatrix}
 1 & c \\
 z & 1
\end{smallmatrix}\right]$
in matrix form.
%(not necessarily symmetric)
This means that
$g_{00} = g_{11} = 1 = f_{0000}=f_{1111}$, $g_{01} =c = f_{0101}$
and $g_{10} =z = f_{1010}$.
%In matrix form $g(x, y)$ is given by
We prove that
$\operatorname{\#CSP}^2(g)\leq^p_{\operatorname{T}}\operatorname{Holant}(\neq_2\mid f)$
in two steps.
In each step, we begin with a signature grid and end with a new signature grid such that the Holant values of both signature grids are the same.

For step one, let $G=(U, V, E)$ be a bipartite graph
representing  an instance of \#CSP$^2(g)$, where each $u \in U$ is a variable,
and each $v \in V$ has degree two and is labeled $g$.
%Suppose $g(u, u')$ is applied to the variables $u$ and $u'$.
For every vertex $u \in U$,
we define a cyclic order of the edges incident to $u$,
 and decompose $u$ into $2k = \deg(u)$ vertices.
 Then we connect the $2k$
edges originally incident to $u$ to these $2k$ new vertices
so that each vertex is incident to exactly one edge.
We also connect these $2k$ new
 vertices in a cycle according to the cyclic order.
%(see Figure~1 (b) in the Appendix).
Thus, in effect we have replaced $u$ by a cycle of length $2k = \deg(u)$.
% by their incident edges.
    Each of  $2k$ vertices has degree 3, and we assign them $(=_3)$.
Clearly this does not change the value of the partition function. The resulting graph has the following properties: (1) every vertex has either degree 2 or degree 3; (2) each degree 2 vertex is connected
to degree 3 vertices;
%%% i avoid talking about two degree 3 vertices because they could be the same.
(3) each degree 3 vertex is connected to exactly one degree 2 vertex.

\input{holantcsp}

Now step two.
We add a vertex on every edge of each cycle  $C_u$
of length $2k =  \deg(u)$, making $C_u$ a cycle of length $4k$.
(This is shown in Figure~\ref{holant-csp-b}).
Name the vertices $1 ,2, \ldots, 4k$ in cyclic order,
with the newly added vertices numbered $1,3, \ldots, 4k-1$. There are
$k$ pairs of these odd numbered vertices
$(1,3), (5,7), \ldots, (4k-3, 4k-1)$.
We will merge each pair $(4i-3, 4i-1)$ ($1 \le i \le k$)
to form a new vertex of degree 4,
and assign a signature $f$ on it.
(This ``pinching'' operation is illustrated by the dotted lines
in Figure~\ref{holant-csp-b}). The input variables of $f$
are carefully assigned so that the two incoming edges originally
at $4i-3$ are named $x_1$ and $x_3$,
and the other two incoming edges originally
at $4i-1$  are named $x_2$ and $x_4$.
Note that the support of $f$ ensures that the values
at $x_1$ and $x_3$ are equal, and
the values $x_2$ and $x_4$ are equal.
For every even numbered vertex $2i$ ($1 \le i \le 2k$) on $C_u$,
it is currently connected to a vertex $v$ of degree 2 labeled $g$.
Suppose in the  instance of \#CSP$^2(g)$,
the constraint $g(u, u')$ is applied to the variables $u$ and $u'$,
in that order.
Then the other adjacent vertex of $v$ is some  even numbered vertex  $2j$
 on the cycle $C_{u'}$ for the variable $u'$.
We will contract the two incident edges at $v$,
merging the vertices  $2i$ on $C_u$ and  $2j$ on $C_{u'}$,
to form a new vertex $v'$ of degree 4, and assign a copy of $f$ on it.
The input variables of $f$
are carefully assigned so that the two incoming edges originally
at $2i$ of $C_u$ are named $x_1$ and $x_3$,
and the other two incoming edges originally
at $2j$  of $C_{u'}$ are named $x_2$ and $x_4$.
The support of $f$ ensures that the values of
 $x_1$ and $x_3$ are equal and the values of
 $x_2$ and $x_4$ are equal.
(This  is illustrated in Figure~\ref{holant-csp-c}).
Finally  we put a $(\ne_2)$ on every edge.
This completes the definition of an instance of
 $\operatorname{Holant}(\neq_2\mid f)$ in this
reduction.

Note that if we traverse the cycle $C_u$, by the support of
$f$ and the $(\ne_2)$ on every edge, there exists some $\epsilon =0, 1$,
such that all four edges for
the  $f$ at any odd numbered pair $(4i-3, 4i-1)$
must take the same value $\epsilon$, and the two adjacent edges
at every even numbered vertex $2i$ must take the same value $1-\epsilon$.
Therefore there is a 1-1 correspondence between
0-1 assignments for the variables in  \#CSP$^2(g)$
and valid configurations in $\operatorname{Holant}(\neq_2\mid f)$.
Furthermore, at every odd numbered pair $(4i-3, 4i-1)$
the value is $f_{0000}=1$ or $f_{1111}=1$.
The value of $f$ at the vertex $v'$ formed by contraction at $v$
reflects perfectly the value of $g(u,u')$.
Hence,
$\operatorname{\#CSP}^2(g)\leq^p_{\operatorname{T}}\operatorname{Holant}(\neq_2\mid f)$.

%Next we determine what is the signature on $v'\in V'$ after this contraction. Clearly there are only two inputs to each original cycle due to the $(\neq_2)$ on it. Therefore its support is $\{0110, 0101, 1010, 1001\}$. Moreover, the weight on $0110$ is $\alpha$,  the weight on $0101$ is $\beta$, the weight on $1010$ is $\gamma$, and the weight on $1001$ is $\delta$. Hence it is exactly the signature $f$. So we have
%$\operatorname{\#CSP}(g)\leq _{\operatorname{T}}
%\operatorname{Holant}(\neq_2\mid f).$
%

If $g\notin\mathscr{P}\cup\mathscr{A}\cup\alpha\mathscr{A}\cup\mathscr{L}$, then
$\operatorname{\#CSP}^2(g)$ is $\#$P-hard by Theorem~\ref{CSP2}.
It follows that $\operatorname{Holant}(\neq_2\mid  f)$ is \#P-hard.
Otherwise, note that $f(x_1,x_2,x_3,x_4)=g(x_1,x_2)\cdot \chi_{x_1 = x_3} \cdot \chi_{x_2 = x_4}$.
Hence, $g \in \mathscr{A} \cup \mathscr{P}$ implies $f \in \mathscr{A} \cup \mathscr{P}$.
Since $g_{00}\neq 0$, if $g\in\mathscr{L}$, then $g\in\mathscr{A}$
by the definition of $\mathscr{L}$, and
therefore $f \in\mathscr{A}$.
Finally,  if $g\in\alpha\mathscr{A}$, i.e., $\left[
\begin{smallmatrix}
1 & 0\\
0 & \alpha
\end{smallmatrix}
\right]^{\otimes 2}g=(1, \alpha c, \alpha z, {\frak i})^T\in \mathscr{A}$,
then after the holographic transformation to $f$ using $\left[\begin{smallmatrix}
1 & 0\\
0 & \beta
\end{smallmatrix}
\right]$, where $\beta^2=\alpha$, we get the signature $\widehat{f}$ with the signature matrix is
$\left[\begin{smallmatrix}
1 & 0 & 0 & 0\\
0 &  \alpha c & 0 & 0\\
0 & 0 &  \alpha z & 0\\
0 & 0 & 0 & {\frak i}\\
\end{smallmatrix}
\right]$.
Since $\widehat{f}\in\mathscr{A}$,
$(\ne_2)\left[
\begin{smallmatrix}
1 & 0\\
0 & \beta
\end{smallmatrix}
\right]^{\otimes 2}
$ is $(\ne_2)$ up to a nonzero scalar, we conclude that
$f$ is $\mathscr{A}$-transformable.
This finishes the proof.
\end{proof}
%\input{holantcsp}

%%%%%%%%%%%%%%%%%%%%%%%%%%%%%%%%%%%%%%%%%%%%%%%%%%%%%%%%%%%%%%%%%%%%%%%%%%%%%%%%%%%%%%%%%%%%%%%%%%%%%%%%%%%%%%%%%%%%%%%%%%%%%%%%%%%%%%%%%%%%%%%%%%%%%
\subsection{Exactly Two Zero Entries  in Inner Pairs: ${\sf N} =2$}\label{exact:four:nonzero}

This subsection deals with the case ${\sf N} =2$.
So
there are exactly four nonzero entries in the inner pairs.
It follows that there is at least one pair with two nonzero entries.
If the two zero entries are in different pairs,  by the symmetry of the three inner pairs, we assume that $f$
has the signature matrix $M(f)=\left[\begin{smallmatrix}
a & 0 & 0 & b\\
0 & c & d & 0\\
0 & 0 & z & 0\\
0 & 0 & 0 & a\\
\end{smallmatrix}\right]$ with $abcdz\neq 0$.
(This uses the full extent of the induced group action of $S_4$
on the three pairs, possibly switching $(c,z)$; see Section~\ref{trac}.)
  Then we have the following lemma.
\begin{lemma}\label{4:nozero}
Let $f$ be a 4-ary signature with the signature matrix  $M(f)=\left[\begin{smallmatrix}
a & 0 & 0 & b\\
0 & c & d & 0\\
0 & 0 & z & 0\\
0 & 0 & 0 & a\\
\end{smallmatrix}\right]$ with $abcdz\neq 0$, then $\operatorname{Holant}(\neq_2|f)$ is $\#\operatorname{P}$-hard.
\end{lemma}
\begin{proof}
Note that $M^{\sf R_{(12)}\sf R_{(34)}\sf R_{T}}(f)=M_{x_4x_3, x_2x_1}(f)=\left[\begin{smallmatrix}
a & 0 & 0 & 0\\
0 & z & d & 0\\
0 & 0 & c & 0\\
b & 0 & 0 & a\\
\end{smallmatrix}\right]$.
By  connecting  two copies of $f$ using $\neq_2$ we get a
 signature $g$ with the signature matrix
\[M(g) =
M(f)NM_{x_4x_3, x_2x_1}(f)=\left[\begin{smallmatrix}
a & 0 & 0 & b\\
0 & c & d & 0\\
0 & 0 & z & 0\\
0 & 0 & 0 & a\\
\end{smallmatrix}\right]N
\left[\begin{smallmatrix}
a & 0 & 0 & 0\\
0 & z & d & 0\\
0 & 0 & c & 0\\
b & 0 & 0 & a\\
\end{smallmatrix}\right]=
\left[\begin{smallmatrix}
ab & 0 & 0 & a^2\\
0 & dz & c^2+d^2 & 0\\
0 & z^2 & dz & 0\\
a^2 & 0 & 0 & 0\\
\end{smallmatrix}\right].\]
By Lemma~\ref{ax=0}
$\operatorname{Holant}(\neq_2|g)\equiv_T^p\operatorname{Holant}(\neq_2|g')$, where
$g'$ has the signature matrix
$M(g')=\left[\begin{smallmatrix}
0 & 0 & 0 & a^2\\
0 & dz & c^2+d^2 & 0\\
0 & z^2 & dz & 0\\
a^2 & 0 & 0 & 0\\
\end{smallmatrix}\right]$.
Note that the support of $g'$ has cardinalty
either 5 or 6, and therefore the support of $g'$ is not
an affine linear subspace over $\mathbb{Z}_2$, hence
$g' \not \in
\mathscr{A} \cup \mathscr{P}$.
By Theorem~\ref{dichotomy-six-vertex}, $\operatorname{Holant}(\neq_2|g')$ is $\#\operatorname{P}$-hard.
Thus $\operatorname{Holant}(\neq_2|f)$ is $\#\operatorname{P}$-hard.
\end{proof}

If the two zero entries are in a single pair,
 by the symmetry of the three inner pairs, we can assume that $f$ has the
signature matrix $M(f)=\left[\begin{smallmatrix}
a & 0 & 0 & 0\\
0 & c & d & 0\\
0 & w & z & 0\\
0 & 0 & 0 & a\\
\end{smallmatrix}\right]$ with $acdzw\neq 0$. Moreover, by a normalization in $f$, we may assume that $a=1$. Then we have the following lemma.
\begin{lemma}\label{4nonzero-twopair}
If $f$ has the signature matrix  $M(f)=\left[\begin{smallmatrix}
1 & 0 & 0 & 0\\
0 & c & d & 0\\
0 & w & z & 0\\
0 & 0 & 0 & 1\\
\end{smallmatrix}\right]$ with $cdzw\neq 0$, then $\operatorname{Holant}(\neq_2|f)$ is $\#\operatorname{P}$-hard.
\end{lemma}
\begin{proof}
We first suppose
$\det \left[\begin{smallmatrix}
 c & d \\
 w & z
\end{smallmatrix}\right] =0$.
If $c+d\neq0$ or $c+w\neq 0$, then
$\operatorname{Holant}(\neq_2|f)$ is $\#\operatorname{P}$-hard by Lemma~\ref{full:out:deg:inner}.
Otherwise, $d = w = -c \not =0$ and since $\det
\left[\begin{smallmatrix}
 c & d \\
 w & z
\end{smallmatrix}\right] =0$,
we have $z = c$.
So $\left[\begin{smallmatrix}
 c & d \\
 w & z
\end{smallmatrix}\right]=\left[\begin{smallmatrix}
 c & -c \\
 -c & c
\end{smallmatrix}\right]$.
 By  connecting  two copies of $f$ using $\neq_2$ we get a
 signature $f_1$ with the signature matrix
\[M(f_1)
=M(f)NM(f)=
\left[\begin{smallmatrix}
0 & 0 & 0 & 1\\
0 & -2c^2 & 2c^2 & 0\\
0 & 2c^2 & -2c^2 & 0\\
1 & 0 & 0 & 0\\
\end{smallmatrix}\right].\]
The support of  $f_1$ is not an affine subspace.
By Theorem~\ref{dichotomy-six-vertex}, $\operatorname{Holant}(\neq_2|f_1)$ is \#$\operatorname{P}$-hard.
Thus $\operatorname{Holant}(\neq_2|f)$ is \#$\operatorname{P}$-hard.

Now we assume $\left[\begin{smallmatrix}
 c & d \\
 w & z
\end{smallmatrix}\right]$ has full rank.
By  connecting  two copies of $f$ using $\neq_2$ we get a signature $f_2$ with the signature matrix
\[M(f_2)
=M(f)NM(f)=\left[\begin{smallmatrix}
0 & 0 & 0 & 1\\
0 & c(d+w) & cz+d^2 & 0\\
0 & cz+w^2 & z(d+w) & 0\\
1 & 0 & 0 & 0\\
\end{smallmatrix}\right].\]
If $d^2\neq w^2$, then there are at least three entries in $\{c(d+w), cz+d^2, cz+w^2, z(d+w)\}$ that are nonzero.
By Theorem~\ref{dichotomy-six-vertex}, $\operatorname{Holant}(\neq_2|f_2)$ is \#$\operatorname{P}$-hard.
Thus $\operatorname{Holant}(\neq_2|f)$ is \#$\operatorname{P}$-hard.
By the symmetry of the pairs $(d, w)$ and $(c, z)$, if $c^2\neq z^2$, $\operatorname{Holant}(\neq_2|f)$ is \#$\operatorname{P}$-hard, too.

Now we can assume that $d^2=w^2$ and $c^2=z^2$, i.e.,
$M(f)=\left[\begin{smallmatrix}
1 & 0 & 0 & 0\\
0 & c & d & 0\\
0 & \epsilon d & \epsilon c & 0\\
0 & 0 & 0 & 1\\
\end{smallmatrix}\right]$ or
$M(f)=\left[\begin{smallmatrix}
1 & 0 & 0 & 0\\
0 & c & d & 0\\
0 & \epsilon d & -\epsilon c & 0\\
0 & 0 & 0 & 1\\
\end{smallmatrix}\right]$, where $\epsilon=\pm 1.$
Claim A deals with a special case of the first of two cases.

\noindent
{\bf Claim A}:
For $M(f)=\left[\begin{smallmatrix}
1 & 0 & 0 & 0\\
0 & c & d & 0\\
0 & \epsilon d & \epsilon c & 0\\
0 & 0 & 0 & 1\\
\end{smallmatrix}\right]$,
 where $cd \not = 0$ and $\epsilon = \pm 1$,
if $c^2+d^2\neq 0$, then $\operatorname{Holant}(\neq_2|f)$ is $\#\operatorname{P}$-hard.

Note that
$M^{\sf R_T}(f)=M_{x_3x_4, x_1x_2}(f)=\left[\begin{smallmatrix}
1 & 0 & 0 & 0\\
0 & c & \epsilon d & 0\\
0 & d & \epsilon c & 0\\
0 & 0 & 0 & 1\\
\end{smallmatrix}\right]$ and
$M^{\sf R_{(12)}\sf R_{(34)}}(f)=M_{x_2x_1, x_4x_3}(f)=\left[\begin{smallmatrix}
1 & 0 & 0 & 0\\
0 & \epsilon c &  \epsilon d & 0\\
0 &  d &  c & 0\\
0 & 0 & 0 & 1\\
\end{smallmatrix}\right].$
By  connecting two copies of $f$ using $\neq_2$, we get a signature $f_3$ with the signature matrix
\[M(f_3)
=M_{x_3x_4, x_1x_2}(f)NM_{x_2x_1, x_4x_3}(f)=\left[\begin{smallmatrix}
1 & 0 & 0 & 0\\
0 & c & \epsilon d & 0\\
0 & d & \epsilon c & 0\\
0 & 0 & 0 & 1\\
\end{smallmatrix}\right]N\left[\begin{smallmatrix}
1 & 0 & 0 & 0\\
0 & \epsilon c &  \epsilon d & 0\\
0 &  d &  c & 0\\
0 & 0 & 0 & 1\\
\end{smallmatrix}\right]=\left[\begin{smallmatrix}
0 & 0 & 0 & 1\\
0 & 2cd & c^2+d^2 & 0\\
0 & c^2+d^2 & 2cd & 0\\
1 & 0 & 0 & 0\\
\end{smallmatrix}\right].\]
Notice that $\epsilon$ has nicely disappeared, because $\epsilon^2 =1$.
If $c^2+d^2\neq 0$, the support of  $f_3$
is not an affine subspace,
 by Theorem~\ref{dichotomy-six-vertex}, $\operatorname{Holant}(\neq_2|f_3)$ is $\#\operatorname{P}$-hard.
Thus $\operatorname{Holant}(\neq_2|f)$ is $\#\operatorname{P}$-hard.
This proves Claim A.

Moreover, if $c^2+d^2=0$ but $16c^8\neq 1$, then
$2cd$ is not a power of $\frak i$, as
$(2cd)^4 = (\pm 2 {\frak i} c^2)^4 \neq 1$, we have $f_3\notin
\mathscr{A}$.
By
 the symmetry of the three inner pairs,
and
$\det
\left[\begin{smallmatrix}
2cd & 1 \\
1 & 2cd
\end{smallmatrix}\right] \not =0$,
we have $ f_3\notin
\mathscr{P}$ by Lemma~\ref{lm:zhiguo}.
Thus $\operatorname{Holant}(\neq_2|f_3)$ is $\#\operatorname{P}$-hard.
It follows that $\operatorname{Holant}(\neq_2|f)$ is $\#\operatorname{P}$-hard.

Now we can assume that $16c^8=1$
in addition to  $c^2+d^2=0$, i.e., $d=\frak{i}^rc$
where $r\equiv 1\pmod 2$.
Note that 
\[M^{\sf R_{(23)}}(f)=M_{x_1x_3, x_2x_4}(f)=\left[\begin{smallmatrix}
1 & 0 & 0 & c\\
0 & 0 & \frak{i}^rc & 0\\
0 & \epsilon \frak{i}^r c  & 0 & 0\\
\epsilon c & 0 & 0 & 1\\
\end{smallmatrix}\right] ~~~\mbox{ and }~~~ M^{\sf R_{(24)}\sf R_{(34)}}(f)=M_{x_1x_4, x_2x_3}(f)=\left[\begin{smallmatrix}
1 & 0 & 0 & \frak{i}^r  c \\
0 & 0 &  c & 0\\
0 & \epsilon c  & 0 & 0\\
\epsilon \frak{i}^r c & 0 & 0 & 1\\
\end{smallmatrix}\right].\]

Then by connecting two copies of $f$, we get a signature $f_4$ with the signature matrix 
\[M(f_4)=M_{x_1x_3, x_2x_4}(f)NM_{x_1x_4, x_2x_3}(f)=
\left[\begin{smallmatrix}
1 & 0 & 0 & c\\
0 & 0 & \frak{i}^rc & 0\\
0 & \epsilon \frak{i}^r c  & 0 & 0\\
\epsilon c & 0 & 0 & 1\\
\end{smallmatrix}\right]N\left[\begin{smallmatrix}
1 & 0 & 0 & \frak{i}^r  c \\
0 & 0 &  c & 0\\
0 & \epsilon c  & 0 & 0\\
\epsilon \frak{i}^r c & 0 & 0 & 1\\
\end{smallmatrix}\right]=\left[\begin{smallmatrix}
c(1+\epsilon\frak{i}^r) & 0 & 0 & 1+\frak{i}^r c^2\\
0 & 0 & \frak{i}^rc^2 & 0\\
0 & \frak{i}^rc^2 & 0 & 0\\
1+\frak{i}^r c^2 & 0 & 0 & c(\epsilon + \frak{i}^r)\\
\end{smallmatrix}\right].\]
Note that $M_{x_1x_3, x_2x_4}(f_4)=\left[\begin{smallmatrix}
c(1+\epsilon\frak{i}^r) & 0 & 0 & 0\\
0 & 1+\frak{i}^r c^2 & \frak{i}^rc^2 & 0\\
0 & \frak{i}^rc^2 & 1+\frak{i}^r c^2 & 0\\
0 & 0 & 0 & c(\epsilon +\frak{i}^r)\\
\end{smallmatrix}\right]$.
By $r\equiv 1\pmod 2$, we have
$c(1+\epsilon\frak{i}^r)\neq 0$ and $c(\epsilon + \frak{i}^r)\neq 0$.
Moreover, there exists $s\in\{0, 1, 2, 3\}$ such that
$\frak i^r c^2=\frac{\frak i^s}{2}$ by $16c^8=1$. Thus  $(1+\frak{i}^r c^2)^2+(\frak{i}^rc^2)^2
=1+\frak i^s+\frac{(-1)^s}{2}\neq 0$ .
%=\left[\begin{smallmatrix}
%a' & 0 & 0 & 0\\
% & c' & d' & 0\\
%%0 & 0 & 0 & x'\\
%\end{smallmatrix}\right],$
%where $a'=c(1+\frak{i}^r\epsilon), x'=c(\frak{i}^r+\epsilon), c'=1+\frak{i}^r c^2, d'=\frak{i}^rc^2$.
%, we have $a'c'd'x'\neq 0, (c')^2+(d')^2\neq 0$.
%Then by connecting two copies of $g'$, we get the signature $g''$ whose signature matrix is
%\[
%M_{x_1x_3, x_2x_4}(g')NM_{x_1x_3, x_2x_4}(g')=
%\left[\begin{smallmatrix}
%0 & 0 & 0 & a'x'\\
%0 & 2c'd' & (c')^2+(d')^2 & 0\\
%0 & (c')^2+(d')^2 & 2c'd' & 0\\
%'x' & 0 & 0 & 0\\
%\end{smallmatrix}\right].
%\]
Note that $f_4$ has the form in Claim A above, up to a nonzero factor.
Then by Claim A, $\operatorname{Holant}(\neq_2|f_4)$ is \#$\operatorname{P}$-hard.
It follows that $\operatorname{Holant}(\neq_2|f)$ is \#$\operatorname{P}$-hard.

So we can strengthen Claim A to

\noindent
{\bf Claim B}:
For $M(f)=\left[\begin{smallmatrix}
1 & 0 & 0 & 0\\
0 & c & d & 0\\
0 & \epsilon d & \epsilon c & 0\\
0 & 0 & 0 & 1\\
\end{smallmatrix}\right]$, where $cd \not = 0$ and $\epsilon = \pm 1$,
$\operatorname{Holant}(\neq_2|f)$ is $\#\operatorname{P}$-hard.

Claim B proves the first case. Now we consider the second case.
 For $M(f)=\left[\begin{smallmatrix}
1 & 0 & 0 & 0\\
0 & c & d & 0\\
0 & d & -c & 0\\
0 & 0 & 0 & 1\\
\end{smallmatrix}\right]$,
by connecting two copies of $f$, we get a signature $f_5$ with the signature matrix 
\[
M(f_5)=
M(f)NM(f)=\left[\begin{smallmatrix}
1 & 0 & 0 & 0\\
0 & c & d & 0\\
0 & d & -c & 0\\
0 & 0 & 0 & 1\\
\end{smallmatrix}\right]N\left[\begin{smallmatrix}
1 & 0 & 0 & 0\\
0 & c & d & 0\\
0 & d & -c & 0\\
0 & 0 & 0 & 1\\
\end{smallmatrix}\right]=\left[\begin{smallmatrix}
0 & 0 & 0 & 1\\
0 & 2cd & d^2-c^2 & 0\\
0 & d^2-c^2 & -2cd & 0\\
1 & 0 & 0 & 0\\
\end{smallmatrix}\right].
\]
If $c^2\neq d^2$, the support of $f_5$ is
not affine,
 by Theorem~\ref{dichotomy-six-vertex}, $\operatorname{Holant}(\neq_2|f_5)$ is $\#\operatorname{P}$-hard.
Thus $\operatorname{Holant}(\neq_2|f)$ is $\#\operatorname{P}$-hard.
Moreover, if $c^2=d^2$ but $16c^8\neq 1$, then
$(2cd)^4 = (\pm 2 c^2)^4 = 16c^8\neq 1$,
we have $f_5\notin \mathscr{A}$.
As well, $\det
\left[\begin{smallmatrix}
2cd & 1 \\
1 & -2cd
\end{smallmatrix}\right]=
-4c^2d^2 -1 \not = 0$ since $(-4c^4)^2 \not = 1$.
Hence by Lemma~\ref{lm:zhiguo},
$f_5\notin  \mathscr{P}$.
Thus $\operatorname{Holant}(\neq_2|f_5)$ is $\#\operatorname{P}$-hard.
It follows that $\operatorname{Holant}(\neq_2|f)$ is $\#\operatorname{P}$-hard.

Now we can assume that $c^2=d^2$, i.e., $d=\pm c$, and $16c^8=1$.
Note that $M^{\sf R_{(24)}}(f)=M_{x_1x_4, x_3x_2}(f)=\left[\begin{smallmatrix}
1 & 0 & 0 & \pm c\\
0 & c & 0 & 0\\
0 & 0  & -c & 0\\
\pm c & 0 & 0 & 1\\
\end{smallmatrix}\right].$
Then by connecting two copies of $f$, we get a signature $f_6$ with the signature matrix 
\[M(f_6)=M_{x_1x_4, x_3x_2}(f)NM_{x_1x_4, x_3x_2}(f)=
\left[\begin{smallmatrix}
1 & 0 & 0 & \pm c\\
0 & c & 0 & 0\\
0 & 0  & -c & 0\\
\pm c & 0 & 0 & 1\\
\end{smallmatrix}\right]N\left[\begin{smallmatrix}
1 & 0 & 0 & \pm c\\
0 & c & 0 & 0\\
0 & 0  & -c & 0\\
\pm c & 0 & 0 & 1\\
\end{smallmatrix}\right]=\left[\begin{smallmatrix}
\pm 2c & 0 & 0 & 1+ c^2\\
0 & 0 & -c^2 & 0\\
0 & -c^2 & 0 & 0\\
1+c^2 & 0 & 0 & \pm 2c\\
\end{smallmatrix}\right].\]
Note that $M_{x_1x_3, x_2x_4}(f_6)=
M^{\sf R_{(23)}}(f_6) = \left[\begin{smallmatrix}
\pm 2c & 0 & 0 & 0\\
0 & 1+ c^2 & -c^2 & 0\\
0 & -c^2 & 1+ c^2 & 0\\
0 & 0 & 0 & \pm 2c\\
\end{smallmatrix}\right]$, and $1+c^2\neq 0$ by $16c^8=1$.
Let $c'=\pm\frac{1+c^2}{2c}, d'=\mp\frac{c}{2}$, then after the nonzero scalar $\pm 2c$, we have
$M_{x_1x_3, x_2x_4}(f_6)=\left[\begin{smallmatrix}
1 & 0 & 0 & 0\\
0 & c' & d' & 0\\
0 & d' & c' & 0\\
0 & 0 & 0 & 1\\
\end{smallmatrix}\right]$.
By Claim B,
the problem
$\operatorname{Holant}(\neq_2|f_6)$
is $\#\operatorname{P}$-hard.
% by the above case
%that $M(f)=\left[\begin{smallmatrix}
%1 & 0 & 0 & 0\\
%0 & c & d & 0\\
%0 & \epsilon d & \epsilon c & 0\\
%0 & 0 & 0 & 1\\
%\end{smallmatrix}\right]$.
It follows that $\operatorname{Holant}(\neq_2|f)$ is $\#\operatorname{P}$-hard.

By the symmetry of the inner pairs, it follows that $\operatorname{Holant}(\neq_2|f)$ is $\#\operatorname{P}$-hard if
 $M(f)=\left[\begin{smallmatrix}
1 & 0 & 0 & 0\\
0 & c & d & 0\\
0 & -d & c & 0\\
0 & 0 & 0 & 1\\
\end{smallmatrix}\right]$.
This finishes the proof for the second
 case that $M(f)=\left[\begin{smallmatrix}
1 & 0 & 0 & 0\\
0 & c & d & 0\\
0 & \epsilon d & -\epsilon c & 0\\
0 & 0 & 0 & 1\\
\end{smallmatrix}\right]$ where $\epsilon=\pm 1$.
\end{proof}

\subsection{${\sf N} \ge 1$ but Not Included in Subsections~\ref{two:(0,0)}
and \ref{exact:four:nonzero}}\label{other:cases}
After the cases in subsections~\ref{two:(0,0)} and \ref{exact:four:nonzero},
we claim that the remaining cases of ${\sf N} \ge 1$
 can be divided  into three cases as follows:
\begin{description}
\item {Case A:} ${\sf N} =4$ and the two nonzero entries are in two distinct
inner pairs,
 or ${\sf N} =3$
and the three nonzero entries are in three
 distinct inner pairs;
\item {Case B:} ${\sf N} =3$ and
two of the three nonzero entries are in the same pair;
\item {Case C:} ${\sf N} = 1$.
% There are exactly five nonzero entries in the inner pairs;
\end{description}
This covers all remaining cases of ${\sf N} \ge 1$, because:
Subsection~\ref{exact:four:nonzero} dealt with ${\sf N} =2$.
Subsection~\ref{two:(0,0)} dealt with
${\sf N} \ge 5$
and the special case of ${\sf N}  =4$  where all four zeros appear
as two pairs of $(0,0)$.
Case A handles the remaining case of ${\sf N}  =4$ (the two nonzero entries
do not form a single pair), or
${\sf N}  =3$ and every inner pair has one zero and one nonzero.
In Case A there are two pairs each of which contains exactly one zero and
one nonzero.
% 3 nonzero entries are in three distinct pairs.
Case B handles the remaining case of ${\sf N}  =3$ (one pair has two 
nonzero entries, one pair is $(0,0)$, 
and the third pair has one zero and one nonzero).
%the 3 zero entries
%are among two pairs, and so one pair is $(0,0)$).
Case C is what remains of ${\sf N} \ge 1$: namely ${\sf N} =1$,
and there are exactly 5  nonzero entries in the inner pairs.

%\begin{description}
%\item {A:} There are exactly two or three nonzero entries in the inner pairs and they are in different inner pairs;
%\item {B:} There are exactly three nonzero entries in the inner pairs and two of them in the same pair;
%\item {C:} There are exactly five nonzero entries in the inner pairs;
%\end{description}

%%%%%%%%%%%%%%%%%%here   %%%%%%%% here

For Case A,   by the symmetry of the three inner pairs,
 we can assume that $z=w=0, cd\neq 0$ and $by=0$.
(We first can assume that the pairs $(c,z)$ and $(d,w)$ each contains
one zero and one nonzero. Then we can flip the pairs
so that $z=w=0$ with possibly also flipping
the pair $(b,y)$. Since ${\sf N} \ge 3$ in Case A,
at least one of $\{b, y\}$ is 0.)
Then we have the following lemma, which proves that
$\operatorname{Holant}(\neq_2| f )$ is \#$\operatorname{P}$-hard
for every $f$ in Case A.
\begin{lemma}\label{no-nonzero-pair}
Let $f$ be a 4-ary signature with the signature matrix $M(f)=\left[\begin{smallmatrix}
a & 0 & 0 & b\\
0 & c & d & 0\\
0 & 0 & 0 & 0\\
y & 0 & 0 & a\\
\end{smallmatrix}\right]$, with $acd\neq 0$ and $by=0$.
%$M(f)=\left[\begin{smallmatrix}
%a & 0 & 0 & 0\\
%0 & c & d & 0\\
%0 & 0 & 0 & 0\\
%y & 0 & 0 & a\\
%\end{smallmatrix}\right]$ with $aycd\neq 0$.
%If $f$ has the signature matrix $M(a, x, b, 0, c, 0, d, 0)$ or $M(a, x, 0, y, c, 0, d, 0)$,
 Then $\operatorname{Holant}(\neq_2| f )$ is \#$\operatorname{P}$-hard.
\end{lemma}
\begin{proof}
Note that
$M^{\sf R_{(12)} R_{(34)}R_{T}}(f) =
M_{x_4x_3, x_2x_1}(f)=
\left[\begin{smallmatrix}
a & 0 & 0 & y\\
0 & 0 & d & 0\\
0 & 0 & c & 0\\
b & 0 & 0 & a\\
\end{smallmatrix}\right]$. Then by  connecting two copies of $f$ using $\neq_2$ we get a signature $f_1$ with the signature matrix
\[M(f_1)
=M_{x_4x_3, x_2x_1}(f)NM(f)=\left[\begin{smallmatrix}
a & 0 & 0 & y\\
0 & 0 & d & 0\\
0 & 0 & c & 0\\
b & 0 & 0 & a\\
\end{smallmatrix}\right]N
\left[\begin{smallmatrix}
a & 0 & 0 & b\\
0 & c & d & 0\\
0 & 0 & 0 & 0\\
y & 0 & 0 & a\\
\end{smallmatrix}\right]=
\left[\begin{smallmatrix}
2ay & 0 & 0 & a^2\\
0 & cd & d^2 & 0\\
0 & c^2 & cd & 0\\
a^2 & 0 & 0 & 2ab\\
\end{smallmatrix}\right].\]
Here we used the fact that $by=0$.
Let $f_1'$ be a 4-ary signature with the signature matrix
$\left[\begin{smallmatrix}
0 & 0 & 0 & a^2\\
0 & cd & d^2 & 0\\
0 & c^2 & cd & 0\\
a^2 & 0 & 0 & 0\\
\end{smallmatrix}\right]$.
Again by $by=0$, in $M(f_1)$
 we have either $2ay=0$ or $2ab=0$, by Lemma~\ref{ax=0},
Holant$(\neq_2|f_1')\equiv_T^p$ Holant$(\neq_2|f_1)$.
By $acd \not =0$, the support of $f_1'$ has cardinality 6,
and so it is
 not an affine subspace. Hence $f_1' \not \in \mathscr{A} \cup \mathscr{P}$.
By Theorem~\ref{dichotomy-six-vertex}, $\operatorname{Holant}(\neq_2| f'_1)$ is $\#$P-hard, thus
$\operatorname{Holant}(\neq_2| f_1)$ is $\#$P-hard.
It follows that $\operatorname{Holant}(\neq_2| f)$ is $\#$P-hard.
\end{proof}

%\subsection{There are exactly three nonzero entries and two of them are in the same pair}
For Case B, by the symmetry of the three inner pairs, we can assume that
$b=y=w=0$, with possibly flipping $(c,z)$. Then we have the following lemma,
which proves that
$\operatorname{Holant}(\neq_2| f )$ is \#$\operatorname{P}$-hard
for every $f$ in Case B.

\begin{lemma}\label{3:nonzero:onepair}
Let $f$ be a 4-ary signature with the signature matrix $M(f)=\left[\begin{smallmatrix}
a & 0 & 0 & 0\\
0 & c & d & 0\\
0 & 0 & z & 0\\
0 & 0 & 0 & a\\
\end{smallmatrix}\right]$, with $acdz\neq 0$, then $\operatorname{Holant}(\neq_2|f)$ is \#$\operatorname{P}$-hard.
\end{lemma}
\begin{proof}
By  connecting  two copies of $f$ using $\neq_2$ we get a
 signature $f_1$ with the
signature matrix
\[M(f_1) =
M(f)NM(f)=\left[\begin{smallmatrix}
a & 0 & 0 & 0\\
0 & c & d & 0\\
0 & 0 & z & 0\\
0 & 0 & 0 & a\\
\end{smallmatrix}\right]N
\left[\begin{smallmatrix}
a & 0 & 0 & 0\\
0 & c & d & 0\\
0 & 0 & z & 0\\
0 & 0 & 0 & a\\
\end{smallmatrix}\right]=
\left[\begin{smallmatrix}
0 & 0 & 0 & a^2\\
0 & cd & cz+d^2 & 0\\
0 & cz & dz & 0\\
a^2 & 0 & 0 & 0\\
\end{smallmatrix}\right].\]
The support of $f_1$ has cardinality either 5 or 6,
hence not an affine subspace.
So $f_1 \not \in \mathscr{A} \cup \mathscr{P}$.
%Note that there are at least five nonzero entries in $f_1$.
By Theorem~\ref{dichotomy-six-vertex}, $\operatorname{Holant}(\neq_2|f_1)$ is \#$\operatorname{P}$-hard.
Thus $\operatorname{Holant}(\neq_2|f)$ is \#$\operatorname{P}$-hard.
\end{proof}

%\subsection{There are exactly five nonzero entries}
%In this subsection, we consider the case that there is exactly one nonzero entry in the inner pairs of $f$.

For Case C, by the symmetry of the three inner pairs, we can assume that
$z=0$,  by normalizing $a=1$, we have the following lemma.
It shows that
$\operatorname{Holant}(\neq_2| f )$ is \#$\operatorname{P}$-hard
for every $f$ in Case C.

\begin{lemma}\label{5:nonzero}
Let $f$ be a 4-ary signature with the signature matrix $M(f)=\left[\begin{smallmatrix}
1 & 0 & 0 & b\\
0 & c & d & 0\\
0 & w & 0 & 0\\
y & 0 & 0 & 1\\
\end{smallmatrix}\right]$, with $bcdwy\neq 0$, then $\operatorname{Holant}(\neq_2|f)$ is $\#$P-hard.
\end{lemma}
\begin{proof}
The inner matrix is
$\left[\begin{smallmatrix}
c & d\\
w & 0 \\
\end{smallmatrix}\right]$.
By Corollary~\ref{inner:matrix:three:nonzero}, $(0, 1, t, 0)^T$ is available  for any $t\in\mathbb{C}$.
We have 
$M^{\sf R_{(23)}}(f) =
M_{x_1x_3, x_2x_4}(f)=\left[\begin{smallmatrix}
1 & 0 & 0 & c\\
0 & b & d & 0\\
0 & w & y & 0\\
0 & 0 & 0 & 1\\
\end{smallmatrix}\right]$.
By doing a binary modification to the variable $x_2$ of $f$ using $(0, 1, t, 0)^T$, we get a signature $f_1$ with the signature matrix
$M_{x_1x_3, x_2x_4}(f_1)=\left[\begin{smallmatrix}
1 & 0 & 0 & ct\\
0 & b & dt & 0\\
0 & w & yt & 0\\
0 & 0 & 0 & t\\
\end{smallmatrix}\right]$.
Note that $M^{\sf R_{(12)}\sf R_{(34)}\sf R_{(23)}}(f)=M_{x_2x_4, x_1x_3}(f)=\left[\begin{smallmatrix}
1 & 0 & 0 & 0\\
0 & b & w & 0\\
0 & d & y & 0\\
c & 0 & 0 & 1\\
\end{smallmatrix}\right].$
Then by  connecting $f_1$ and $f$ using $\neq_2$ we get a
 signature $f_2$ with the signature matrix
\[M(f_2)
= M_{x_1x_3, x_2x_4}(f_1)NM_{x_2x_4, x_1x_3}(f)=\left[\begin{smallmatrix}
1 & 0 & 0 & ct\\
0 & b & dt & 0\\
0 & w & yt & 0\\
0 & 0 & 0 & t\\
\end{smallmatrix}\right]N\left[\begin{smallmatrix}
1 & 0 & 0 & 0\\
0 & b & w & 0\\
0 & d & y & 0\\
c & 0 & 0 & 1\\
\end{smallmatrix}\right]=
\left[\begin{smallmatrix}
(1+t)c & 0 & 0 & 1\\
0 & bd(1+t) & by+dwt & 0\\
0 & dw+byt & wy(1+t) & 0\\
t & 0 & 0 & 0\\
\end{smallmatrix}\right].\]
The purpose of this construction is to create  $f_2(1,1,1,1)=0$.
By Lemma~\ref{ax=0},
\[\operatorname{Holant}(\neq_2|f_2)\equiv_{T}\operatorname{Holant}(\neq_2|f_2'),\]
where $f_2'$ has the signature matrix
$\left[\begin{smallmatrix}
0 & 0 & 0 & 1\\
0 & bd(1+t) & by+dwt & 0\\
0 & dw+byt & wy(1+t) & 0\\
t & 0 & 0 & 0\\
\end{smallmatrix}\right].$
Then we can choose $t\neq -1, -\frac{by}{dw}$
and $-\frac{dw}{by}$, so that $f_2'$ does not have affine support,
hence $f_2'  \not \in \mathscr{A} \cup \mathscr{P}$.
% such that
%$(bw+dyt)(b^2+d^2t)(w^2+y^2t)( bw+dyt)\neq 0$.
 By Theorem~\ref{dichotomy-six-vertex}, $\operatorname{Holant}(\neq_2|f_2')$ is $\#\operatorname{P}$-hard.
Thus $\operatorname{Holant}(\neq_2|f)$ is $\#\operatorname{P}$-hard.
\end{proof}

 Lemma~\ref{spin}, Lemma~\ref{4:nozero},
Lemma~\ref{4nonzero-twopair},
Lemma~\ref{no-nonzero-pair}
Lemma~\ref{3:nonzero:onepair} and Lemma~\ref{5:nonzero} show that if there is at least one zero in
$\{b, c, d, y, z, w\}$, then Theorem~\ref{main:theorem} holds.
More precisely, we have the following theorem.
%main theorem hold
\begin{theorem}\label{at:least:one:zero}
Let $f$ be a 4-ary signature with the signature matrix
$M(f)=\left[\begin{smallmatrix}
a & 0 & 0 & b\\
0 & c & d & 0\\
0 & w & z & 0\\
y & 0 & 0 & a
\end{smallmatrix}\right]$, where $a\neq 0$ and there is at least one zero in
$\{b, c, d, y, z, w\}$. Then $\operatorname{Holant}(\neq_2|f)$
is $\#\operatorname{P}$-hard except for the case that there are two
 $(0, 0)$  pairs 
in $\{(b, y), (c, z), (d, w)\}$
and
$f$ is $\mathscr{A}$-transformable or
$f\in\mathscr{P}$ (all of which are discussed in Lemma~\ref{spin}),
  in which case the problem is computable in polynomial time.
\end{theorem}

Moreover,
we have the following corollary,
\begin{corollary}\label{affine:support}
Let $f$ be a 4-ary signature with the signature matrix $M(f)=\left[\begin{smallmatrix}
a & 0 & 0 & b\\
0 & c & d & 0\\
0 & w & z & 0\\
y & 0 & 0 & a\\
\end{smallmatrix}\right]$ with $a\neq 0$.
If the support of $f$ is not affine, then $\operatorname{Holant}(\neq_2|f)$ is $\#$P-hard.
In particular, if there are exactly three nonzero entries in the inner matrix
$\left[\begin{smallmatrix}
 c & d \\
 w & z
\end{smallmatrix}\right]$, then $\operatorname{Holant}(\neq_2|f)$ is $\#$P-hard.
\end{corollary}
\begin{proof}
If the support of $f$ is not affine, then ${\sf N} \ge 1$.
Lemma~\ref{spin}, Lemma~\ref{4:nozero},
Lemma~\ref{4nonzero-twopair},
Lemma~\ref{no-nonzero-pair}
Lemma~\ref{3:nonzero:onepair}, and Lemma~\ref{5:nonzero}
cover all cases for  ${\sf N} \ge 1$.
Only in the case covered by Lemma~\ref{spin} there are
tractable problems $\operatorname{Holant}(\neq_2|f)$;
all the other lemmas lead to  $\#$P-hardness.
However $f$ in the case covered in  Lemma~\ref{spin}
has affine support. The Corollary follows.
\end{proof}

\section{Three Equal or Three Opposite Pairs}\label{no:binary}
To use  M\"{o}bius transformations to generate binary signatures
as in Lemma~\ref{construct-binary} or Corollary~\ref{construct:(0,1,0,0)},
we need at least three distinct binary signatures to start the process.
We use Lemma~\ref{construct:basic:binary} to
get a starting binary signature of the form
$(0, 1, t, 0)^T$ for some $t\neq 0$ and $t\neq \pm 1$, which can
give us either at least three distinct  binary signatures
$(0, 1, t^i, 0)^T$ ($ 1 \le i \le 3$),  or $(0, 1, 0, 0)^T$.
But to use Lemma~\ref{construct:basic:binary},
there is an exceptional case where one cannot construct such
starting binary signatures.
This exceptional case in Lemma~\ref{construct:basic:binary} is that
the signature matrix is of the form
$\left[\begin{smallmatrix}
a & 0 & 0 & b\\
0 & c & d & 0\\
0 & \epsilon d & \epsilon c & 0\\
\epsilon b & 0 & 0 & a
\end{smallmatrix}\right]$, for some $\epsilon=\pm 1$.
%$\left[\begin{smallmatrix}
%a & 0 & 0 & \epsilon x\\
%0 & \epsilon z & \epsilon w & 0\\
%0 & w & z & 0\\
%x & 0 & 0 & a
%\end{smallmatrix}\right]$, where $\epsilon=\pm 1$.
In this section, we consider this case, namely
% that
$y=\epsilon b, z=\epsilon c$, and $w=\epsilon d$ for some
$\epsilon=\pm 1$.

Firstly, we assume that $\epsilon=1$.
After Section~\ref{zero:entry}, we may assume $abcd \not =0$.
By normalizing $a=1$, we can assume that
 $f$ has the signature matrix
$M(f)=\left[\begin{smallmatrix}
1 & 0 & 0 & b\\
0 & c & d & 0\\
0 & d & c & 0\\
b & 0 & 0 & 1\\
\end{smallmatrix}\right]$ with $bcd\neq 0$.

\begin{lemma}\label{symmelization}
Let $\mathcal{F}$ be a set of  signatures,  and let $g=[t, 0, 1, 0, \frac{1}{t}]$
 for some $t \not =0$,
then either
$\operatorname{Holant}$$(\neq_2|\mathcal{F}, g)$ is $\#\operatorname{P}$-hard, or
 $\mathcal{F}$ is $\mathscr{P}$-transformable, or $\mathscr{A}$-transformable, or $\mathscr{L}$-transformable.
\end{lemma}
\begin{proof}
Let $T=\left[\begin{smallmatrix}
1 & \sqrt{t} \\
\frak{i} & -\frak{i}\sqrt{t}\\
\end{smallmatrix}\right]=\left[\begin{smallmatrix}
1 & 0 \\
0 & \frak{i}\\
\end{smallmatrix}\right]\left[\begin{smallmatrix}
1 & 1 \\
1 & -1\\
\end{smallmatrix}\right]\left[\begin{smallmatrix}
1 & 0 \\
0 & \sqrt{t}\\
\end{smallmatrix}\right]$.
Note that $(\neq_2)(T^{-1})^{\otimes 2}$ is $(=_2)$,
and  $T^{\otimes 4}g$ is $(=_4)$, both up to a
nonzero
 scalar factor.
% 2 and $T^{\otimes 4}g$ is $(=_4)$ after the scalar $8t$.
%%% not quote 2, 8t. the 2 is also 2t..
By a holographic transformation using $T$, we have
\[
\operatorname{Holant}(\neq_2|\mathcal{F}, g)\equiv^p_{T}\operatorname{Holant}(=_2|=_4, \widehat{\mathcal{F}}),
\]
where $\widehat{\mathcal{F}}=T\mathcal{F}$.
By Lemma~\ref{equality-4-csp2}
\[
\#\operatorname{CSP}^2(\widehat{\mathcal{F}})\leq^p_T\operatorname{Holant}(=_2|=_4, \widehat{\mathcal{F}}).
\]
By Theorem~\ref{CSP2}, $\#\operatorname{CSP}^2(\widehat{\mathcal{F}})$ is $\#\operatorname{P}$-hard or
$\widehat{\mathcal{F}}\subseteq{\mathscr{P}}$, or $\widehat{\mathcal{F}}\subseteq{\mathscr{A}}$,
or $\widehat{\mathcal{F}}\subseteq{\alpha\mathscr{A}}$, or $\widehat{\mathcal{F}}\subseteq{\mathscr{L}}$.
This implies that $\operatorname{Holant}$$(\neq_2|\mathcal{F}, g)$ is $\#\operatorname{P}$-hard or
$\mathcal{F}$ is $\mathscr{P}$-transformable, or $\mathscr{A}$-transformable, or $\mathscr{L}$-transformable.
\end{proof}

\begin{lemma}\label{pm1}\label{pmi}
Let $f$ be a 4-ary signature with the signature matrix $M(f)=\left[\begin{smallmatrix}
1 & 0 & 0 & \frak i^r\\
0 & c & \frak{i}^s c & 0\\
0 & \frak{i}^s c & c & 0\\
\frak i^r & 0 & 0 & 1\\
\end{smallmatrix}\right]$, where $s, r\in\{0, 1, 2, 3\}$ and $c\neq 0$,
then
\begin{itemize}
\item Either $\operatorname{Holant}(\neq_2|f)$ is \#$\operatorname{P}$-hard;
\item or $f\in\mathscr{A}$;
\item or
for some $t\neq 0$,
$\operatorname{Holant}(\neq_2|f, [t, 0, 1, 0, \frac{1}{t}])
\leq_T^p
\operatorname{Holant}(\neq_2|f)$.
\end{itemize}
%, or
% $f$ is $\mathscr{P}$-transformable, or $\mathscr{A}$-transformable, or $\mathscr{L}$-transformable.
\end{lemma}
\begin{proof}
If $r \not \equiv s \pmod 2$, then by connecting two copies of $f$, we get a signature $f_1$ with the signature matrix
\[M(f_1) =
M(f)NM(f)=\left[\begin{smallmatrix}
1 & 0 & 0 & \frak i^r\\
0 & c & \frak{i}^s c & 0\\
0 & \frak{i}^s c & c & 0\\
\frak i^r & 0 & 0 & 1\\
\end{smallmatrix}\right]N\left[\begin{smallmatrix}
1 & 0 & 0 & \frak i^r\\
0 & c & \frak{i}^s c & 0\\
0 & \frak{i}^s c & c & 0\\
\frak i^r & 0 & 0 & 1\\
\end{smallmatrix}\right]=\left[\begin{smallmatrix}
2\frak i^r & 0 & 0 & 1+\frak i^{2r}\\
0 & 2\frak{i}^s c^2 &  (1+\frak i^{2s})c^2 & 0\\
0 &  (1+\frak i^{2s})c^2 & 2\frak{i}^s c^2 & 0\\
1+\frak i^{2r} & 0 & 0 & 2\frak i^r\\
\end{smallmatrix}\right].\]
Note that there are exactly one zero in $\{1+\frak i^{2s}, 1+\frak i^{2r}\}$.
By the symmetry of 3 inner pairs, and Lemma~\ref{4nonzero-twopair}, $\operatorname{Holant}(\neq_2|f_1)$ is $\#$P-hard.
Thus $\operatorname{Holant}(\neq_2|f)$ is $\#$P-hard.

Now we may assume that $r\equiv s\pmod 2$, i.e, $\frak i^s=\epsilon\frak i^r  $, where $\epsilon=\pm 1$,
and $M(f)=\left[\begin{smallmatrix}
1 & 0 & 0 & \frak i^r\\
0 & c & \epsilon\frak{i}^r  c & 0\\
0 & \epsilon\frak{i}^r  c & c & 0\\
\frak i^r & 0 & 0 & 1\\
\end{smallmatrix}\right]$.
If $c^4=1$, then $f\in\mathscr{A}$. Indeed,
let $c=\frak i^k, \epsilon=\frak i^{2\ell}$, where $k, \ell\in\{0, 1, 2, 3\}$,
and
\[Q(x_1, x_2, x_3)=2(k+\ell)x_1x_2+2(r+\ell)x_1x_3+2\ell x_2x_3+(r+k+2\ell)x_1+kx_2+rx_3,\]
then $f(x_1, x_2, x_3, x_4)=\frak i^{Q(x_1, x_2, x_3)}$
on the support of $f$: $x_1+  x_2+  x_3+  x_4 \equiv 0 \pmod 2$.
Then by Definition~\ref{definition-affine}, $f\in\mathscr{A}$.

Now we suppose $c^4\neq 1$. We are also given that $c \not = 0$.
We claim that either  $\operatorname{Holant}(\neq_2|f)$ is $\#$P-hard
or
we can get a signature $\tilde{f}$ with the signature matrix
$M(\tilde{f})=\left[\begin{smallmatrix}
1 & 0 & 0 & \frak i^r\\
0 & c' & \epsilon \frak i^r c' & 0\\
0 & \epsilon\frak i^r c' & c' & 0\\
\frak i^r & 0 & 0 & 1\\
\end{smallmatrix}\right]$, where $c'\neq \pm 1$ and $c'\notin\frak i\mathbb{R}.$
% is not a pure imagine number and $(c')^4\neq 0, 1$.
 Clearly if
 $c\notin\frak i\mathbb{R}$, then we can take  $\tilde{f}$ to be $f$.
Suppose $c \in\frak i\mathbb{R}$.
 Then $c=\rho i$, where $\rho\in\mathbb{R}$ and $\rho\neq\pm 1$ by $c^4\neq 1$. By connecting two copies of $f$,
we get a signature $\widehat{f}$ with the signature matrix
\[M(\widehat{f}) =
M(f)NM(f)=\left[\begin{smallmatrix}
1 & 0 & 0 & \frak i^r\\
0 & c & \epsilon \frak i^r c & 0\\
0 & \epsilon \frak i^r c & c & 0\\
\frak i^r & 0 & 0 & 1\\
\end{smallmatrix}\right]N\left[\begin{smallmatrix}
1 & 0 & 0 & \frak i^r\\
0 & c & \epsilon \frak i^r c & 0\\
0 & \epsilon \frak i^r c & c & 0\\
\frak i^r & 0 & 0 & 1\\
\end{smallmatrix}\right]=\left[\begin{smallmatrix}
2\frak i^r & 0 & 0 & 1+\frak i^{2r}\\
0 & 2 \epsilon \frak i^r c^2 & (1+\frak i^{2r})c^2 & 0\\
0 & (1+\frak i^{2r})c^2 & 2  \epsilon \frak i^r c^2 & 0\\
1+\frak i^{2r} & 0 & 0 & 2\frak i^r\\
\end{smallmatrix}\right].\]
Note that $c^2=-\rho^2$ is a real number and $|c^2|\neq 1$.
\begin{itemize}
\item If $r\equiv 1\pmod 2$, then
$1+\frak i^{2r} =0$, and $M(\widehat{f})=2\frak i^r\left[\begin{smallmatrix}
1 & 0 & 0 & 0\\
0 & \epsilon c^2 & 0 & 0\\
0 & 0 & \epsilon c^2 & 0\\
0 & 0 & 0 & 1\\
\end{smallmatrix}\right]$.
By Lemma~\ref{lm:zhiguo},  $\widehat{f}\notin\mathscr{P}$. Moreover, $\left[\begin{smallmatrix}
1 & 0 \\
 0 & \beta\\
\end{smallmatrix}\right]^{\otimes 4}\widehat{f}\notin\mathscr{A}$ by $|\epsilon c^2|\neq 1$, where $\beta^{16}=1$.
By Lemma~\ref{spin}, $\operatorname{Holant}(\neq_2|\widehat{f})$ is $\#$P-hard.
It follows that $\operatorname{Holant}(\neq_2|f)$ is $\#$P-hard.
\item If $r\equiv 0\pmod 2$, then $1+\frak i^{2r}=2$ and $\frak i^{-r}=\frak i^{r}$.
Thus we have $M(\widehat{f})=2\frak i^r\left[\begin{smallmatrix}
1 & 0 & 0 & \frak i^r\\
0 & \epsilon c^2 & \frak i^r c^2 & 0\\
0 & \frak i^r c^2 &  \epsilon c^2 & 0\\
\frak i^r & 0 & 0 & 1\\
\end{smallmatrix}\right]$.
Hence $\widehat{f}$ has the form $\tilde{f}$ with $c'=\epsilon c^2$.
This finishes the proof of the claim.
\end{itemize}
With $\tilde{f}$ in hand,
we have $M^{\sf R_{(23)}}(\tilde{f})
 = M_{x_1x_3, x_2x_4}(\tilde{f})=2\frak i^r\left[\begin{smallmatrix}
1 & 0 & 0 & c'\\
0 & \frak i^r &  \epsilon \frak i^r  c' & 0\\
0 &  \epsilon \frak i^r  c' &  \frak i^r  & 0\\
c' & 0 & 0 & 1\\
\end{smallmatrix}\right]$,
% (with the swap of $x_2, x_3$),
where $c'\neq\pm 1$ and $c'\notin\frak i\mathbb{R}$.
By Lemma~\ref{eigenvalue-interpolation}, we have
$g_{\lambda}$ that is a 4-ary signature with the signature matrix
$M(g_{\lambda})=\left[\begin{smallmatrix}
1+\epsilon\lambda & 0 & 0 & 1-\epsilon\lambda\\
0 & 1+\lambda & \epsilon(1-\lambda) & 0\\
0 & \epsilon(1-\lambda) & 1+\lambda & 0\\
1-\epsilon\lambda & 0 & 0 & 1+\epsilon\lambda\\
\end{smallmatrix}\right]$ for any $\lambda\in\mathbb{C}$.
In particular we have $g_0$ with the signature matrix
$M(g_{0})=\left[\begin{smallmatrix}
1& 0 & 0 & 1\\
0 & 1 & \epsilon & 0\\
0 & \epsilon & 1 & 0\\
1 & 0 & 0 & 1\\
\end{smallmatrix}\right]$.
If $\epsilon=1$, then $g_0=[1, 0, 1, 0, 1]$,
and the lemma is proved with $t=1$.
%Thus we are done by Lemma~\ref{symmelization}.

If $\epsilon=-1$, i.e., $M(g_{0})=\left[\begin{smallmatrix}
1& 0 & 0 & 1\\
0 & 1 & -1 & 0\\
0 & -1 & 1 & 0\\
1 & 0 & 0 & 1\\
\end{smallmatrix}\right]$ and
% $M_{x_3x_2, x_1x_4}(g_0)=
%%% JYC: i am using 23 , 14. because easier to explain, the they are
% really the same matrix, b'c flipping middle 2 col do not change it.
% from 12 34 ==> flip 12 to 21 34 ==> switch 13 to 23 14
 we have $M^{\sf R_{(24)}}(g_0)=M_{x_1x_4, x_3x_2}(g_0)=
\left[\begin{smallmatrix}
1& 0 & 0 & -1\\
0 & 1 & 1 & 0\\
0 & 1 & 1 & 0\\
-1 & 0 & 0 & 1\\
\end{smallmatrix}\right]$.
%This can be seen by the operations of flipping middle rows $(12)$ and
% then the "2<-->3"  JYC we need a name for these.
By connecting two copies of $g_0$ via $N$, we get a signature $h$ with the signature matrix
\[M(h) =
M_{x_1x_4, x_3x_2}(g_0)NM_{x_1x_4, x_3x_2}(g_0)=
2\left[\begin{smallmatrix}
-1& 0 & 0 & 1\\
0 & 1 & 1 & 0\\
0 & 1 & 1 & 0\\
1 & 0 & 0 & -1\\
\end{smallmatrix}\right].\]
This $h$ is a symmetric signature
$[-1, 0, 1, 0, -1]$, up to a factor 2,
satisfying the lemma with $t = -1$.
%
%After the scalar factor 2, $h=[-1, 0, 1, 0, -1]$.
%Then we are also done by Lemma~\ref{symmelization}
\end{proof}

%%%%%%%%%%%%%%%%%%%%%%%%%%%%%%%%%%%{{{{{{{{{{{{{{{{{{{{{{{{

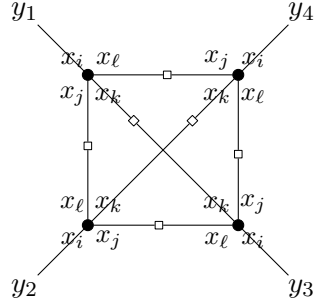
\begin{figure}
\centering
 \begin{tikzpicture}[scale=0.4]
\node [external] (1) at (0, 0) {};
\node [external] (2) at (9, 0) {};
\node [external] (4) at (0, 9) {};
\node [external] (3) at (9, 9) {};
\node at (-0.1, -0.1) {$y_2$};
\node at (9.1, 9.1) {$y_4$};
\node at (-0.1, 9.1) {$y_1$};
\node at (9.1, -0.1) {$y_3$};
\node [internal, scale=0.4] (5) at (2, 2) {};
\node at (1.5, 1.5) {\small{$x_i$}};
\node at (2.7, 2.7) {\small{$x_k$}};
\node at (2.7, 1.5) {\small{$x_j$}};
\node at (1.5, 2.7) {\small{$x_{\ell}$}};
\node [internal, scale=0.4] (6) at (7, 2) {};
\node at (7.5, 1.5) {\small{$x_i$}};
\node at (6.3, 2.7) {\small{$x_k$}};
\node at (7.5, 2.7) {\small{$x_j$}};
\node at (6.3, 1.5) {\small{$x_{\ell}$}};
\node [internal, scale=0.4] (7) at (7, 7) {};
\node at (7.5, 7.5) {\small{$x_i$}};
\node at (6.3, 6.3) {\small{$x_k$}};
\node at (6.3, 7.5) {\small{$x_j$}};
\node at (7.5, 6.3) {\small{$x_{\ell}$}};
\node [internal, scale=0.4] (8) at (2, 7) {};
\node at (1.5, 7.5) {\small{$x_i$}};
\node at (2.7, 6.3) {\small{$x_k$}};
\node at (1.5, 6.3) {\small{$x_j$}};
\node at (2.7, 7.5) {\small{$x_{\ell}$}};
\draw  (1) to  (3) [postaction={decorate, decoration={
                                        markings,
                                        mark=at position 0.63 with {\arrow[>=square,white, scale=0.7] {>}; },
                                        mark=at position 0.63 with {\arrow[>=open square, scale=0.7]  {>}; } } }];
\draw  (2) to  (4) [postaction={decorate, decoration={
                                        markings,
                                        mark=at position 0.63 with {\arrow[>=square,white, scale=0.7] {>}; },
                                        mark=at position 0.63 with {\arrow[>=open square, scale=0.7]  {>}; } } }];
\draw  (5) to  (6) [postaction={decorate, decoration={
                                        markings,
                                        mark=at position 0.5 with {\arrow[>=square,white, scale=0.7] {>}; },
                                        mark=at position 0.5 with {\arrow[>=open square, scale=0.7]  {>}; } } }];
\draw  (6) to  (7)[postaction={decorate, decoration={
                                        markings,
                                        mark=at position 0.5 with {\arrow[>=square,white, scale=0.7] {>}; },
                                        mark=at position 0.5 with {\arrow[>=open square, scale=0.7]  {>}; } } }];
\draw  (7) to  (8)[postaction={decorate, decoration={
                                        markings,
                                        mark=at position 0.5 with {\arrow[>=square,white, scale=0.7] {>}; },
                                        mark=at position 0.5 with {\arrow[>=open square, scale=0.7]  {>}; } } }];
\draw  (8) to  (5)[postaction={decorate, decoration={
                                        markings,
                                        mark=at position 0.5 with {\arrow[>=square,white, scale=0.7] {>}; },
                                        mark=at position 0.5 with {\arrow[>=open square, scale=0.7]  {>}; } } }];
\end{tikzpicture}
\caption{The circles are assigned $f$, and the squares are assigned $(\neq_2)$.
The gadget produces a rotational symmetric signature.}
\label{gadget:rotational:symmetry}
\end{figure}

\begin{lemma}\label{all:case:three:equal:pair}
Let $f$ be a 4-ary signature with the signature matrix
$M(f)=\left[\begin{smallmatrix}
1 & 0 & 0 & b\\
0 & c & d & 0\\
0 & d & c & 0\\
b & 0 & 0 & 1\\
\end{smallmatrix}\right]$ with $bcd\neq 0$,
then we have
\begin{itemize}
\item Either $\operatorname{Holant}(\neq_2|f)$ is \#$\operatorname{P}$-hard;
\item or $f\in\mathscr{A}$;
\item or for some $t\neq 0$,
 $\operatorname{Holant}(\neq_2|f, [t, 0, 1, 0, \frac{1}{t}])\leq_T^p
\operatorname{Holant}(\neq_2|f)$;
\item or $\operatorname{Holant}(\neq_2|f, \mathcal{EQ}_2)\leq_T^p
\operatorname{Holant}(\neq_2|f)$.
\end{itemize}
\end{lemma}
\begin{proof}
Let $t(b, c, d)=b^2c^2+2d^2+(1+d^2)(b^2+c^2)$
and $M(b, c, d)=\left[\begin{smallmatrix}
4d(b^2+c^2) & 4bc(1+d^2)\\
4bc(1+d^2)  & 4d(b^2+c^2)\\
\end{smallmatrix}\right]$.
Both $t(b, c, d)$ and $M(b, c, d)$ are symmetric in its first two
arguments.
The gadget in Figure~\ref{gadget:rotational:symmetry} is
rotationally symmetric. We have:
\begin{itemize}
\item If the ordered tuple
 $(i, j, k, l)=(1, 2, 3, 4)$,
then the gadget gives a signature $g_{bcd}$ with the signature matrix
\[M_{y_1y_3, y_2y_4}(g_{bcd})=\left[\begin{smallmatrix}
4d(b^2+c^2) & 0 & 0 & 4bc(1+d^2)\\
0 & t(b, c, d) & t(b, c, d) & 0\\
0 & t(b, c, d) & t(b, c, d) & 0\\
4bc(1+d^2) & 0 & 0 & 4d(b^2+c^2)\\
\end{smallmatrix}\right].\]
Note that the outer matrix is $M(b, c, d)$.
\item
If $(i, j, k, l)=(1, 4, 3, 2)$, then
 the gadget gives a signature $g_{dcb}$ with the signature matrix
\[M_{y_1y_3, y_2y_4}(g_{dcb})=\left[\begin{smallmatrix}
4b(d^2+c^2) & 0 & 0 & 4dc(1+b^2)\\
0 & t(d, c, b) & t(d, c, b) & 0\\
0 & t(d, c, b) & t(d, c, b) & 0\\
4dc(1+b^2) & 0 & 0 & 4b(d^2+c^2)\\
\end{smallmatrix}\right].\] Note that the outer matrix is $M(d, c, b)$.
\item
If $(i, j, k, l)=(1, 2, 4, 3)$, the gadget gives a signature $g_{bdc}$ with the
signature matrix
\[M_{y_1y_3, y_2y_4}(g_{bdc})=\left[\begin{smallmatrix}
4c(b^2+d^2) & 0 & 0 & 4bd(1+c^2)\\
0 & t(b, d, c) & t(b, d, c) & 0\\
0 & t(b, d, c) & t(b, d, c) & 0\\
4bd(1+c^2) & 0 & 0 & 4c(b^2+d^2)\\
\end{smallmatrix}\right].\] Note that the outer matrix is $M(b, d, c)$.
\end{itemize}

Note that $g_{bcd}, g_{dcb}, g_{bdc}$ are all redundant signatures.
If the compressed signature matrix of any
 one of them has full-rank, then by Theorem~\ref{redundant},
Holant$(\neq_2|f)$ is \#P-hard.
Otherwise, all three compressed signature matrices
of $g_{bcd}, g_{dcb}, g_{bdc}$ are degenerate.
We divide the proof into four cases as follows.

\begin{description}
\item {Case A:}
If $\det(M(b, c, d))=\det(M(d, c, b))=\det(M(b, d, c))=0$, i.e.,
\begin{equation*}
  \begin{aligned}
    d^2(b^2+c^2)^2=b^2c^2(1+d^2)^2, \\
    b^2(d^2+c^2)^2=d^2c^2(1+b^2)^2, \\
    c^2(b^2+d^2)^2=b^2d^2(1+c^2)^2,
  \end{aligned}
\end{equation*}
dividing both sides by $b^2c^2d^2$, we have
\begin{equation}\label{equ1}
    \frac{b^2}{c^2}+\frac{c^2}{b^2}=\frac{1}{d^2}+d^2,
\end{equation}
\begin{equation}\label{equ2}
    \frac{b^2}{d^2}+\frac{d^2}{b^2}=\frac{1}{c^2}+c^2,
    \end{equation}
\begin{equation}\label{equ3}
    \frac{d^2}{c^2}+\frac{c^2}{d^2}=\frac{1}{b^2}+b^2.
\end{equation}
If $c^2= d^2$, then by (\ref{equ3}),  we have $b^2=1$, i.e., $b^2=1, c^2=d^2$. This case was treated  in Lemma~\ref{pm1}.
Thus we can assume that $c^2\neq d^2$.
By a symmetric argument, we can assume that $b^2\neq d^2$ and $b^2\neq c^2$.
Moreover,
if $b^2=-1$, then by (\ref{equ3}),  we have $c^2=-d^2$. This case was
also treated  in Lemma~\ref{pm1}.
Thus we may assume that $b^2\neq -1$.
By a  symmetric argument, we can assume that $c^2\neq -1$ and $d^2\neq -1$.

Now subtracting (\ref{equ2}) from  (\ref{equ1}), we have
\[\frac{c^2-d^2}{b^2}+\frac{b^2(d^2-c^2)}{c^2d^2}=\frac{c^2-d^2}{c^2d^2}+(d^2-c^2).\]
Since $c^2-d^2\neq 0$, we have $\frac{1}{b^2}-\frac{b^2}{c^2d^2}=\frac{1}{c^2d^2}-1,$
i.e,
\[
\frac{1}{b^2}+1=\frac{b^2}{c^2d^2}+\frac{1}{c^2d^2}.
\]
Then by $1+b^2\neq 0$, we have
\begin{equation}\label{equu1}
c^2d^2=b^2.\\
\end{equation}
Similarly by subtracting (\ref{equ3}) from (\ref{equ1}),
% and (\ref{equ2})$-$(\ref{equ3}),
 we have
 \begin{equation}\label{equu2}
b^2d^2=c^2.\\
\end{equation}
%\begin{equation}\label{equu3}
%b^2c^2=d^2.\\
%\end{equation}
By (\ref{equu1}) and (\ref{equu2}),  we have $d^4=1$.
By $d^2\neq -1$,  we have
 $d^2=1$. Then we have $b^2=c^2$ by (\ref{equu1}). This contradicts that $b^2\neq c^2$.

\item {Case B:}
If $t(b, c, d)=t(d, c, b)=t(b, d, c)=0$, i.e.,
\begin{eqnarray}
   && 2d^2+2b^2c^2+(1+d^2)(b^2+c^2)=0 \label{equation1-anti-symmetry}\\
   && 2b^2+2d^2c^2+(1+b^2)(d^2+c^2)=0 \label{equation2-anti-symmetry}\\
   && 2c^2+2b^2d^2+(1+c^2)(b^2+d^2)=0 \label{equation3-anti-symmetry},
\end{eqnarray}
by subtracting (\ref{equation2-anti-symmetry})
and (\ref{equation3-anti-symmetry}) from (\ref{equation1-anti-symmetry})
respectively,
%$-$(\ref{equation2-anti-symmetry}),
%(\ref{equation1-anti-symmetry})$-$(\ref{equation3-anti-symmetry}),
%(\ref{equation2-anti-symmetry})$-$(\ref{equation3-anti-symmetry}),
 we have
\begin{eqnarray}
    &&(1-c^2)(d^2-b^2)=0, \label{equation4-anti-symmetry}\\
    &&(1-b^2)(d^2-c^2)=0. \label{equation5-anti-symmetry}
\end{eqnarray}
    %\begin{equation}\label{equation6-anti-symmetry}
    %(1-d^2)(b^2-c^2)=0.
    %\end{equation}
%Then $b^2, c^2, d^2$
%have to satisfy one of the following conditions:
By (\ref{equation4-anti-symmetry}) and (\ref{equation5-anti-symmetry}),
there are the following possibilities:
 \begin{itemize}
\item $b^2=1, c^2=1$. Then with (\ref{equation1-anti-symmetry}), we have $b^2=c^2=1, d^2=-1$;
\item or $c^2=1, c^2=d^2$. Then with (\ref{equation1-anti-symmetry}), we have $c^2=d^2=1, b^2=-1$;
\item or $b^2=d^2, b^2=1$. Then with (\ref{equation1-anti-symmetry}), we have $b^2=d^2=1, c^2=-1$;
\item or $b^2=d^2, c^2=d^2$. Then with (\ref{equation1-anti-symmetry}), we have $b^2=c^2=d^2=-1$.
\end{itemize}
The  above four cases were  all treated in Lemma~\ref{pm1}.

\item {Case C:} If there are exactly
 two zeros in $\{t(b, c, d), t(d, c, b), t(b, d, c)\}$, by the symmetry of the inner pairs,
we assume that $t(d, c, b) \neq 0$ and $t(b, c, d) = t(b, d, c)=0$. If
% $t(b, c, d)\neq 0, t(d, c, b)=t(b, d, c)=0$. If
$M(d, c, b)=\left[\begin{smallmatrix}
4b(d^2+c^2) & 4dc(1+b^2)\\
4dc(1+b^2) & 4b(d^2+c^2)\\
\end{smallmatrix}\right]$ has full rank, then
$g_{dcb}$ is a redundant signature and its compressed signature matrix has full rank.
Thus Holant$(\neq_2 \mid g_{dcb})$ is $\#$P-hard by Theorem~\ref{redundant}.
So Holant$(\neq_2|f)$ is $\#$P-hard.
Otherwise, we have the following equations
from  $t(b, c, d) = t(b, d, c)=0$
 and $\det M(d, c, b)=0$,
\begin{eqnarray}
&&    2d^2+2c^2b^2+(1+d^2)(c^2+b^2)=0, \label{equuu1} \\
&&    2c^2+2d^2b^2+(1+c^2)(d^2+b^2)=0, \label{equuu2} \\
&&    b^2(d^2+c^2)^2-d^2c^2(1+b^2)^2=0. \label{equuu3}
\end{eqnarray}
By subtracting (\ref{equuu2}) from (\ref{equuu1}), we have
\[(1-b^2)(d^2-c^2)=0.\]
\begin{itemize}
\item If $b^2=1$,
together with (\ref{equuu3}) we have $d^2=c^2$.
%with $b^2(d^2+c^2)^2-d^2c^2(1+b^2)^2=0$, we have $d^2=c^2$.
%Then $M_{y_1y_4, y_3y_2}(f)=\left[\begin{smallmatrix}
%1 & 0 & 0 & \pm 1\\
%0 & c & \frak{i}^r c & 0\\
%0 & \frak{i}^r c &  c & 0\\
%\pm 1 & 0 & 0 & 1\\
%\end{smallmatrix}\right]$, where $r=1, 3$.
\item If $d^2=c^2$,
%with $b^2(d^2+c^2)^2-d^2c^2(1+b^2)^2=0$,
together with (\ref{equuu3}) and $d \not  =0$, we have $b^2=1$.
%Then $M_{y_1y_4, y_3y_2}(f)=\left[\begin{smallmatrix}
%1 & 0 & 0 & \pm i\\
%0 & c & \frak{i}^r c & 0\\
%0 & \frak{i}^r c &  c & 0\\
%\pm i & 0 & 0 & 1\\
%\end{smallmatrix}\right]$, where $r=0, 2$.
\end{itemize}
Hence we have both $b^2=1$ and $d^2=c^2$.
This  case was treated in Lemma~\ref{pm1}.

\item {Case D:} If there is exactly one zero in $\{t(b, c, d), t(d, c, b), t(b, d, c)\}$, by the symmetry of the inner pairs,
we assume that $t(b, c, d)=0$ and $t(d, c, b)t(b, d, c)\neq 0$. If
$\det(M(d, c, b))\neq 0$,
then $g_{dcb}$ is a redundant signature and its compressed signature matrix has full rank.
Thus Holant$(\neq_2|g_{dcb})$ is $\#$P-hard. So Holant$(\neq_2|f)$ is $\#$P-hard.
Thus we can assume that $\det(M(d, c, b))=0$.
By the same argument, we can assume that $\det(M(b, d, c))=0$  by $t(b, d, c)\neq 0$.
If $\det(M(b, c, d))=0$, this is the case that was treated in Case A.
Otherwise, the outer matrix $M(b, c, d)$ of
$M_{y_1y_3, y_2y_4}(g_{bcd})=\left[\begin{smallmatrix}
d(b^2+c^2) & 0 & 0 & bc(1+d^2)\\
0 & 0 & 0 & 0\\
0 & 0 & 0 & 0\\
bc(1+d^2) & 0 & 0 & d(b^2+c^2)\\
\end{smallmatrix}\right]$  has full rank.

\noindent
{\bf Claim A}:
If we have a 4-ary signature $h$ with the signature matrix
$M(h)=\left[\begin{smallmatrix}
1 & 0 & 0 & 0\\
0 & 0 & 0 & 0\\
0 & 0 & 0 & 0\\
0 & 0 & 0 & 1\\
\end{smallmatrix}\right]$, then we can finish the proof
of this lemma.
(This $h$ is $(=_4)$.)

By connecting $h$ and $f$ via $N$, we get a signature $h'$ with the signature matrix
\[
M(h)NM(f)=b\left[\begin{smallmatrix}
1 & 0 & 0 & \frac{1}{b}\\
0 & 0 & 0 & 0\\
0 & 0 & 0 & 0\\
\frac{1}{b} & 0 & 0 & 1\\
\end{smallmatrix}\right].
\]
If $b^4\neq 1$, then by Lemma~\ref{inter:2[1,0,1]}, we have
\[
 \operatorname{Holant}(\neq_2|f, \mathcal{EQ}_2)\leq_T^p\operatorname{Holant}(\neq_2|f).
 \]
 %
 %If $f\notin\mathscr{P}\cup\mathscr{A}\cup\alpha\mathscr{A}\cup\mathscr{L}$, then $\#\operatorname{CSP}^2(f)$
%is \#P-hard by Theorem~\ref{CSP2}. Thus
% $\operatorname{Holant}(\neq_2|f)$ is \#P-hard.
% Otherwise, $f$ is $\mathscr{P}$-transformable or $\mathscr{A}$-transformable or $\mathscr{L}$-transformable.
Now we may assume that $b^4=1$. Note that
\[
M(h)NM_{x_1x_3, x_2x_4}(f)=c\left[\begin{smallmatrix}
1 & 0 & 0 & \frac{1}{c}\\
0 & 0 & 0 & 0\\
0 & 0 & 0 & 0\\
\frac{1}{c} & 0 & 0 & 1\\
\end{smallmatrix}\right], ~~ M(h)NM_{x_1x_4, x_2x_3}(f)=d\left[\begin{smallmatrix}
1 & 0 & 0 & \frac{1}{d}\\
0 & 0 & 0 & 0\\
0 & 0 & 0 & 0\\
\frac{1}{d} & 0 & 0 & 1\\
\end{smallmatrix}\right].
\]
By the same argument, we can assume that $c^4=d^4=1$.
Then we can finish the proof by Lemma~\ref{pm1}.
This proves the Claim A.

Similarly, we have

\noindent
{\bf Claim B}:
If we have a 4-ary signature  $h$ with the signature matrix
$M(h)=\left[\begin{smallmatrix}
0 & 0 & 0 & 1\\
0 & 0 & 0 & 0\\
0 & 0 & 0 & 0\\
1 & 0 & 0 & 0\\
\end{smallmatrix}\right]$, then we can also finish the proof of the lemma.

Note that
\begin{eqnarray*}
M(h)NM(f)
&=&
\left[\begin{smallmatrix}
1 & 0 & 0 & b\\
0 & 0 & 0 & 0\\
0 & 0 & 0 & 0\\
b & 0 & 0 & 1\\
\end{smallmatrix}\right],\\
M(h)NM_{x_1x_3, x_2x_4}(f)
&=&
\left[\begin{smallmatrix}
1 & 0 & 0 & c\\
0 & 0 & 0 & 0\\
0 & 0 & 0 & 0\\
c & 0 & 0 & 1\\
\end{smallmatrix}\right],\\
M(h)NM_{x_1x_4, x_2x_3}(f)
&=&
\left[\begin{smallmatrix}
1 & 0 & 0 & d\\
0 & 0 & 0 & 0\\
0 & 0 & 0 & 0\\
d & 0 & 0 & 1\\
\end{smallmatrix}\right].
\end{eqnarray*}
By the same argument as in the above case, we can assume that $b^4=c^4=d^4=1$.
This case was treated in Lemma~\ref{pm1}.
This proves Claim B.

If one of $\{1+d^2, b^2+c^2\}$ is zero, then exactly one of them is 0, since
$\det (M(b,c,d)) \not =0$.
 Thus we have
$M_{y_1y_3, y_2y_4}(g_{bcd})=4d(b^2+c^2)\left[\begin{smallmatrix}
1 & 0 & 0 & 0\\
0 & 0 & 0 & 0\\
0 & 0 & 0 & 0\\
0 & 0 & 0 & 1\\
\end{smallmatrix}\right]$ or $M_{y_1y_3, y_2y_4}(g_{bcd})=4bc(1+d^2)\left[\begin{smallmatrix}
0 & 0 & 0 & 1\\
0 & 0 & 0 & 0\\
0 & 0 & 0 & 0\\
1 & 0 & 0 & 0\\
\end{smallmatrix}\right]$. Then we can finish the proof of the lemma
 by the above Claims.

If $(1+d^2)(b^2+c^2)\neq 0$, after the scalar $d(b^2+c^2) \neq 0$,
we have $M_{y_1y_3, y_2y_4}(g_{bcd})=\left[\begin{smallmatrix}
1 & 0 & 0 & t\\
0 & 0 & 0 & 0\\
0 & 0 & 0 & 0\\
t & 0 & 0 & 1\\
\end{smallmatrix}\right]$, where $t=\frac{bc(1+d^2)}{d(b^2+c^2)}
\neq 0$, and furthermore $t^2\neq 1$ since $M(b, c, d)$ has full rank.
\begin{itemize}
\item If $t^2\neq -1$, i.e., $t^4\neq 0, 1$,
then by Lemma~\ref{inter:2[1,0,1]}, we have
\[
 \operatorname{Holant}(\neq_2|f, \mathcal{EQ}_2)\leq_T^p\operatorname{Holant}(\neq_2|f).
 \]
 %If $f\notin\mathscr{P}\cup\mathscr{A}\cup\alpha\mathscr{A}\cup\mathscr{L}$, then $\#\operatorname{CSP}^2(f)$
%is \#P-hard by Theorem~\ref{CSP2}. Thus
% $\operatorname{Holant}(\neq_2|f)$ is \#P-hard.
% Otherwise, $f$ is $\mathscr{P}$-transformable or $\mathscr{A}$-transformable or $\mathscr{L}$-transformable.

\item If $t^2=-1$, by connecting two copies of $g_{bcd}$, we have a signature $h_1$ with the signature matrix
\[M_{y_1y_3, y_2y_4}(g_{bcd})NM_{y_1y_3, y_2y_4}(g_{bcd})=\left[\begin{smallmatrix}
1 & 0 & 0 & t\\
0 & 0 & 0 & 0\\
0 & 0 & 0 & 0\\
t & 0 & 0 & 1\\
\end{smallmatrix}\right]N\left[\begin{smallmatrix}
1 & 0 & 0 & t\\
0 & 0 & 0 & 0\\
0 & 0 & 0 & 0\\
t & 0 & 0 & 1\\
\end{smallmatrix}\right]=2t
\left[\begin{smallmatrix}
1 & 0 & 0 & 0\\
0 & 0 & 0 & 0\\
0 & 0 & 0 & 0\\
0 & 0 & 0 & 1\\
\end{smallmatrix}\right].\]
Then we finish the proof by Claim A.
%By connecting $h_1$ to $f$, we get the signature $h_2$ whose signature matrix is
%\[M(h_1)NM(f)=
%\left[\begin{smallmatrix}
%1 & 0 & 0 & 0\\
%0 & 0 & 0 & 0\\
%0 & 0 & 0 & 0\\
%0 & 0 & 0 & 1\\
%\end{smallmatrix}\right]N\left[\begin{smallmatrix}
%1 & 0 & 0 & b\\
%0 & c & d & 0\\
%0 & d & c & 0\\
%b & 0 & 0 & 1\\
%\end{smallmatrix}\right]=b\left[\begin{smallmatrix}
%1 & 0 & 0 & \frac{1}{b}\\
%0 & 0 & 0 & 0\\
%0 & 0 & 0 & 0\\
%\frac{1}{b} & 0 & 0 & 1\\
%\end{smallmatrix}\right].\]
%If $b^4\neq 1$, i.e., $\frac{1}{b^4}\neq 1$, then
%then by Lemma~\ref{inter:2[1,0,1]}, we have
%\[
% \#\operatorname{CSP}^2(f)\leq_T^p\operatorname{Holant}(\neq_2|f).
% \]
%If $f\notin\mathscr{P}\cup\mathscr{A}\cup\alpha\mathscr{A}\cup\mathscr{L}$, then $\#\operatorname{CSP}^2(f)$
%s \#P-hard by Theorem~\ref{CSP2}. Thus
% $\operatorname{Holant}(\neq_2|f)$ is \#P-hard.
% Otherwise, $f$ is $\mathscr{P}$-transformable or $\mathscr{A}$-transformable or $\mathscr{L}$-transformable.
%So we can assume that $b^4=1$. By the symmetry of the inner pairs, we can assume that $c^4=d^4=1$.
%Then we are done by Lemma~\ref{pm1}.
\end{itemize}
%%
%
%If one of $\{1+d^2, b^2+c^2\}$
%is zero,
%we prove the lemma for the case $1+d^2=0$. The proof for the case that $b^2+c^2=0$ is similar and we omit it here.
%By connecting $g_{bcd}$ and $f$ using $\neq_2$ we get the signature $h_3$ whose signature matrix is
%\[M_{y_1y_3, y_2y_4}(g_{bcd})NM(f)=\left[\begin{smallmatrix}
%d(b^2+c^2) & 0 & 0 & 0\\
%0 & 0 & 0 & 0\\
%%0 & 0 & 0 & 0\\
%0 & 0 & 0 & d(b^2+c^2)\\
%\end{smallmatrix}\right]N
%\left[\begin{smallmatrix}
%1 & 0 & 0 & b\\
%0 & c & d & 0\\
%0 & d & c & 0\\
%b & 0 & 0 & 1\\
%\end{smallmatrix}\right]=bd(b^2+c^2)
%\left[\begin{smallmatrix}
%1 & 0 & 0 & \frac{1}{b}\\
%0 & 0 & 0 & 0\\
%0 & 0 & 0 & 0\\
%\frac{1}{b} & 0 & 0 & 1\\
%\end{smallmatrix}\right].\]
%If $b^4\neq 1$, i.e., $\frac{1}{b^4}\neq 1$,
%then by Lemma~\ref{inter:2[1,0,1]}, we have
%\[
%% \#\operatorname{CSP}^2(f)\leq_T^p\operatorname{Holant}(\neq_2|f).
% \]
% If $f\notin\mathscr{P}\cup\mathscr{A}\cup\alpha\mathscr{A}\cup\mathscr{L}$, then $\#\operatorname{CSP}^2(f)$
%is \#P-hard by Theorem~\ref{CSP2}. Thus
 %$\operatorname{Holant}(\neq_2|f)$ is \#P-hard.
 %Otherwise, $f$ is $\mathscr{P}$-transformable or $\mathscr{A}$-transformable or $\mathscr{L}$-transformable.
%So we can assume that $b^4=1$. By symmetry, we can assume that $c^4=d^4=1$.
%Then we are done by Lemma~\ref{pm1} and Lemma~\ref{pmi}.
\end{description}

To summarize, if any of Cases B,  C, or D applies,
then we have proved the lemma.
Suppose Cases B,  C, or D do not apply,
then we have $t(b, c, d)t(d, c, b)t(b, d, c)\neq 0$.
If there is one matrix in $\{M(b, c, d), M(d, c, b), M(b, d, c)\}$ that has full rank,
then we have a redundant signature and its compressed signature matrix has full rank.
Thus Holant$(\neq_2|f)$ is \#P-hard by Theorem~\ref{redundant}.
Otherwise, all of $\{M(b, c, d), M(d, c, b), M(b, d, c)\}$ are degenerate.
So we are in Case A, and the lemma has been proved in that case.
\end{proof}

\begin{lemma}\label{just:one:binary}
Let $f$ be a 4-ary signature with the signature matrix
$M(f)=\left[\begin{smallmatrix}
1 & 0 & 0 & b\\
0 & c & d & 0\\
0 & d & c & 0\\
b & 0 & 0 & 1\\
\end{smallmatrix}\right]$ with $bcd\neq 0$,
then $\operatorname{Holant}(\neq_2|f)$ is
$\#\operatorname{P}$-hard, or $f$ is
 $\mathscr{P}$-transformable, or $\mathscr{A}$-transformable,
 or $\mathscr{L}$-transformable.
\end{lemma}
\begin{proof}
By Lemma~\ref{all:case:three:equal:pair},
we have
\begin{itemize}
\item Either $\operatorname{Holant}(\neq_2|f)$ is \#$\operatorname{P}$-hard;
\item or $f\in\mathscr{A}$;
\item or for some $t \not  =0$,  
$\operatorname{Holant}(\neq_2|f, [t, 0, 1, 0, \frac{1}{t}])\leq_T^p
\operatorname{Holant}(\neq_2|f)$;
\item or $\operatorname{Holant}(\neq_2|f, \mathcal{EQ}_2)\leq_T^p
\operatorname{Holant}(\neq_2|f)$.
\end{itemize}
If $\operatorname{Holant}(\neq_2|f)$ is \#$\operatorname{P}$-hard or $f\in\mathscr{A}$,
then we are done.

If $\operatorname{Holant}(\neq_2|f, [t, 0, 1, 0, \frac{1}{t}])\leq_T^p
\operatorname{Holant}(\neq_2|f)$ for some $t\neq 0$,
by Lemma~\ref{symmelization}, then $f$ is
 $\mathscr{P}$-transformable, or $\mathscr{A}$-transformable,
 or $\mathscr{L}$-transformable.

 If $\operatorname{Holant}(\neq_2|f, \mathcal{EQ}_2)\leq_T^p
\operatorname{Holant}(\neq_2|f)$,
 then we have $\#\operatorname{CSP}^2(f)\leq_T^p
\operatorname{Holant}(\neq_2|f)$.
 By Theorem~\ref{CSP2}, either
 $\operatorname{Holant}(\neq_2|f)$ is \#$\operatorname{P}$-hard, or
 $f\in\mathscr{P}\cup\mathscr{A}\cup\alpha\mathscr{A}\cup\mathscr{L}$.
 This latter condition implies that $f$ is
 $\mathscr{P}$-transformable, or $\mathscr{A}$-transformable,
 or $\mathscr{L}$-transformable.

\end{proof}

%By the proof of Lemma~\ref{just:one:binary}, we have the following corollary.
%\begin{corollary}\label{equality4}
%Let $f$ be a signature whose signature matrices are
%$M(f)=\left[\begin{smallmatrix}
%1 & 0 & 0 & b\\
%0 & c & d & 0\\
%0 & d & c & 0\\
%b & 0 & 0 & 1\\
%\end{smallmatrix}\right]$,
%then $\operatorname{Holant}(\neq_2|f)$ is
%$\#\operatorname{P}$-hard, or $f\in\mathscr{A}$ or
%\[
%\operatorname{Holant}(\neq_2|[t, 0, 1, 0, \frac{1}{t}], f)\leq_T^p\operatorname{Holant}(\neq_2|f),
%\]
%where $t\neq 0$.
%\end{corollary}

\begin{lemma}\label{even}
Let $f$ be a signature with the signature matrix
$M(f)=\left[\begin{smallmatrix}
1 & 0 & 0 & b\\
0 & c & d & 0\\
0 & -d & -c & 0\\
-b & 0 & 0 & 1\\
\end{smallmatrix}\right]$ with $bcd\neq 0$,
% and  $M(f)=\left[\begin{smallmatrix}
%1 & 0 & 0 & b\\
%0 & c & d & 0\\
%0 & d & c & 0\\
%b & 0 & 0 & 1\\
%\end{smallmatrix}\right]$ where $bcd\neq 0$,
then $\operatorname{Holant}(\neq_2|f)$ is
$\#\operatorname{P}$-hard, or $f$ is
 $\mathscr{P}$-transformable, or $\mathscr{A}$-transformable,
 or $\mathscr{L}$-transformable.
\end{lemma}
\begin{proof}
By the holographic transformation using
$M(f)=\left[\begin{smallmatrix}
1 & 0\\
0 & \alpha\\
\end{smallmatrix}\right]$, we have
\[
\operatorname{Holant}(\neq_2|f)\equiv_T^p\operatorname{Holant}(\neq_2|\tilde{f}),
\]
where $M(\tilde{f})=\left[\begin{smallmatrix}
 1 & 0 & 0 & \frak i b\\
 0 & \frak i c & \frak i d & 0\\
 0 & -\frak i d & -\frak i c & 0\\
 -\frak i b & 0 & 0 & -1\\
\end{smallmatrix}\right]$.
Note that
$\operatorname{Holant}(\neq_2|f)$ is \#P-hard iff $\operatorname{Holant}(\neq_2|\tilde{f})$ is \#P-hard,
and $f$ is $\mathscr{C}$-transformable iff $\tilde{f}$ is $\mathscr{C}$-transformable where
 $\mathscr{C}$ is $\mathscr{P}$, or $\mathscr{A}$, or $\mathscr{L}$.

Now we consider $\operatorname{Holant}(\neq_2|\tilde{f})$.
Since $bcd\neq 0$, there is at least one nonzero in $\{b+c, b+d, c+d\}$.
By the symmetry of the inner pairs, we may assume that $c+d\neq 0$.
Then by connecting the variables $x_3, x_4$ of $\tilde{f}$ using $(\neq_2)$, we get the binary signature
$\frak i(c+d)(0,1, -1, 0)^T$, i.e., we have $(0, 1, -1, 0)^T$ after the nonzero scalar $\frak i(c+d)$.
By doing a binary modification to the variable $x_1$ of $\tilde{f}$ using $(0, 1, -1, 0)^T$,
we get a signature $\tilde{f}_1$ with the signature matrix
$M(\tilde{f}_1)=\left[\begin{smallmatrix}
 1 & 0 & 0 & \frak i b\\
 0 & \frak i c & \frak i d & 0\\
 0 & \frak i d & \frak i c & 0\\
 \frak i b & 0 & 0 & 1\\
\end{smallmatrix}\right]$.
Note that doing the same modification on  the variable $x_1$ of
 $\tilde{f}_1$ reverts it back to $\tilde{f}$.

By Lemma~\ref{all:case:three:equal:pair}, we have the following alternatives:
\begin{itemize}
\item if Holant$(\neq_2|\tilde{f}_1)$ is \#P-hard, then Holant$(\neq_2|\tilde{f})$ is \#P-hard.
It follows that Holant$(\neq_2|f)$ is \#P-hard.
\item if $\tilde{f}_1\in\mathscr{A}$, then $\tilde{f}\in\mathscr{A}$.
Thus $f$ is $\mathscr{A}$-transformable.
\item if \[
\operatorname{Holant}(\neq_2|[t, 0, 1, 0, \frac{1}{t}], \tilde{f}_1)\leq_T^p\operatorname{Holant}(\neq_2|\tilde{f}_1),
\]
then \[\operatorname{Holant}(\neq_2|[t, 0, 1, 0, \frac{1}{t}], \tilde{f})\leq_T^p\operatorname{Holant}(\neq_2|\tilde{f}).\]
Then by Lemma~\ref{symmelization}, $\operatorname{Holant}(\neq_2|\tilde{f})$ is
$\#\operatorname{P}$-hard, or $\tilde{f}$ is
 $\mathscr{P}$-transformable, or $\mathscr{A}$-transformable,
 or $\mathscr{L}$-transformable.
 It follows that $\operatorname{Holant}(\neq_2|f)$ is
$\#\operatorname{P}$-hard, or $f$ is
 $\mathscr{P}$-transformable, or $\mathscr{A}$-transformable,
 or $\mathscr{L}$-transformable.
 \item if \[
\operatorname{Holant}(\neq_2|\mathcal{EQ}_2, \tilde{f}_1)\leq_T^p\operatorname{Holant}(\neq_2|\tilde{f}_1),
\]
then \[
\operatorname{Holant}(\neq_2|\mathcal{EQ}_2, \tilde{f})\leq_T^p\operatorname{Holant}(\neq_2|\tilde{f}).
\]
This implies that
\[
\#\operatorname{CSP}^2(\tilde{f})\leq_T^p\operatorname{Holant}(\neq_2|\tilde{f}).
\]
By Lemma~\ref{CSP2},
either $\operatorname{Holant}(\neq_2|\tilde{f})$ is
$\#\operatorname{P}$-hard, in which case  $\operatorname{Holant}(\neq_2|f)$ is
$\#\operatorname{P}$-hard,
or
 $\tilde{f}\in\mathscr{P}\cup\mathscr{A}\cup\alpha\mathscr{A}\cup\mathscr{L}$.
This latter condition implies that $\tilde{f}$ is
 $\mathscr{P}$-transformable, or $\mathscr{A}$-transformable,
 or $\mathscr{L}$-transformable.
 Thus $f$ is
 $\mathscr{P}$-transformable, or $\mathscr{A}$-transformable,
 or $\mathscr{L}$-transformable.
\end{itemize}
\end{proof}

\section{Degenerate Inner Matrices}\label{dege:innceser}
In this section, we consider the case that $by=cz=dw$.
This is precisely when the inner matrix of the  signature matrix
is always degenerate under all relabelings of the variables, i.e., the inner matrix of $M_{x_ix_j, x_kx_{\ell}}(f)$ is degenerate for any permutation $(i, j, k, \ell)$
of $(1, 2, 3, 4)$ where $f$ has the signature matrix $M(f)=\left[\begin{smallmatrix}
a & 0 & 0 & b\\
0 & c & d & 0\\
0 & w & z & 0\\
y & 0 & 0 & a\\
\end{smallmatrix}\right]$.
Note that to use  M\"{o}bius transformations to generate binary signatures
using Lemma~\ref{construct-binary} or Corollary~\ref{construct:(0,1,0,0)},
we need a signature matrix having a  full-rank inner matrix.
%with that its signature matrix has full-rank inner matrix.
So we have to treat this case separately.

\begin{lemma}\label{by=cz=dw}
Let $f$ be a 4-ary signature with the signature matrix
$M(f)=\left[\begin{smallmatrix}
a & 0 & 0 & b\\
0 & c & d & 0\\
0 & w & z & 0\\
y & 0 & 0 & a\\
\end{smallmatrix}\right]$ satisfying
 $abcdyzw\neq 0$ and $by=cz=dw$, then  $f$ is either
$\mathscr{P}$-transformable, or $\mathscr{A}$-transformable, or $\mathscr{L}$-transformable, or
$\operatorname{Holant}$$(\neq_2|f)$ is $\#\operatorname{P}$-hard.
\end{lemma}
\begin{proof}
By normalizing $a=1$, $M(f)=\left[\begin{smallmatrix}
1 & 0 & 0 & b\\
0 & c & d & 0\\
0 & w & z & 0\\
y & 0 & 0 & 1\\
\end{smallmatrix}\right]$.
\begin{itemize}
\item  Suppose  $by\neq 1$.
If $c+d\neq 0$ or $c+w\neq 0$,
then $\operatorname{Holant}(\neq_2|f)$ is $\#\operatorname{P}$-hard
by Lemma~\ref{full:out:deg:inner}.
Otherwise, $d = -c = w$, $z = \frac{dw}{c}=c$, and  we have
$M(f)=\left[\begin{smallmatrix}
1 & 0 & 0 & b\\
0 & c & -c & 0\\
0 & -c & c & 0\\
y & 0 & 0 & 1\\
\end{smallmatrix}\right]$. Here from $by=cz$ we have $by=c^2$.
%By $bc\neq 0$, we have $b\neq c$ or $b\neq -c$.
%Without loss of generality, we assume that $b\neq c$ and
Now we consider
$M^{\sf R_{(24)}}(f) = M_{x_1x_4, x_3x_2}(f)=\left[\begin{smallmatrix}
1 & 0 & 0 & -c\\
0 & c & b & 0\\
0 & y & c & 0\\
-c & 0 & 0 & 1\\
\end{smallmatrix}\right]$.
By $(-c)^2 = by \neq 1$ we can repeat the argument above
to get either  $\operatorname{Holant}(\neq_2|f)$ is $\#\operatorname{P}$-hard
or $b=y =-c$, and $M(f)$ takes the form
$M(f)=\left[\begin{smallmatrix}
1 & 0 & 0 & -c\\
0 & c & -c & 0\\
0 & -c & c & 0\\
-c & 0 & 0 & 1\\
\end{smallmatrix}\right]$.
However now we consider
$M^{\sf R_{(23)}}(f) = M_{x_1x_3, x_2x_4}(f)=\left[\begin{smallmatrix}
1 & 0 & 0 & c\\
0 & -c & -c & 0\\
0 & -c & -c & 0\\
c & 0 & 0 & 1\\
\end{smallmatrix}\right]$, and repeat the argument
we get a contradiction that the nonzero $-c = c$.
%Note that $c^2\neq 1$, $-2c\neq 0$.
%Thus $\operatorname{Holant}$$(\neq_2|f)$ is $\#\operatorname{P}$-hard
%by
%Lemma~\ref{full:out:deg:inner}.
\item Suppose $by=1$. Then $M(f)=\left[\begin{smallmatrix}
1 & 0 & 0 & b\\
0 & c & d & 0\\
0 & \frac{1}{d} & \frac{1}{c} & 0\\
\frac{1}{b} & 0 & 0 & 1\\
\end{smallmatrix}\right]$.
By connecting two copies of $f$, we get a signature $f_1$ with the signature matrix
\[
M(f_1)=M(f)NM_{x_3x_4, x_1x_2}(f)=\left[\begin{smallmatrix}
1 & 0 & 0 & b\\
0 & c & d & 0\\
0 & \frac{1}{d} & \frac{1}{c} & 0\\
\frac{1}{b} & 0 & 0 & 1\\
\end{smallmatrix}\right]N
\left[\begin{smallmatrix}
1 & 0 & 0 & \frac{1}{b}\\
0 & c & \frac{1}{d} & 0\\
0 & d & \frac{1}{c} & 0\\
b & 0 & 0 & 1\\
\end{smallmatrix}\right]=2\left[\begin{smallmatrix}
b & 0 & 0 & 1\\
0 & cd & 1 & 0\\
0 & 1 & \frac{1}{cd} & 0\\
1 & 0 & 0 & \frac{1}{b}\\
\end{smallmatrix}\right].
\]
Note that $M^{\sf R_{(23)}\sf R_{T}}(f_1)=M_{x_2x_4, x_1x_3}(f_1)=
2\left[\begin{smallmatrix}
b & 0 & 0 & \frac{1}{cd}\\
0 & 1 & 1 & 0\\
0 & 1 & 1 & 0\\
cd & 0 & 0 & \frac{1}{b}\\
\end{smallmatrix}\right].$
Then by connecting two copies of $f_1$, we get a signature $f_2$ with the signature matrix
\[
M(f_2) =
M_{x_1x_3, x_2x_4}(f_1)NM_{x_2x_4, x_1x_3}(f_1)=
4
\left[\begin{smallmatrix}
b & 0 & 0 & cd\\
0 & 1 & 1 & 0\\
0 & 1 & 1 & 0\\
\frac{1}{cd} & 0 & 0 & \frac{1}{b}\\
\end{smallmatrix}\right]N
\left[\begin{smallmatrix}
b & 0 & 0 & \frac{1}{cd}\\
0 & 1 & 1 & 0\\
0 & 1 & 1 & 0\\
cd & 0 & 0 & \frac{1}{b}\\
\end{smallmatrix}\right]=
8
\left[\begin{smallmatrix}
bcd & 0 & 0 & 1\\
0 & 1 & 1 & 0\\
0 & 1 & 1 & 0\\
1 & 0 & 0 & \frac{1}{bcd}\\
\end{smallmatrix}\right].
\]
Note that $f_2$ is symmetric and $f_2=[t, 0, 1, 0, \frac{1}{t}]$ with $t=bcd$.
Then we are done by Lemma~\ref{symmelization}.
\end{itemize}
\end{proof}

\section{Using M\"{o}bius Transformations to Achieve Dichotomy}
Let $f$ be a 4-ary signature with the signature matrix
$M(f)=\left[\begin{smallmatrix}
a & 0 & 0 & b\\
0 & c & d & 0\\
0 & w & z & 0\\
y & 0 & 0 & a\\
\end{smallmatrix}\right]$.
After Section~\ref{zero:entry}
we may assume that  $abcdyzw\neq 0$.
% Section~\ref{no:binary} and
Also after Section~\ref{dege:inner},
we can assume that the inner matrix $\left[\begin{smallmatrix}
 c & d \\
 w & z \\
\end{smallmatrix}\right]$ has full rank by the symmetry of the inner pairs.
So if we have five distinct binary signatures $(0, 1, t_i, 0)^T$ for $1\leq i\leq 5$, then we can get $(0, 1, u, 0)^T$ for
any $u\in\mathbb{C}$ by Lemma~\ref{construct-binary}.
Then by carefully choosing binary signatures $(0, 1, u, 0)^T$ and doing binary modifications to $f$ using these binary signatures,
$f$ can be simplified greatly.

\begin{lemma}\label{with-any-binary}
Let $f$ be a 4-ary signature with the signature matrix
$M(f)=\left[\begin{smallmatrix}
a & 0 & 0 & b\\
0 & c & d & 0\\
0 & w & z & 0\\
y & 0 & 0 & a\\
\end{smallmatrix}\right]$ where $abcdyzw\neq 0$ and
$\left[\begin{smallmatrix}
 c & d \\
 w & z \\
\end{smallmatrix}\right]$ has full rank.
For any $t$, if  $t^i$ are distinct for $1\leq i\leq 5$, then
the problem
$\operatorname{Holant}(\neq_2|f, (0, 1, t, 0)^T)$
is \#$\operatorname{P}$-hard.
\end{lemma}
\begin{proof}
By Lemma~\ref{construct-binary},
 $(0, 1, u, 0)^T$ is available for any $u\in\mathbb{C}$.
 %Thus we have $(0, 1, w^{-1}, 0)^T$, $(0, 1, d^{-1}, 0)^T$.
 By normalizing $c=1$, $f$ has the signature matrix
$M(f)=\left[\begin{smallmatrix}
a & 0 & 0 & b\\
0 & 1 & d & 0\\
0 & w & z & 0\\
y & 0 & 0 & a\\
\end{smallmatrix}\right]$.
Then by doing binary modifications to the variables $x_1, x_3$ of $f$ by $(0, 1, w^{-1}, 0)^T$
and $(0, 1, d^{-1}, 0)^T$ respectively, we get a signature $f_1$ with the signature matrix
$M(f_1)=\left[\begin{smallmatrix}
a & 0 & 0 & \frac{b}{d}\\
0 & 1 & 1 & 0\\
0 & 1 & \frac{z}{dw} & 0\\
\frac{y}{w} & 0 & 0 & \frac{a}{dw}\\
\end{smallmatrix}\right]$.
Note that $\frac{z}{dw}\neq 1$ since the inner matrix of $M(f)$ has full rank.

If $\frac{z}{dw}\neq -1$, by doing a binary modification to the variable $x_3$ of $f_1$ using  $(0, 1, -\frac{z}{dw}, 0)^T$, we have
a signature $f_2$ with the signature matrix
$M(f_2)=\left[\begin{smallmatrix}
a & 0 & 0 & -\frac{bz}{d^2w}\\
0 & 1 & -\frac{z}{dw} & 0\\
0 & 1 & -\frac{z^2}{d^2w^2} & 0\\
\frac{y}{w} & 0 & 0 & -\frac{az}{d^2w^2}\\
\end{smallmatrix}\right]$.
Then by connecting $f_2$ and $f_1$ using $\neq_2$ we get a signature $f_3$ with the signature matrix $M(f_3)$ to be

\[
M(f_2)NM(f_1)=\left[\begin{smallmatrix}
a & 0 & 0 & -\frac{bz}{d^2w}\\
0 & 1 & -\frac{z}{dw} & 0\\
0 & 1 & -\frac{z^2}{d^2w^2} & 0\\
\frac{y}{w} & 0 & 0 & -\frac{az}{d^2w^2}\\
\end{smallmatrix}\right]N\left[\begin{smallmatrix}
a & 0 & 0 & \frac{b}{d}\\
0 & 1 & 1 & 0\\
0 & 1 & \frac{z}{dw} & 0\\
\frac{y}{w} & 0 & 0 & \frac{a}{dw}\\
\end{smallmatrix}\right]=\left[\begin{smallmatrix}
\frac{a}{w}(y-\frac{bz}{d^2}) & 0 & 0 & \frac{1}{dw}(a^2-\frac{b^2z}{d^2})\\
0 & 1-\frac{z}{dw} & 0 & 0\\
0 & 1-\frac{z^2}{d^2w^2} & \frac{z}{dw}(1-\frac{z}{dw}) & 0\\
\frac{1}{w^2}(y^2-\frac{a^2z}{d^2}) & 0 & 0 & \frac{a}{dw^2}(y-\frac{bz}{d^2})\\
\end{smallmatrix}\right].\]
The key to this construction is to produce the zero entry
$f_3(0,1,1,0) =0$.
Since $\frac{z}{dw}\neq \pm 1$, there are exactly three nonzero entries in the inner matrix of $M(f_3)$.
This implies that the support of $f$ is
not an affine subspace over $\mathbb{Z}_2$,
since $0101 \oplus 1001  \oplus 1010 = 0110$.
%we have $(1-\frac{z}{dw})(1-\frac{z^2}{d^2w^2})(\frac{z}{dw}(1-\frac{z}{dw}))\neq 0$.
By Corollary~\ref{affine:support}
%Lemma~\ref{4:nozero}, Lemma~\ref{3:nonzero:onepair} and Lemma~\ref{5:nonzero},
Holant$(\neq_2|f_3)$ is $\#$P-hard.
Thus Holant$(\neq_2|f)$ is $\#$P-hard.

Now we may assume that $\frac{z}{dw}=-1$,
i.e., $M(f_1)=\left[\begin{smallmatrix}
a & 0 & 0 & \frac{b}{d}\\
0 & 1 & 1 & 0\\
0 & 1 & -1 & 0\\
\frac{y}{w} & 0 & 0 & \frac{a}{dw}\\
\end{smallmatrix}\right]$.
\begin{itemize}
\item For $\frac{by}{dw}\neq \pm1$, note that
$M_{x_1x_2,x_3x_4}(f_1')
=M^{\sf R_{(24)}}(f_1) 
=M_{x_1x_4, x_3x_2}(f_1)=\left[\begin{smallmatrix}
a & 0 & 0 & 1\\
0 & 1 & \frac{b}{d} & 0\\
0 & \frac{y}{w} & -1 & 0\\
1 & 0 & 0 & \frac{a}{dw}\\
\end{smallmatrix}\right]$. By doing a binary modification to $x_3$ of $f'_1$ using $(0, 1, \frac{d^2}{b^2}, 0)^T$, we get a signature $f_4$
with the signature matrix
$M(f_4)=\left[\begin{smallmatrix}
a & 0 & 0 & \frac{d^2}{b^2}\\
0 & 1 & \frac{d}{b} & 0\\
0 & \frac{y}{w} & -\frac{d^2}{b^2} & 0\\
1 & 0 & 0 & \frac{ad}{b^2w}\\
\end{smallmatrix}\right]$.
Then by connecting $f_4$ and $f'_1$ using $\neq_2$, we get a signature $f_5$ with the signature matrix
\[
%M(f_4)NM_{x_1x_4, x_3x_2}(f_1)=\left[\begin{smallmatrix}
M(f_5) =
M(f_4)NM(f'_1) =\left[\begin{smallmatrix}
a & 0 & 0 & \frac{d^2}{b^2}\\
0 & 1 & \frac{d}{b} & 0\\
0 & \frac{y}{w} & -\frac{d^2}{b^2} & 0\\
1 & 0 & 0 & \frac{ad}{b^2w}\\
\end{smallmatrix}\right]N\left[\begin{smallmatrix}
a & 0 & 0 & 1\\
0 & 1 & \frac{b}{d} & 0\\
0 & \frac{y}{w} & -1 & 0\\
1 & 0 & 0 & \frac{a}{dw}\\
\end{smallmatrix}\right]=
\left[\begin{smallmatrix}
a(1+\frac{d^2}{b^2}) & 0 & 0 & \frac{a^2}{dw}+\frac{d^2}{b^2}\\
0 & \frac{y}{w}+\frac{d}{b} & 0 & 0\\
0 & \frac{y^2}{w^2}-\frac{d^2}{b^2} & -(\frac{y}{w}+\frac{d}{b}) & 0\\
1+\frac{a^2d}{b^2w} & 0 & 0 & \frac{a}{dw}+\frac{ad}{b^2w}\\
\end{smallmatrix}\right].
\]
Again the key point of this construction is to produce the zero entry
$f_5(0,1,1,0) =0$.
Since $\frac{by}{dw}\neq \pm 1$, there are exactly three nonzero entries in the inner matrix of $M(f_5)$.
%we have $(1-\frac{z}{dw})(1-\frac{z^2}{d^2w^2})(\frac{z}{dw}(1-\frac{z}{dw}))\neq 0$.
By
Corollary~\ref{affine:support}
%Lemma~\ref{4:nozero}, Lemma~\ref{3:nonzero:onepair} and Lemma~\ref{5:nonzero},
Holant$(\neq_2|f_5)$ is $\#$P-hard.
Thus Holant$(\neq_2|f)$ is $\#$P-hard.
\item For $\frac{by}{dw}=-1$, the inner matrix of $M(f_1')
=M_{x_1x_4, x_3x_2}(f_1)$ is degenerate.
Note that there is at most one entry
 of $\{\frac{y}{w}, \frac{b}{d}\}$ is $-1$.
If $\frac{a^2}{dw}\neq 1$, then
$\det
\left[\begin{smallmatrix}
a & 1\\
1 & \frac{a}{dw}
\end{smallmatrix}\right]
\not =0$.
Then Holant$(\neq_2|f'_1)$ is \#P-hard by Lemma~\ref{full:out:deg:inner}.
So Holant$(\neq_2|f)$ is \#P-hard.

Now we may assume that $\frac{a^2}{dw}= 1$.
Then
$M(f_1)
=\left[\begin{smallmatrix}
a & 0 & 0 & \frac{b}{d}\\
0 & 1 & 1 & 0\\
0 & 1 & -1 & 0\\
-\frac{d}{b} & 0 & 0 & \frac{1}{a}
\end{smallmatrix}\right]$ and
$M(f'_1) = M_{x_1x_4,x_3x_2}(f_1) = \left[\begin{smallmatrix}
a & 0 & 0 & 1\\
0 & 1 & \frac{b}{d} & 0\\
0 & -\frac{d}{b} & -1 & 0\\
1 &  0 & 0 & \frac{1}{a}
\end{smallmatrix}\right]$.
By doing binary modifications to the variables $x_1, x_3$ of
this function $f'_1$
 using $(0, 1, -\frac{b}{d}, 0)^T$ and $(0, 1, \frac{d}{b}, 0)^T$
respectively, we get a signature $f_6$ with the signature matrix
$M(f_6)=\left[\begin{smallmatrix}
a & 0 & 0 & \frac{d}{b}\\
0 & 1 & 1 & 0\\
0 & 1 & 1 & 0\\
-\frac{b}{d} & 0 & 0 & -\frac{1}{a}\\
\end{smallmatrix}\right]$.
By connecting two copies of $f_6$ via $N$, we get a signature $f_7$ with the signature matrix
\[M(f_7)=M(f_6)NM_{x_3x_4, x_1x_2}(f_6)=\left[\begin{smallmatrix}
a & 0 & 0 & \frac{d}{b}\\
0 & 1 & 1 & 0\\
0 & 1 & 1 & 0\\
-\frac{b}{d} & 0 & 0 & -\frac{1}{a}\\
\end{smallmatrix}\right]N\left[\begin{smallmatrix}
a & 0 & 0 & -\frac{b}{d}\\
0 & 1 & 1 & 0\\
0 & 1 & 1 & 0\\
\frac{d}{b} & 0 & 0 & -\frac{1}{a}\\
\end{smallmatrix}\right]=2\left[\begin{smallmatrix}
\frac{ad}{b} & 0 & 0 & -1\\
0 & 1 & 1 & 0\\
0 & 1 & 1 & 0\\
-1 & 0 & 0 & \frac{b}{ad}
\end{smallmatrix}\right].\]
%By Lemma~\ref{normalize:a=x}, we can assume that $f_7$ has the signature matrix $M(f_7)=\left[\begin{smallmatrix}
%1 & 0 & 0 & -1\\
% & 1 & 1 & 0\\
%0 & 1 & 1 & 0\\
%-1 & 0 & 0 & 1
%\end{smallmatrix}\right]$.
Then by connecting two copies of $f_7$ via $N$, we get a signature $f_8$ with the signature matrix
\[
M(f_8)=
M(f_7)NM(f_7)=4\left[\begin{smallmatrix}
\frac{ad}{b} & 0 & 0 & -1\\
0 & 1 & 1 & 0\\
0 & 1 & 1 & 0\\
-1 & 0 & 0 & \frac{b}{ad}
\end{smallmatrix}\right]N\left[\begin{smallmatrix}
\frac{ad}{b} & 0 & 0 & -1\\
0 & 1 & 1 & 0\\
0 & 1 & 1 & 0\\
-1 & 0 & 0 & \frac{b}{ad}
\end{smallmatrix}\right]=8\left[\begin{smallmatrix}
-\frac{ad}{b} & 0 & 0 & 1\\
0 & 1 & 1 & 0\\
0 & 1 & 1 & 0\\
1 & 0 & 0 & -\frac{b}{ad}
\end{smallmatrix}\right].
\]
Now the important point is that $f_8$ is a \emph{symmetric}
signature where all  weight 2 entries are equal.
Thus,  after the nonzero scalar 8,
$f_8=[s, 0, 1, 0, \frac{1}{s}]$ written as a symmetric
signature listing the values of $f_8$ according to its Hamming weight,
and where $s=-\frac{ad}{b}\neq 0$.

Now since we already have $(0, 1, u, 0)^T$ for any $u\in\mathbb{C}$
by Lemma~\ref{construct-binary},
 we have the reduction Holant$(\neq_2|f_8, (0, 1, u, 0)^T)\leq_T^p$ Holant$(\neq_2|f, (0, 1, t, 0)^T)$.
% for any $u\in\mathbb{C}$.
We finish the proof for this case by proving that  Holant$(\neq_2|f_8, (0, 1, u, 0)^T)$ is \#P-hard for a carefully
chosen $u$.

Let $T=\left[\begin{smallmatrix}
1 & \sqrt{s}\\
 \frak i  & -\frak i\sqrt{s}
\end{smallmatrix}\right]
= \left[\begin{smallmatrix}
1 & 1\\
 \frak i  & -\frak i
\end{smallmatrix}\right]
\left[\begin{smallmatrix}
1 & 0 \\
0 & \sqrt{s}
\end{smallmatrix}\right]
=\left[\begin{smallmatrix}
1 & 0 \\
0 & \frak i
\end{smallmatrix}\right]
\left[\begin{smallmatrix}
1 & 1\\
1 & -1
\end{smallmatrix}\right]
\left[\begin{smallmatrix}
1 & 0 \\
0 & \sqrt{s}
\end{smallmatrix}\right]$.
%=\sqrt{2}Z
%\left[\begin{smallmatrix}
%1 & 0 \\
%0 & \sqrt{s}
%\end{smallmatrix}\right]$.
By the holographic transformation using $T$,
note that $(\ne_2) (T^{-1})^{\otimes 2}$
and $T^{\otimes 4} f_8$ are
nonzero scalar multiples of $(=_2)$ and $(=_4)$ respectively,
 we have
\[
\operatorname{Holant}(\neq_2|f_8, (0, 1, u, 0)^T)\equiv^p_{T}\operatorname{Holant}(=_2|=_4, T^{\otimes 2}(0, 1, u, 0)^T)
.\]
We can calculate that
 $g=T^{\otimes 2}(0, 1, u, 0)^T=\sqrt{s}(u+1, (u-1)\frak i, (1-u)\frak i, u+1)^T$.
By Lemma~\ref{equality-4-csp2}, we have
%$=_2$ on the left side and $=_4$ on the right side, we can construct $=_{2k}$ for any $k\in\mathbb{N}$.
%This implies that
\[
\operatorname{\#CSP}^2(g)\leq_T^p\operatorname{Holant}(=_2|=_4, g).
\]
We can choose any $u$.
Let $u=3$, then  $g\notin\mathscr{P}\cup\mathscr{A}\cup\alpha\mathscr{A}\cup\mathscr{L}$. Thus \#CSP$^2(g)$ is \#P-hard
by Theorem~\ref{CSP2}.
Hence Holant$(\neq_2|f_8, (0, 1, u, 0)^T)$ is \#P-hard and
therefore we conclude that Holant$(\neq_2|f, (0, 1, t, 0)^T)$ is \#P-hard.
\item For $\frac{by}{dw}=1$, note that the inner matrix of $M(f''_1)
=M^{\sf R_{(23)}}(f_1)
= M_{x_1x_3, x_2x_4}(f_1)=\left[\begin{smallmatrix}
a & 0 & 0 & 1\\
0 & \frac{b}{d} & 1 & 0\\
0 & 1 & \frac{d}{b} & 0\\
-1 & 0 & 0 & \frac{a}{dw}
\end{smallmatrix}\right]$ is degenerate.
By doing binary modifications to the variables $x_1$ and $x_4$ of $f''_1$
using the binary signatures
 $(0, 1, \frac{b}{d}, 0)^T$ and $(0, 1, \frac{d}{b}, 0)^T$
in succession, we get a signature $h_1$ with the signature matrix
$M(h_1)=\left[\begin{smallmatrix}
a & 0 & 0 & \frac{d}{b}\\
0 & 1 & 1 & 0\\
0 & 1 & 1 & 0\\
-\frac{b}{d} & 0 & 0 & \frac{a}{dw}
\end{smallmatrix}\right]$.
Note that $h_1$ is a redundant signature.
If $\frac{a^2}{dw}\neq -1$,
then the compressed signature matrix of $h_1$ has full rank.
Thus Holant$(\neq_2|h_1)$ is \#P-hard by Theorem~\ref{redundant} and Holant$(\neq_2|f)$ is \#P-hard.

Now we can assume that $\frac{a^2}{dw}=-1$.
Then $M(h_1)=\left[\begin{smallmatrix}
a & 0 & 0 & \frac{d}{b}\\
0 & 1 & 1 & 0\\
0 & 1 & 1 & 0\\
-\frac{b}{d} & 0 & 0 & -\frac{1}{a}
\end{smallmatrix}\right]$.
Then by connecting two copies of $h_1$ via $N$, we get a signature $h_2$ with the signature matrix
\[M(h_2)=M(h_1)NM_{x_3x_4, x_1x_2}(h_1)=
\left[\begin{smallmatrix}
a & 0 & 0 & \frac{d}{b}\\
0 & 1 & 1 & 0\\
0 & 1 & 1 & 0\\
-\frac{b}{d} & 0 & 0 & -\frac{1}{a}
\end{smallmatrix}\right]N
\left[\begin{smallmatrix}
a & 0 & 0 & -\frac{b}{d}\\
0 & 1 & 1 & 0\\
0 & 1 & 1 & 0\\
\frac{d}{b} & 0 & 0 & -\frac{1}{a}
\end{smallmatrix}\right]=2
\left[\begin{smallmatrix}
\frac{ad}{b} & 0 & 0 & -1\\
0 & 1 & 1 & 0\\
0 & 1 & 1 & 0\\
-1 & 0 & 0 & \frac{b}{ad}
\end{smallmatrix}\right].\]
Then by connecting two copies of $h_2$ via $N$, we get a signature $h_3$ with the signature matrix
\[M(h_3)=M(h_2)NM(h_2)=4\left[\begin{smallmatrix}
\frac{ad}{b} & 0 & 0 & -1\\
0 & 1 & 1 & 0\\
0 & 1 & 1 & 0\\
-1 & 0 & 0 & \frac{b}{ad}
\end{smallmatrix}\right]N
\left[\begin{smallmatrix}
\frac{ad}{b} & 0 & 0 & -1\\
0 & 1 & 1 & 0\\
0 & 1 & 1 & 0\\
-1 & 0 & 0 & \frac{b}{ad}
\end{smallmatrix}\right]=8\left[\begin{smallmatrix}
-\frac{ad}{b} & 0 & 0 & 1\\
0 & 1 & 1 & 0\\
0 & 1 & 1 & 0\\
1 & 0 & 0 & -\frac{b}{ad}
\end{smallmatrix}\right].\]

Note that $h_3=8[s, 0, 1, 0, \frac{1}{s}]$ is a symmetric signature where $s=-\frac{ad}{b}\neq 0$.
The remaining proof is similar with the case that $\frac{by}{dw}=-1$ and we omit it here.
\end{itemize}
\end{proof}

\begin{theorem}\label{with:(0,1,0,0)}
Let $f$ be a 4-ary signature with the signature matrix
$M(f)=\left[\begin{smallmatrix}
a & 0 & 0 & b\\
0 & c & d & 0\\
0 & w & z & 0\\
y & 0 & 0 & a\\
\end{smallmatrix}\right]$ where $abcdyzw\neq 0$ and
$\left[\begin{smallmatrix}
 c & d \\
 w & z \\
\end{smallmatrix}\right]$ has full rank, then
$\operatorname{Holant}(\neq_2|f, (0, 1, 0, 0)^T)$ is $\#\operatorname{P}$-hard, or $f$ is  $\mathscr{A}$-transformable.
\end{theorem}
\begin{proof}
By a normalization in $f$, we may assume that $c=1$.

Note that we have $(0, 1, 0, 0)^T=(1, 0)^T\otimes (0, 1)^T$. Thus we
can set $x_i=0$ and $x_j=1$ for any two variables $x_i, x_j$.
By setting $x_1=0$ and $x_4=1$, $x_3=0$ and $x_2=1$, $x_1=0$ and $x_2=1$, $x_3=0$ and $x_4=1$ respectively using $(0, 1, 0, 0)^T$
%doing loops to $f$ using $(0, 1, 0, 0)^T$,
% $x_1$ and $x_4$, $x_2$ and $x_3$, $x_1$ and $x_2$, $x_3$ and $x_4$ using $(0, 1, 0 ,0)$ via $N$ respectively,
we get the binary signatures
$(0, 1, b, 0)^T$,
$(0, 1, y, 0)^T$,
$(0, 1, d, 0)^T$, and
$(0, 1, w, 0)^T$,
i.e., for any $u\in\{b, y, d, w\}$, we can get $(0, 1, u, 0)^T$.
Thus if  there exists $u\in\{b, y, d, w\}$ such that $u^i$ are distinct for $1\leq i\leq 5$, then
$\operatorname{Holant}(\neq_2|f, (0, 1, 0, 0)^T)$ is $\#\operatorname{P}$-hard by Lemma~\ref{with-any-binary}.
So we may assume  all of $\{b, y, d, w\}$ are in $\{\pm 1, \pm \frak i, \omega, \omega^2\}$,
where $\omega = e^{2 \pi {\frak i}/3}$.
%where $\omega^3=1$ and $\omega\neq 1$.
%Then by doing loops to $f$ using $g=(0, 1, 0, 0)^{T}$,
%we can get binary signatures
% $h_1=M_{x_3x_4, x_1x_2}(f)g=(0, 1, d, 0)^T,$
%$ h_2=M(f)g=(0, 1, w, 0),$
% $ h_3=M_{x_1x_2, x_4x_3}(f)g=d(0, 1,\frac{z}{d}, 0), $
%$h_4=M_{x_2x_3, x_1x_4}(f)g=(0, 1, b, 0),$
%h_5=M_{x_1x_4, x_2x_3}(f)g=(0, 1, y, 0)$.

%Note that  for any $u\neq 0$, if we have $g(x_1, x_2)=(0, 1, u, 0)^T$, then
%we also
%have $g'(x_1, x_2)=g(x_2, x_1)=u(0, 1, u^{-1}, 0)^T$, i.e., we have $(0, 1, u^{-1}, 0)^T$ after the nonzero scalar $u$.
%%
%%can construct $(0, 1, u^{-1}, 0)^T$.
%%If $u$ is not a root of unity, then we can get $(0, 1, u^{-1}, 0)^T$ by interpolation.
%%If $u$ is a root of unity of order $d$, by linking $d-1$ copies of $(0, 1, u, 0)^T$,
%%we get $(0, 1, u^{d-1}, 0)^T=(0, 1, u^{-1}, 0)^T$.%
%%
%%Since we have $(0, 1, d, 0)^T$, $(0, 1, w, 0)^T$  by doing loops to $f$ using $(0, 1, 0, 0)^T$,
By Lemma~\ref{inverse}, we have $(0, 1, d^{-1}, 0)^T$
and $(0, 1, w^{-1}, 0)^T$.
By binary modifications using  $(0, 1, w^{-1}, 0)^T$
and $(0, 1, d^{-1}, 0)^T$ to the variables
$x_1, x_3$ of $f$ respectively, we get a signature $f_1$ with the signature matrix
$M(f_1)=\left[\begin{smallmatrix}
a & 0 & 0 & \frac{b}{d}\\
0 & 1 & 1 & 0\\
0 & 1 & \frac{z}{dw} & 0\\
\frac{y}{w} & 0 & 0 & \frac{a}{dw}\\
\end{smallmatrix}\right]$.
Since  $\left[\begin{smallmatrix}
 1 & d\\
 w & z \\
\end{smallmatrix}\right]$ has full rank, we have $\frac{z}{dw}\neq 1$.
By doing a loop to $f_1$ using $\neq_2$, we get the
 binary signature  $h=M(f)(\neq_2)=(0, 2, 1+\frac{z}{dw}, 0)^T$.

We claim that $\frac{z}{dw}=-1$ or we are done.
By pinning $x_4=0$ and $x_3=1$ using $(0, 1, 0, 0)^T$, we get the binary signature $(0, 1, \frac{z}{dw}, 0)^T$.
If $(\frac{z}{dw})^i$ are distinct for $1\leq i\leq 5$, then
the problem
$\operatorname{Holant}(\neq_2|f, (0, 1, \frac{z}{dw}, 0)^T)$ is $\#\operatorname{P}$-hard by Lemma~\ref{with-any-binary}.
 Thus $\operatorname{Holant}(\neq_2|f, (0, 1, 0, 0)^T)$ is $\#\operatorname{P}$-hard.
 So we may assume
 that $\frac{z}{dw}\in\{-1, \frak i, -\frak i, \omega, \omega^2\}$.
 If $\frac{z}{dw}=\frak i$ or $-\frak i$, then $|1+\frac{z}{dw}|=\sqrt{2}$.
 If $\frac{z}{dw}=\omega$ or $\omega^2$, then $|1+\frac{z}{dw}|=1 \not = 2$.
 So
 $\operatorname{Holant}(\neq_2|f, h)$ is $\#\operatorname{P}$-hard by Lemma~\ref{with-any-binary}.
 Thus $\operatorname{Holant}(\neq_2|f, (0, 1, 0, 0))^T$ is $\#\operatorname{P}$-hard.
 So we may assume that $(0, 1, \frac{z}{dw}, 0)^T=(0, 1, -1, 0)^T$.

%If there exists $u\in\{b, y, \frac{z}{d}, d, w\}$ such that $u^i$ are distinct for $1\leq i\leq 5$, then
%$\operatorname{Holant}(\neq_2|f, (0, 1, 0, 0)^T)$ is $\#\operatorname{P}$-hard by Lemma~\ref{with-any-binary}.
%Thus all of them are roots of unity and the orders are less than 5.

Suppose there exists  $u\in\{b, d, y,  w\}$
such that $u=\omega$ or $\omega^2$.
By linking $(0, 1, u, 0)^T$ and $(0, 1, -1, 0)^T$ using $\neq_2$, we get
the binary signature $(0, 1, -u, 0)^T$.
Note that $(-u)^i$ are distinct for $1\leq i\leq 6$, therefore
$\operatorname{Holant}(\neq_2|f, (0, 1, -u, 0)^T)$ is $\#\operatorname{P}$-hard by Lemma~\ref{with-any-binary}.
It follows that  $\operatorname{Holant}(\neq_2|f, (0, 1, 0, 0)^T)$ is $\#\operatorname{P}$-hard.

Now we can assume
$\{b, y, d,  w\}\subseteq\{1, -1, \frak{i}, -\frak{i}\}$.
Then by $\frac{z}{dw}=-1$, we have $z^4=1$.
Thus $\{b, y, z, d,  w\}\subseteq\{1, -1, \frak{i}, -\frak{i}\}$.
  There exist
  $j, k,  \ell, m, n\in\{0, 1, 2, 3\}$, such that
\[b=\frak{i}^{j},~~~ y=\frak{i}^{k},~~~ z=\frak{i}^{\ell}, ~~~
  d=\frak{i}^{m}, ~~~ w=\frak{i}^{n}.\]
%So $h_i$ are affine signatures for $1\leq i\leq 5$.
The signature matrices of $f$ and $f_1$ are respectively
$$M(f)=\left[\begin{smallmatrix}
a & 0 & 0 & \frak{i}^{j}\\
0 & 1 & \frak{i}^{m} & 0\\
0 & \frak{i}^{n} & \frak{i}^{\ell} & 0\\
\frak{i}^{k}& 0 & 0 & a
\end{smallmatrix}\right],~~~~\mbox{and}~~~~
M(f_1)=\left[\begin{smallmatrix}
a & 0 & 0 & \frak{i}^{j-m}\\
0 & 1 & 1 & 0\\
0 & 1 & \frak{i}^{\ell-m-n} & 0\\
\frak{i}^{k-n}& 0 & 0 & a\frak i^{-m-n}
\end{smallmatrix}\right].$$
We have $\ell-m-n\equiv 2 \pmod 4$ by $\frac{z}{dw}=\frak{i}^{\ell-m-n}=-1$.

Note that 
$M^{\sf R_{(24)}}(f) = M_{x_1x_4, x_3x_2}(f)=
\left[\begin{smallmatrix}
a & 0 & 0 & \frak{i}^{m}\\
0 & 1 & \frak{i}^{j} & 0\\
0 & \frak{i}^{k} & \frak{i}^{\ell} & 0\\
\frak{i}^{n}& 0 & 0 & a
\end{smallmatrix}\right]$.
By doing a loop to $f$ using $(0, 1, \frak i^{-k}, 0)^T$, we get the binary signature
\[
M_{x_3x_2, x_1x_4}(f)N(0, \frak i^{-k}, 1, 0)^T=(0, 2, \frak i^{j}(1+\frak i^{\ell-j-k}), 0)^T.
\]
If $\ell-j-k\equiv 1 \pmod 2$, then $|\frak{i}^{j}(1+\frak{i}^{\ell-j-k})|
= \sqrt{2}$ and
%\neq 2, 0$ and
Holant$(\neq_2|f, (0, 2, \frak{i}^{j}(1+\frak{i}^{\ell-j-k}), 0)^T)$ is $\#$P-hard
by Lemma~\ref{with-any-binary}.
Thus Holant$(\neq_2|f, (0, 1, 0, 0)^T)$ is $\#$P-hard.
Hence  we may assume $\ell-j-k\equiv 0 \pmod 2$.
By $\ell-m-n\equiv 2 \pmod 4$, we have
\[ j+k + m+n \equiv 0 \pmod 2.\]

By connecting two copies of $f$, we get a signature $f_2$ with the signature matrix
\[M(f_2)=M(f)NM_{x_3x_4, x_1x_2}(f)=
\left[\begin{smallmatrix}
2a\frak{i}^{j} & 0 & 0 & a^2+\frak{i}^{j+k}\\
0 & 2\frak{i}^{m} & \frak{i}^{\ell}+\frak{i}^{m+n} & 0\\
0 & \frak{i}^{\ell}+\frak{i}^{m+n} & 2\frak{i}^{n+\ell} & 0\\
a^2+\frak{i}^{j+k} & 0 & 0 & 2a\frak{i}^{k}
\end{smallmatrix}\right].
\]
Note that $\frak{i}^{\ell}+\frak{i}^{m+n}=0$ by $\ell-m-n\equiv 2 \pmod 4$.
If $a^2+\frak{i}^{j+k}\neq 0$, then
Holant$(\neq_2|f_2)$ is $\#$P-hard by Lemma~\ref{4nonzero-twopair}
and the symmetry of the three inner pairs.
Thus Holant$(\neq_2|f)$ is $\#$P-hard.

Now we assume $a^2+\frak{i}^{j+k}=0$. Then $M(f)=\left[\begin{smallmatrix}
\frak{i}^{\frac{j+k}{2}+\epsilon} & 0 & 0 & \frak{i}^{j}\\
0 & 1 & \frak{i}^{m} & 0\\
0 & \frak{i}^{n} & \frak{i}^{\ell} & 0\\
\frak{i}^{k}& 0 & 0 & \frak{i}^{\frac{j+k}{2}+\epsilon}
\end{smallmatrix}\right]$, where $\epsilon=\pm 1$.
Let $r =j+m$.
We apply a holographic transformation
defined by
$\left[\begin{smallmatrix}
 1 & 0\\
 0 & \gamma
\end{smallmatrix}\right]$, where $\gamma^2=\frak{i}^{\frac{j+k}{2}+r+\epsilon}$, then we get
$
\operatorname{Holant}(\neq_2|f)\equiv^p_{T}\operatorname{Holant}(\neq_2|\widehat{f}),
$
where $\widehat{f}=\left[\begin{smallmatrix}
 1 & 0\\
 0 & \gamma
\end{smallmatrix}\right]^{\otimes 4}f$, and its signature matrix is
$M(\widehat{f})=\frak{i}^{\frac{j+k}{2}+\epsilon}\left[\begin{smallmatrix}
1 & 0 & 0 & \frak{i}^{j+r}\\
0 & \frak i^{r} & \frak{i}^{m+r} & 0\\
0 & \frak{i}^{n+r} & \frak{i}^{\ell+r} & 0\\
\frak{i}^{k+r}& 0 & 0 & -\frak{i}^{j+k+2r}
\end{smallmatrix}\right]$.
%%% that - in -\frak{i}^{j+k+2r} comes from i^{2 epsilin}
By $\ell \equiv m+n+ 2 \pmod 4$, we have
% $\ell+r+1\equiv r+m+n+3 \pmod 4$.
%Let $k\equiv n+s\pmod 4$, where $s\in\{0, 1, 2, 3\}$.
%Then
\[M(\widehat{f})=\frak{i}^{\frac{j+k}{2}+\epsilon}\left[\begin{smallmatrix}
1 & 0 & 0 & \frak{i}^{j+r}\\
0 & \frak i^{r} & \frak{i}^{m+r} & 0\\
0 & \frak{i}^{n+r} & -\frak{i}^{m+n+r} & 0\\
\frak{i}^{k+r}& 0 & 0 & -\frak{i}^{j+k+2r}
\end{smallmatrix}\right].\]
This function is an affine function; indeed
 let
\[Q(x_1, x_2, x_3)=(k-n-r)x_1x_2+(2j+2)x_1x_3+(2j)x_2x_3+(n+r)x_1+rx_2+(j+r)x_3,\]
then
 $\widehat{f}(x_1, x_2, x_3, x_4)=\frak{i}^{\frac{j+k}{2}+\epsilon}\cdot
\frak{i}^{Q(x_1, x_2, x_3)}$ on the support
of $\widehat{f}$: $x_1 + x_2 + x_3 + x_4 \equiv 0 \pmod 2$.
 Moreover, $k-n-r \equiv j + k + n + m \equiv 0 \pmod 2$,
all cross terms have even coefficients.
% we have $s-r\equiv k+n+j+m\pmod 2$ by $j+k\equiv m+n\pmod 2$ and
% $m-j-r+2\equiv -2j+2\pmod 2$.
% This implies
% that the coefficients of the cross terms of $Q$ are even.
Thus $\widehat{f}\in\mathscr{A}$.
\end{proof}

Finally we are ready to prove the main theorem: Theorem~\ref{main:theorem}.
%\begin{theorem}\label{main:theorem}
%Let $f$ be a 4-ary signature with the signature matrix
%$M(f)=\left[\begin{smallmatrix}
%a & 0 & 0 & b\\
%0 & c & d & 0\\
%0 & w & z & 0\\
% & 0 & 0 & x
%%If  $ax=0$, then  $\operatorname{Holant}(\neq_2|f)$ is
%equivalent to the six-vertex model $\operatorname{Holant}(\neq_2|f')$
%where $f'$ is obtained from $f$ by setting $a=x=0$,  i.e.,
%$M(f')=\left[\begin{smallmatrix}
%0 & 0 & 0 & b\\
%0 & c & d & 0\\
%0 & w & z & 0\\
%y & 0 & 0 & 0
%\end{smallmatrix}\right]$. Explicitly,
%$\operatorname{Holant}(\neq_2|f)$
%is $\#\operatorname{P}$-hard except in the following cases:
%\begin{itemize}
%\item $f'\in\mathscr{P}$,
%\item $f'\in\mathscr{A}$,
%\item there is at least one zero in each pair $(b, y), (c, z), (d, w)$.
%end{itemize}
%If $ax\neq 0$, then $\operatorname{Holant}(\neq_2|f)$
%is $\#\operatorname{P}$-hard except in the following cases:
%\begin{itemize}
%\item $f$ is $\mathscr{P}$-transformable;
%\item $f$ is $\mathscr{A}$-transformable;
%\item $f$ is $\mathscr{L}$-transformable.
%\end{itemize}
%n all listed cases, $\operatorname{Holant}(\neq_2|f)$ is computable in polynomial time.
%\end{theorem}%%end of theorem

\begin{proof}
 By Lemma~\ref{normalize:a=x}, we can assume that $f$ has the signature matrix
$M(f)=\left[\begin{smallmatrix}
a & 0 & 0 & b\\
0 & c & d & 0\\
0 & w & z & 0\\
y & 0 & 0 & a
\end{smallmatrix}\right]$.

If $a=0$, this is the six-vertex model.
The theorem  follows from  Theorem~\ref{dichotomy-six-vertex}.

If there is at least one zero in $\{b, y, c, z, d, w\}$, then we are done by Theorem~\ref{at:least:one:zero}.
%Lemma~\ref{spin}, Lemma~\ref{4:nozero}, Lemma~\ref{4nonzero-twopair}, Lemma~\ref{no-nonzero-pair},
%Lemma~\ref{3:nonzero:onepair} and Lemma~\ref{5:nonzero}.

Below we assume there are no zeros in  $\{b, y, c, z, d, w\}$.
If $y=\epsilon b, z=\epsilon c, w=\epsilon d,$ where $\epsilon=\pm 1$, the theorem follows Lemma~\ref{just:one:binary}
and Lemma~\ref{even}.

If $by=cz=dw$, then we are done by Lemma~\ref{by=cz=dw}.

Otherwise, by the symmetry of the three inner pairs
$(b, y), (c, z), (d, w)$,
we can assume that $\left[\begin{smallmatrix}
c & d\\
w & z
\end{smallmatrix}\right]$ has full rank and $\left[\begin{smallmatrix}
a & 0 & 0 & b\\
0 & c & d & 0\\
0 & w & z & 0\\
y & 0 & 0 & a
\end{smallmatrix}\right]$ does not have the form
$\left[\begin{smallmatrix}
a & 0 & 0 & b\\
0 & c & d & 0\\
0 & \epsilon d & \epsilon c & 0\\
\epsilon b & 0 & 0 & a
\end{smallmatrix}\right]$ for any $\epsilon=\pm 1$.
By Lemma~\ref{construct:basic:binary} we have $(0, 1, t, 0)^T$, where $t
\neq  \pm 1$.
 %or $(0, 1, 0, 0)^T$.
If $t=0$, we have $(0, 1, 0, 0)^T$, then we are done by Theorem~\ref{with:(0,1,0,0)}.

Otherwise, we have $(0, 1, t, 0)^T$, where $t\neq 0$ and $t \not = \pm  1$.
If $t^i$ are distinct for $1\leq i\leq 5$, then we have $(0, 1, u, 0)^T$ for any $u\in\mathbb{C}$ by Lemma~\ref{construct-binary}.
Thus we have $(0, 1, 0, 0)^T$ by
letting $u=0$. Then we are done by Theorem~\ref{with:(0,1,0,0)} again.
Otherwise, $t$ is an $n$-th primitive root of unity with $n=3$ or 4.
Then we have $(0, 1, 0, 0)^T$ by
Corollary~\ref{construct:(0,1,0,0)}.
And we are done by Theorem~\ref{with:(0,1,0,0)} once again.
\end{proof}

\section{A Sample of Problems}
We illustrate the scope of Theorem~\ref{main:theorem} by several concrete problems.

\vspace{0.3in}

$\mathbf{Problems:}$ \#{\sc EO} on 4-Regular Graphs.

$\mathbf{Input:}$ A 4-regular graph $G$.

$\mathbf{Output:}$ The number of Eulerian orientations of $G$,
i.e., the number of orientations of $G$ such that at every vertex
the in-degree and out-degree are equal.

This problem can be expressed as Holant$(\neq_2|f)$, where $f$
has the signature matrix
$M(f)=\left[\begin{smallmatrix}
0 & 0 & 0 & 1\\
 0 & 1 & 1 & 0\\
 0 & 1 & 1 & 0\\
 1 & 0 & 0 & 0
 \end{smallmatrix}\right]$.
Huang and Lu proved that this problem is \#P-complete~\cite{huang-lu-2016}.
 Theorem~\ref{main:theorem} confirms this fact.

\vspace{0.3in}

$\mathbf{Problems:}$ $T(G; 3, 3)$.

$\mathbf{Input:}$ A graph $G$.

$\mathbf{Output:}$ The value of the Tutte polynomial $T(G; x, y)$  at $(3, 3)$.

Let $G_m$ be the medial graph of $G$, then $G_m$ is a 4-regular graph.
Las Vergnas proved the following theorem.
%\bibitem{tutte}Michel Las Vergnas: On the evaluation at (3, 3) of the Tutte polynomial of a graph. J. Comb. Theory, Ser. B 45(3): 367-372 (1988).
 \begin{theorem}\cite{tutte}
 Let $G$ be a connected graph and $\mathcal{EO}(G_m)$ be the set of all Eulerian Orientations
 of the medial graph $G_m$ of $G$. Then
 \[
 \sum_{O\in\mathcal{EO}(G_m)}2^{\beta(O)}=2T(G; 3, 3),
 \]
 where $\beta(O)$ is the number of saddle vertices in the orientation $O$, i.e., vertices in
 which the edges are oriented ``in, out, in, out'' in cyclic order.
 \end{theorem}
 Note that $\sum_{O\in\mathcal{EO}(G_m)}2^{\beta(O)}$
 can be expressed as Holant$(\neq_2|f)$, where $f$
has the signature matrix
$M(f)=\left[\begin{smallmatrix}
0 & 0 & 0 & 1\\
 0 & 1 & 2 & 0\\
 0 & 2 & 1 & 0\\
 1 & 0 & 0 & 0
 \end{smallmatrix}\right]$.
 Theorem~\ref{main:theorem} confirms that this problem is \#P-hard.

\vspace{0.3in}

$\mathbf{Problems:}$ Holant$(f)$, where $f$ has the signature matrix
$M(f)=
\left[\begin{smallmatrix}
 5 & \frak i & \frak i & -1\\
 \frak i & -1 & 3 & -\frak i\\
 \frak i & 3 & -1 & -\frak i\\
 -1 & -\frak i & -\frak i & 5
 \end{smallmatrix}\right]$.

$\mathbf{Input:}$ An instance of Holant$(f)$.

$\mathbf{Output:}$ The evaluation of this instance.

By the holographic transformation $Z=\frac{1}{\sqrt{2}}
\left[\begin{smallmatrix}
 1 & 1\\
 \mathfrak i & -\mathfrak i
 \end{smallmatrix}\right]$, we have
 \[
 \operatorname{Holant}(f)\equiv_{T} \operatorname{Holant}(\neq_2|\widehat{f}),
 \]
 where $M(\widehat{f})=
\left[\begin{smallmatrix}
 1 & 0 & 0 & 1\\
 0 & 1 & 0 & 0\\
 0 & 0 & 1 & 0\\
 1 & 0 & 0 & 0
 \end{smallmatrix}\right]$.
 By  Theorem~\ref{main:theorem},
 Holant$(f)$ can be computed in polynomial time.
 %Note that Holant$(f)$ is \#P-hard without the planar restriction.
It can be shown that both $f$ and $\widehat{f}$ are neither in
$\mathscr{P}$--transformable nor $\mathscr{A}$-transformable,
so this tractable case is not covered by previously known dichotomy results.

\setlength{\unitlength}{5mm}
\begin{picture}(24,15)(-6.5,-2.5)\label{figure:8-vertex}
\put(0,3){\line(1,0){12}}
\put(0,6){\line(1,0){12}}
\put(0,9){\line(1,0){12}}
\put(3,0){\line(0, 1){12}}
\put(6,0){\line(0, 1){12}}
\put(9,0){\line(0, 1){12}}
\put(3.8, 4.3){$\sigma_{i,j}$}
\put(6.3, 4.3){$\sigma_{i,j+1}$}
\put(3.8, 7.5){$\sigma_{i+1,j}$}
\put(6.3, 7.5){$\sigma_{i+1,j+1}$}
%%%%%%%%%%%%%%%%%%%%%%%%%%%%%%%%%%%%
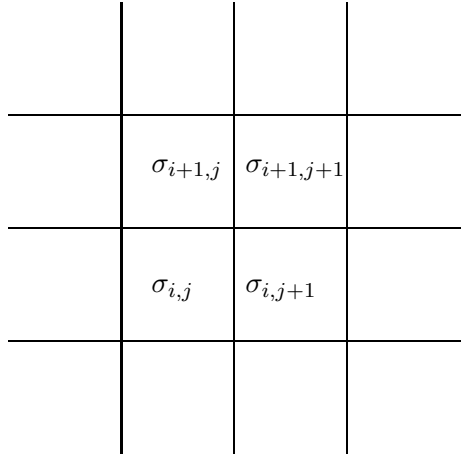
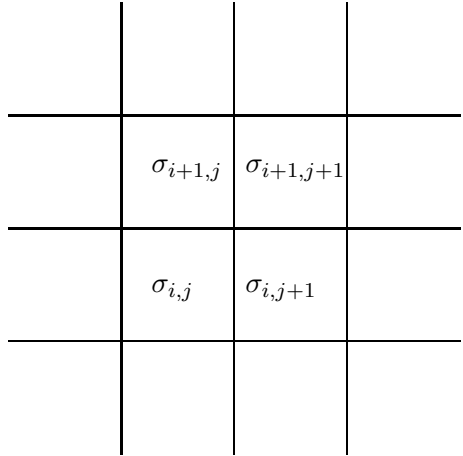
\put(-3.4,-1.5){Connection between the eight-vertex model and 2,4-spin Ising.}
%\caption{from vertex to spin}
\end{picture}

\input{2-4-spin-Ising.tex}

\end{document}